\documentclass[12pt]{article}

  \usepackage{graphicx}
  \usepackage{epsfig}
  \usepackage{amssymb}
  \usepackage{subfigure}
    \usepackage{feynmf}
\evensidemargin - 1.truecm
\begin{document}

\newcommand{\be}{\begin{equation}}
\newcommand{\ee}{\end{equation}}
\newcommand{\nn}{\nonumber}
\newcommand{\bea}{\begin{eqnarray}}
\newcommand{\eea}{\end{eqnarray}}
\newcommand{\bfig}{\begin{figure}}
\newcommand{\efig}{\end{figure}}
\newcommand{\bc}{\begin{center}}
\newcommand{\ec}{\end{center}}
\newcommand{\bd}{\begin{displaymath}}
\newcommand{\ed}{\end{displaymath}}


\begin{fmffile}{2MffaUNO}

\begin{titlepage}
\nopagebreak
{\flushright{
        \begin{minipage}{5cm}
	Rome1-1367/03\\
        Freiburg-THEP 04/02\\
        {\tt hep-ph/0401193}\\
        \end{minipage}        }

}
\renewcommand{\thefootnote}{\fnsymbol{footnote}}
\vspace*{0.6cm} 
\begin{center}
\boldmath
{\Large\bf Master integrals with 2 and 3 massive \\[1mm]
propagators for the 2-loop electroweak \\[3mm] 
form factor --- planar case}\unboldmath
\vskip 1.cm
{\large U. Aglietti\footnote{Email: Ugo.Aglietti@roma1.infn.it}},
\vskip .2cm
{\it Dipartimento di Fisica Universit\`a di Roma 
``La Sapienza" and INFN \\ 
Sezione di Roma, 
I-00185 Roma, Italy} 
\vskip .2cm
{\large  R. Bonciani\footnote{Email:
Roberto.Bonciani@physik.uni-freiburg.de}
\footnote{This work was supported by the European Union under
contract HPRN-CT-2000-00149},}
\vskip .2cm
{\it Physikalisches Institut,
Albert-Ludwigs-Universit\"at
Freiburg, \\ 
D-79104 Freiburg, Germany} 
\end{center}
\vskip 0.8cm

\begin{abstract}
\noindent
We compute the master integrals containing 2 and 3 massive propagators 
entering the planar amplitudes of the 2-loop electroweak form factor. 
The masses of the $W$, $Z$ and Higgs bosons are assumed to be degenerate.
This work is a continuation of our previous evaluation of master 
integrals containing at most 1 massive propagator. The $1/\epsilon$ 
poles and the finite parts are computed exactly in terms of a {\it new} 
class of 1-dimensional harmonic polylogarithms of the variable $x$, with
$\epsilon=2-D/2$ and $D$ the space-time dimension. 
Since thresholds and pseudothresholds in $s=\pm 4m^2$ do appear 
in addition to the old ones in $s=0,\pm m^2$,
an extension of the basis function set involving 
complex constants and radicals is 
introduced, together with a set of recursion equations to reduce 
integrals with semi-integer powers. It is shown that the basic 
properties of the harmonic polylogarithms are maintained by the 
generalization. We derive small-momentum expansions $|s| \ll m^2$
of all the 6-denominator amplitudes. We also present large momentum 
expansions $|s| \gg m^2$ of all the 6-denominator amplitudes which can 
be represented in terms of ordinary harmonic polylogarithms. Comparison 
with previous results in the literature is performed finding complete 
agreement.


\end{abstract}
\vfill
\end{titlepage}

\section{Introduction \label{Intro}}

We present in this paper the computation of the
master integrals containing 2 and 3
massive propagators entering the planar amplitudes
of the 2-loop electroweak form factor.
This work is the natural continuation of our previous
evaluation of master integrals containing at most
1 massive propagator.
The model process we consider is: 
\be
f(p_{1}) + \bar{f}(p_{2})  \rightarrow X(q),
\label{basic}
\ee  
where $f\bar{f}$ is an on-shell massless fermion pair,
$p_{1}^{2}=p_{2}^{2} = 0$, and $X$ is a singlet under the 
electroweak gauge group $SU(2)_{L} \times U(1)_{Y}$.

At the 2-loop level, the annihilation in Eq.~(\ref{basic}) involves the 
emission of 2 virtual bosons among $\gamma$, $W$, $Z$ and $H$'s.
The cases of emission of $(i)$ 2 photons and $(ii)$ 1 photon
and 1 massive boson, have already been treated in our previous
work \cite{UgoRo}. Since the above amplitudes have only thresholds 
in $s=0,m^2$ and pseudothreshold in $s=-m^2$, we succeeded to express 
them in terms of 1-dimensional harmonic polylogarithms 
\cite{Polylog,Polylog3} with maximum weight up to 4 included. 

In this paper we consider the amplitudes with 2 or 3 bosons of 
mass $m$ exchanged: the masses of $W$, $Z$ and Higgs bosons are 
assumed to be degenerate, $m_W\approx m_Z\approx m_H \approx m.$\footnote{
Small corrections of order $g^4 (m_{Z,H}^2-m_W^2)^n$ can be included by
means of expansions of the denominators of the form
\be
\frac{1}{k^2+m_{Z,H}^2}=\frac{1}{k^2+m_W^2}-\frac{m_{Z,H}^2-m_W^2}{(k^2+m_W^2)^2}+\cdots.
\ee
The amplitudes with powers of the $W$ denominators can be reduced to the
master integrals by means of the integration-by-parts identities (see Section 2).}
In general, these amplitudes have thresholds in $s=0,m^2,4m^2$ and 
pseudothresholds in $s=-m^2,-4m^2$.
As a consequence of this more complicated structure, the basis function set
of the harmonic polylogarithms considered in \cite{Polylog} is no more sufficient 
and a generalization is presented.
New basis functions such as $1/(4 \pm x)$, $1/\sqrt{x(4 \pm x)}$, etc. are introduced 
and recursion relations to reduce integrals with semi-integer powers coming
from the evaluation of the master integrals are derived. 
The basic properties of the harmonic polylogarithms, such as the uniqueness of 
representation as repeated integration, the algebra structure, the closure under
the inverse transformation $x\rightarrow 1/x$, etc. are all maintained.

We use dimensional regularization \cite{DimReg} to regulate both ultraviolet and 
infrared divergences, which then appear as poles in $1/\epsilon$, with
$\epsilon=2-D/2$ and $D$ the space-time dimension.
The ultraviolet poles are related to coupling constant renormalization and
are subtracted with  ordinary renormalization  prescriptions, 
such as for example the $\overline{MS}$ scheme.
The infrared poles are not physical, and in the physical cross-section, 
are canceled by the corresponding ones appearing in the real 
photon emission contributions or are factorized into QED 
structure functions.
The use of dimensional regularization as double 
regulator for both infrared and ultraviolet divergences makes
a priori impossible to trace the dynamical origin of the 
$1/\epsilon$ poles. This is however possible
by using power counting estimates for the ultraviolet and the infrared regions
together with some physical intuition (see sec.~(\ref{Results})).
In the ultraviolet region one can usually neglect masses and external momenta
while in the infrared region massive lines can be shrunk to a point.
Furthermore leading $1/\epsilon$ poles often originate from ordered regions such as
$k_1^2\ll k_2^2$, with $k_1$ and $k_2$ the loop momenta.
This (rather qualitative) analysis offers some checks of the results. 

The paper is organized as follows.

In Section~\ref{compff} we outline the strategy for the exact analytical 
evaluation of the 2-loop electro-weak form factor.
The presentation is rather sketchy and we refer to our previous work \cite{UgoRo} 
for a detailed discussion of the various steps of the computation.

In Section~\ref{HPLs} we discuss the properties of the harmonic polylogarithms. 
An overview of the ordinary harmonic polylogarithms is presented, in order 
to introduce later the generalization necessary to represent the master
integrals.
As already anticipated, the basic tool is a set of recursion equations
to reduce the integrals coming from the evaluation of the master integrals
to a unique form, specified by the choice of the basis functions.

Section~\ref{Results} contains the main results of our work. We present 
exact analytical results for the 21 non-trivial master integrals
containing 2 and 3 massive denominators. All the master integrals are 
represented in terms of generalized harmonic polylogarithms.

In Section~\ref{red6den} we give the results for the reducible 
6-denominator diagrams. The latter are 3 amplitudes of vertex-insertion 
type (see later), which are interesting by themselves or for reference use.

In Section~\ref{PiccoliP} the small momentum expansion $|s|\ll m^2$
of all the 6-denominator amplitudes is presented. We compare our results
with those present in the literature finding complete agreement.

In Section~\ref{GrandiP} we compute the large momentum expansion 
$|s|\gg m^2$ of all the 6-denominator amplitudes which can be 
represented in terms of ordinary harmonic polylogarithms. 
We could not give the
large momentum expansion of the amplitudes containing generalized
harmonic polylogarithms $H(\vec{w};x)$,
because the expansion of the latter for $|x|\gg 1$ is still an
open problem.

In Section~\ref{concl} we draw our conclusions and we discuss the open 
problems in the computation of the master integrals.

In order to make the paper as clear as possible, we added 4
appendices.

In Appendix \ref{app1} we report the 1-loop master integrals 
with 2 massive propagators which occur in our computation.
They are written in terms of generalized harmonic polylogarithms, 
like the 2-loop amplitudes.

In Appendix \ref{app2} we present the 4 factorized 2-loop master integrals,
which are the product of one-loop master integrals. Their derivation is rather
trivial and we give them for completeness.

In Appendix \ref{app3} we compute some interesting 2-loop amplitudes 
containing up to 5 denominators included which can be reduced to simpler
topologies by means of the integration-by-parts identities.

Finally, in Appendix \ref{app4} we give the
reducible 6-denominator scalar amplitudes of self-energy 
insertion type.

\bfig
\bc
\subfigure[]{
\begin{fmfgraph*}(35,35)
\fmfleft{i1,i2}
\fmfright{o}
\fmf{photon}{i1,v1}
\fmf{photon}{i2,v2}
\fmf{fermion}{v5,o}
\fmflabel{$p_{2}$}{i1}
\fmflabel{$p_{1}$}{i2}
\fmfv{l=$p_{1} \! - \! k_{1} \! + \! k_{2}$,l.a=15,l.d=.1w}{v3}
\fmfv{l=$p_{2} \! + \! k_{1} \! - \! k_{2}$,l.a=-15,l.d=.1w}{v4}
\fmf{photon,tension=.3,label=$p_{1} \! - \! k_{1}$,
                       label.side=left}{v2,v3}
\fmf{photon,tension=.3}{v3,v5}
\fmf{photon,tension=.3,label=$p_{2} \! + \! k_{1}$,
                       label.side=right}{v1,v4}
\fmf{photon,tension=.3}{v4,v5}
\fmf{fermion,tension=0,label=$k_{1}$,label.side=right}{v2,v1}
\fmf{fermion,tension=0,label=$k_{2}$,label.side=left}{v4,v3}
\end{fmfgraph*} }
%
%
\hspace{4mm}
\subfigure[]{
\begin{fmfgraph*}(35,35)
\fmfleft{i1,i2}
\fmfright{o}
\fmf{photon}{i1,v1}
\fmf{photon}{i2,v2}
\fmf{fermion}{v5,o}
\fmflabel{$p_{2}$}{i1}
\fmflabel{$p_{1}$}{i2}
\fmfv{l=$p_{1} \! - \! k_{1} \! + \! k_{2}$,l.a=15,l.d=.1w}{v3}
\fmfv{l=$p_{2} \! + \! k_{1} \! - \! k_{2}$,l.a=-15,l.d=.1w}{v4}
\fmf{fermion,tension=.3,label=$p_{1} \! - \! k_{1}$,
                       label.side=left}{v2,v3}
\fmf{photon,tension=.3}{v3,v5}
\fmf{fermion,tension=.3,label=$p_{2} \! + \! k_{1}$,
                       label.side=right}{v1,v4}
\fmf{photon,tension=.3}{v4,v5}
\fmf{photon,tension=0,label=$k_{1}$,label.side=right}{v2,v1}
\fmf{photon,tension=0,label=$k_{2}$,label.side=left}{v4,v3}
\end{fmfgraph*} }
%
%
\hspace{4mm}
\subfigure[]{
\begin{fmfgraph*}(35,35)
\fmfleft{i1,i2}
\fmfright{o}
\fmfforce{0.2w,0.9h}{v2}
\fmfforce{0.2w,0.1h}{v1}
\fmfforce{0.2w,0.5h}{v3}
\fmfforce{0.8w,0.5h}{v5}
\fmf{photon}{i1,v1}
\fmf{photon}{i2,v2}
\fmf{fermion}{v5,o}
\fmflabel{$p_{2}$}{i1}
\fmflabel{$p_{1}$}{i2}
\fmfv{l=$k_{1} \! + \! k_{2}$,l.a=10,l.d=.1w}{v3}
\fmf{photon,tension=0,label=$p_{1} \! - \! k_{1}$,
                      label.side=left}{v2,v5}
\fmf{photon,tension=0}{v3,v4}
\fmf{fermion,tension=.4,label=$p_{2} \! - \! k_{2}$,
                       label.side=right}{v1,v4}
\fmf{photon,tension=.4,label=$p_{2} \! + \! k_{1}$,
                       label.side=right}{v4,v5}
\fmf{photon,tension=0,label=$k_{2}$,label.side=left}{v1,v3}
\fmf{fermion,tension=0,label=$k_{1}$,label.side=right}{v2,v3}
\end{fmfgraph*} } \\
%
%
\hspace{4mm}
\subfigure[]{
\begin{fmfgraph*}(35,35)
\fmfleft{i1,i2}
\fmfright{o}
\fmfforce{0.2w,0.9h}{v2}
\fmfforce{0.2w,0.1h}{v1}
\fmfforce{0.2w,0.5h}{v3}
\fmfforce{0.8w,0.5h}{v5}
\fmf{photon}{i1,v1}
\fmf{photon}{i2,v2}
\fmf{fermion}{v5,o}
\fmflabel{$p_{2}$}{i1}
\fmflabel{$p_{1}$}{i2}
\fmfv{l=$k_{1} \! + \! k_{2}$,l.a=10,l.d=.1w}{v3}
\fmf{photon,tension=0,label=$p_{1} \! - \! k_{1}$,
                      label.side=left}{v2,v5}
\fmf{photon,tension=0}{v3,v4}
\fmf{photon,tension=.4,label=$p_{2} \! - \! k_{2}$,
                       label.side=right}{v1,v4}
\fmf{photon,tension=.4,label=$p_{2} \! + \! k_{1}$,
                       label.side=right}{v4,v5}
\fmf{fermion,tension=0,label=$k_{2}$,label.side=left}{v1,v3}
\fmf{fermion,tension=0,label=$k_{1}$,label.side=right}{v2,v3}
\end{fmfgraph*} }
%
%
\hspace{4mm}
\subfigure[]{
\begin{fmfgraph*}(35,35)
\fmfleft{i1,i2}
\fmfright{o}
\fmfforce{0.2w,0.9h}{v2}
\fmfforce{0.2w,0.1h}{v1}
\fmfforce{0.2w,0.5h}{v3}
\fmfforce{0.8w,0.5h}{v5}
\fmf{photon}{i1,v1}
\fmf{photon}{i2,v2}
\fmf{fermion}{v5,o}
\fmflabel{$p_{2}$}{i1}
\fmflabel{$p_{1}$}{i2}
\fmfv{l=$k_{1} \! + \! k_{2}$,l.a=10,l.d=.1w}{v3}
\fmf{photon,tension=0,label=$p_{1} \! - \! k_{1}$,
                      label.side=left}{v2,v5}
\fmf{fermion,tension=0}{v3,v4}
\fmf{photon,tension=.4,label=$p_{2} \! - \! k_{2}$,
                       label.side=right}{v1,v4}
\fmf{photon,tension=.4,label=$p_{2} \! + \! k_{1}$,
                       label.side=right}{v4,v5}
\fmf{photon,tension=0,label=$k_{2}$,label.side=left}{v1,v3}
\fmf{fermion,tension=0,label=$k_{1}$,label.side=right}{v2,v3}
\end{fmfgraph*} }
%
%
\subfigure[]{
\begin{fmfgraph*}(35,35)
\fmfleft{i1,i2}
\fmfright{o}
\fmfforce{0.2w,0.9h}{v2}
\fmfforce{0.2w,0.1h}{v1}
\fmfforce{0.2w,0.5h}{v3}
\fmfforce{0.8w,0.5h}{v5}
\fmf{photon}{i1,v1}
\fmf{photon}{i2,v2}
\fmf{fermion}{v5,o}
\fmflabel{$p_{2}$}{i1}
\fmflabel{$p_{1}$}{i2}
\fmfv{l=$k_{1} \! + \! k_{2}$,l.a=10,l.d=.1w}{v3}
\fmf{photon,tension=0,label=$p_{1} \! - \! k_{1}$,
                      label.side=left}{v2,v5}
\fmf{fermion,tension=0}{v3,v4}
\fmf{photon,tension=.4,label=$p_{2} \! - \! k_{2}$,
                       label.side=right}{v1,v4}
\fmf{photon,tension=.4,label=$p_{2} \! + \! k_{1}$,
                       label.side=right}{v4,v5}
\fmf{fermion,tension=0,label=$k_{2}$,label.side=left}{v1,v3}
\fmf{photon,tension=0,label=$k_{1}$,label.side=right}{v2,v3}
\end{fmfgraph*} } \\
%
%
\hspace{4mm}
\subfigure[]{
\begin{fmfgraph*}(35,35)
\fmfleft{i1,i2}
\fmfright{o}
\fmfforce{0.2w,0.9h}{v2}
\fmfforce{0.2w,0.1h}{v1}
\fmfforce{0.2w,0.5h}{v3}
\fmfforce{0.8w,0.5h}{v5}
\fmf{photon}{i1,v1}
\fmf{photon}{i2,v2}
\fmf{fermion}{v5,o}
\fmflabel{$p_{2}$}{i1}
\fmflabel{$p_{1}$}{i2}
\fmfv{l=$k_{1} \! + \! k_{2}$,l.a=10,l.d=.1w}{v3}
\fmf{photon,tension=0,label=$p_{1} \! - \! k_{1}$,
                      label.side=left}{v2,v5}
\fmf{fermion,tension=0}{v3,v4}
\fmf{photon,tension=.4,label=$p_{2} \! - \! k_{2}$,
                       label.side=right}{v1,v4}
\fmf{photon,tension=.4,label=$p_{2} \! + \! k_{1}$,
                       label.side=right}{v4,v5}
\fmf{fermion,tension=0,label=$k_{2}$,label.side=left}{v1,v3}
\fmf{fermion,tension=0,label=$k_{1}$,label.side=right}{v2,v3}
\end{fmfgraph*} }
%
\vspace*{8mm}
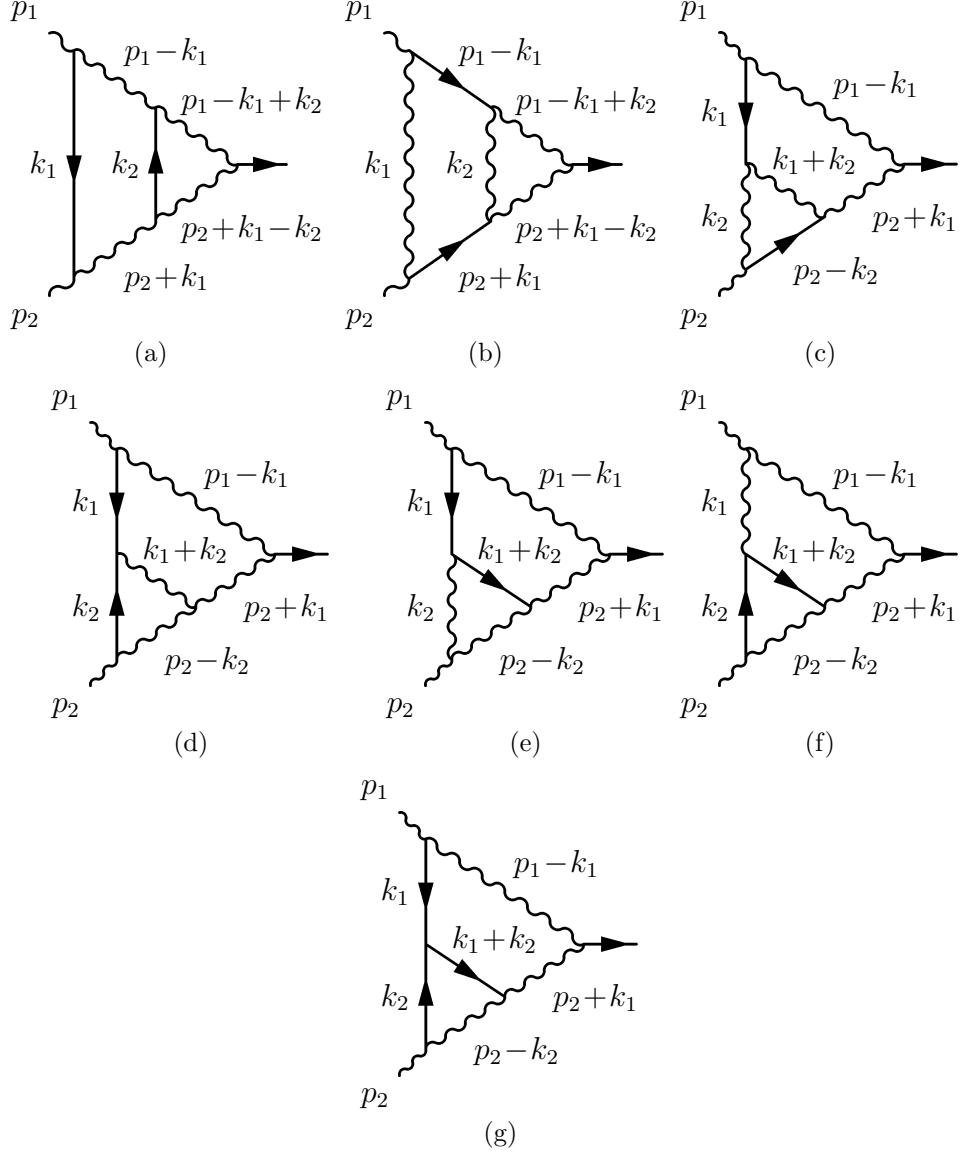
\caption{\label{fig1} Vertex-correction diagrams with 2 and 3 massive
propagators. The topologies related to these diagrams are {\it real}
6-denominator topologies (see text). The graphical conventions are the same
as in our previous paper \cite{UgoRo}.}
\ec
\efig

\bfig
\bc
\subfigure[]{
\begin{fmfgraph*}(35,35)
\fmfleft{i1,i2}
\fmfright{o}
\fmfforce{0.2w,0.93h}{v2}
\fmfforce{0.2w,0.07h}{v1}
\fmfforce{0.8w,0.5h}{v5}
\fmfforce{0.2w,0.4h}{v9}
\fmfforce{0.5w,0.45h}{v10}
\fmfforce{0.2w,0.5h}{v11}
\fmf{photon}{i1,v1}
\fmf{photon}{i2,v2}
\fmf{fermion}{v5,o}
\fmflabel{$p_{2}$}{i1}
\fmflabel{$p_{1}$}{i2}
\fmfv{l=$k_{2}$,l.a=180,l.d=0.05w}{v10}
\fmfv{l=$k_{1}$,l.a=180,l.d=.06w}{v11}
\fmf{photon,label=$p_{1} \! - \! k_{1}$,label.side=left}{v2,v3}
\fmf{fermion,tension=.25,right}{v3,v4}
\fmf{photon,tension=.25,label=$p_{1} \! - \! k_{1} \! - \! k_{2}
$,label.side=left}{v3,v4}
\fmf{photon,label=$p_{1} \! - \! k_{1}$,label.side=left}{v4,v5}
\fmf{photon,label=$p_{2} \! + \! k_{1}$,label.side=right}{v1,v5}
\fmf{fermion}{v2,v1}
\end{fmfgraph*} }
%
%
\hspace{6mm}
\subfigure[]{
\begin{fmfgraph*}(35,35)
\fmfleft{i1,i2}
\fmfright{o}
\fmfforce{0.2w,0.93h}{v2}
\fmfforce{0.2w,0.07h}{v1}
\fmfforce{0.2w,0.3h}{v3}
\fmfforce{0.2w,0.7h}{v4}
\fmfforce{0.8w,0.5h}{v5}
\fmf{photon}{i1,v1}
\fmf{photon}{i2,v2}
\fmf{fermion}{v5,o}
\fmflabel{$p_{2}$}{i1}
\fmflabel{$p_{1}$}{i2}
\fmf{photon,label=$p_{1} \! - \! k_{1}$,label.side=left}{v2,v5}
\fmf{photon,label=$k_{1}$,label.side=left}{v1,v3}
\fmf{photon,label=$k_{1}$,label.side=right}{v2,v4}
\fmf{photon,label=$p_{2} \! + \! k_{1}$,label.side=right}{v1,v5}
\fmf{fermion,right,label=$k_{1} \! \! + \! \! k_{2}$,label.side=left,
     l.d=0.03w}{v4,v3}
\fmf{fermion,right,label=$k_{2}$,label.side=right}{v3,v4}
\end{fmfgraph*} }
%
%
\hspace{6mm}
\subfigure[]{
\begin{fmfgraph*}(35,35)
\fmfleft{i1,i2}
\fmfright{o}
\fmfforce{0.2w,0.93h}{v2}
\fmfforce{0.2w,0.07h}{v1}
\fmfforce{0.2w,0.3h}{v3}
\fmfforce{0.2w,0.7h}{v4}
\fmfforce{0.8w,0.5h}{v5}
\fmf{photon}{i1,v1}
\fmf{photon}{i2,v2}
\fmf{fermion}{v5,o}
\fmflabel{$p_{2}$}{i1}
\fmflabel{$p_{1}$}{i2}
\fmf{photon,label=$p_{1} \! - \! k_{1}$,label.side=left}{v2,v5}
\fmf{fermion,label=$k_{1}$,label.side=right}{v3,v1}
\fmf{fermion,label=$k_{1}$,label.side=right}{v2,v4}
\fmf{photon,label=$p_{2} \! + \! k_{1}$,label.side=right}{v1,v5}
\fmf{photon,right,label=$k_{1} \! \! + \! \! k_{2}$,label.side=left,
     l.d=0.03w}{v4,v3}
\fmf{photon,right,label=$k_{2}$,label.side=right}{v3,v4}
\end{fmfgraph*} } \\
%
%
\subfigure[]{
\begin{fmfgraph*}(35,35)
\fmfleft{i1,i2}
\fmfright{o}
\fmfforce{0.2w,0.93h}{v2}
\fmfforce{0.2w,0.07h}{v1}
\fmfforce{0.2w,0.3h}{v3}
\fmfforce{0.2w,0.7h}{v4}
\fmfforce{0.8w,0.5h}{v5}
\fmf{photon}{i1,v1}
\fmf{photon}{i2,v2}
\fmf{fermion}{v5,o}
\fmflabel{$p_{2}$}{i1}
\fmflabel{$p_{1}$}{i2}
\fmf{photon,label=$p_{1} \! - \! k_{1}$,label.side=left}{v2,v5}
\fmf{fermion,label=$k_{1}$,label.side=right}{v3,v1}
\fmf{photon,label=$k_{1}$,label.side=right}{v2,v4}
\fmf{photon,label=$p_{2} \! + \! k_{1}$,label.side=right}{v1,v5}
\fmf{fermion,right,label=$k_{1} \! \! + \! \! k_{2}$,label.side=left,
     l.d=0.03w}{v4,v3}
\fmf{fermion,right,label=$k_{2}$,label.side=right}{v3,v4}
\end{fmfgraph*} }
%
%
\hspace{6mm}
\subfigure[]{
\begin{fmfgraph*}(35,35)
\fmfleft{i1,i2}
\fmfright{o}
\fmfforce{0.2w,0.93h}{v2}
\fmfforce{0.2w,0.07h}{v1}
\fmfforce{0.2w,0.3h}{v3}
\fmfforce{0.2w,0.7h}{v4}
\fmfforce{0.8w,0.5h}{v5}
\fmf{photon}{i1,v1}
\fmf{photon}{i2,v2}
\fmf{fermion}{v5,o}
\fmflabel{$p_{2}$}{i1}
\fmflabel{$p_{1}$}{i2}
\fmf{photon,label=$p_{1} \! - \! k_{1}$,label.side=left}{v2,v5}
\fmf{fermion,label=$k_{1}$,label.side=right}{v3,v1}
\fmf{fermion,label=$k_{1}$,label.side=right}{v2,v4}
\fmf{photon,label=$p_{2} \! + \! k_{1}$,label.side=right}{v1,v5}
\fmf{fermion,right,label=$k_{1} \! \! + \! \! k_{2}$,label.side=left,
     l.d=0.03w}{v4,v3}
\fmf{photon,right,label=$k_{2}$,label.side=right}{v3,v4}
\end{fmfgraph*} }
%
%
\hspace{6mm}
\subfigure[]{
\begin{fmfgraph*}(35,35)
\fmfleft{i1,i2}
\fmfright{o}
\fmfforce{0.2w,0.93h}{v2}
\fmfforce{0.2w,0.07h}{v1}
\fmfforce{0.2w,0.3h}{v3}
\fmfforce{0.2w,0.7h}{v4}
\fmfforce{0.8w,0.5h}{v5}
\fmf{photon}{i1,v1}
\fmf{photon}{i2,v2}
\fmf{fermion}{v5,o}
\fmflabel{$p_{2}$}{i1}
\fmflabel{$p_{1}$}{i2}
\fmf{photon,label=$p_{1} \! - \! k_{1}$,label.side=left}{v2,v5}
\fmf{fermion,label=$k_{1}$,label.side=right}{v3,v1}
\fmf{fermion,label=$k_{1}$,label.side=right}{v2,v4}
\fmf{photon,label=$p_{2} \! + \! k_{1}$,label.side=right}{v1,v5}
\fmf{fermion,right,label=$k_{1} \! \! + \! \! k_{2}$,label.side=left,
     l.d=0.03w}{v4,v3}
\fmf{fermion,right,label=$k_{2}$,label.side=right}{v3,v4}
\end{fmfgraph*} } \\
%
%
\subfigure[]{
\begin{fmfgraph*}(35,35)
\fmfleft{i1,i2}
\fmfright{o}
\fmfforce{0.2w,0.9h}{v2}
\fmfforce{0.2w,0.1h}{v1}
\fmfforce{0.2w,0.5h}{v3}
\fmfforce{0.8w,0.5h}{v5}
\fmf{photon}{i1,v1}
\fmf{photon}{i2,v2}
\fmf{fermion}{v5,o}
\fmflabel{$p_{2}$}{i1}
\fmflabel{$p_{1}$}{i2}
\fmf{photon,tension=0,label=$p_{1}-k_{1}$,label.side=left}{v2,v5}
\fmf{fermion,left=90,label=$k_{2}$,label.side=right}{v3,v3}
\fmf{photon,tension=0,label=$p_{2}+k_{1}$,label.side=right}{v1,v5}
\fmf{fermion,tension=0,label=$k_{1}$,label.side=right}{v3,v1}
\fmf{photon,tension=0,label=$k_{1}$,label.side=right}{v2,v3}
\end{fmfgraph*} }
%
%
\hspace{6mm}
\subfigure[]{
\begin{fmfgraph*}(35,35)
\fmfleft{i1,i2}
\fmfright{o}
\fmfforce{0.2w,0.9h}{v2}
\fmfforce{0.2w,0.1h}{v1}
\fmfforce{0.2w,0.5h}{v3}
\fmfforce{0.8w,0.5h}{v5}
\fmf{photon}{i1,v1}
\fmf{photon}{i2,v2}
\fmf{fermion}{v5,o}
\fmflabel{$p_{2}$}{i1}
\fmflabel{$p_{1}$}{i2}
\fmf{photon,tension=0,label=$p_{1}-k_{1}$,label.side=left}{v2,v5}
\fmf{fermion,left=90,label=$k_{2}$,label.side=right}{v3,v3}
\fmf{photon,tension=0,label=$p_{2}+k_{1}$,label.side=right}{v1,v5}
\fmf{fermion,tension=0,label=$k_{1}$,label.side=right}{v3,v1}
\fmf{fermion,tension=0,label=$k_{1}$,label.side=right}{v2,v3}
\end{fmfgraph*} }
%
\vspace*{8mm}
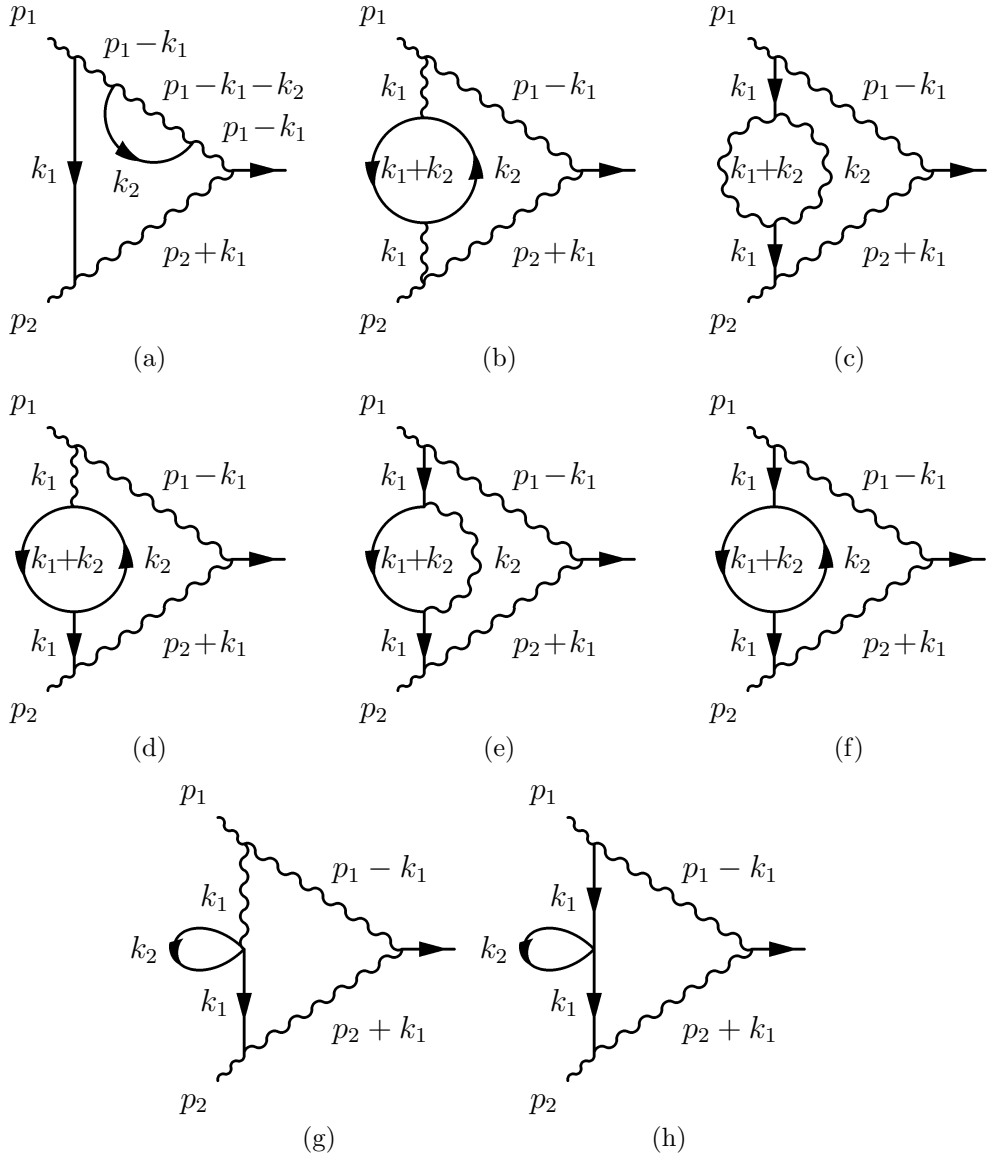
\caption{\label{fig1bis} Self-energy correction diagrams with 2 and 3 
massive propagators. The topologies related to these diagrams are 5 and
4 denominator topologies (see text).}
\ec
\efig


\section{The computation of the form factor \label{compff}}

The planar diagrams containing 2 and 3 massive propagators 
entering the 2-loop electroweak form factor are plotted in
Figs.~\ref{fig1} and \ref{fig1bis}. We have omitted 1-particle-reducible contributions,
related to field and mass renormalization of the external particles
$f$, $\overline{f}$ and $X$.
The amplitudes can be obtained, for example, by means of a Dyson-Scwhinger 
expansion of the  $\overline{f}fX$ vertex function \cite{itzzub}.
Let us look at the general structure of the diagrams:

\begin{itemize}
\item
1-loop: one has a triangle diagram with a vector particle, 
i.e. a $\gamma$, a $W$ or a $Z^0$, exchanged between the fermions;
\item
2-loops: the triangle diagram has its bare propagators 
and vertices replaced by the corresponding one-loop 
quantities\footnote{
We neglect in this paper the crossed ladder topology, 
which is a non-planar one.}.
\end{itemize}
According to the kind of correction,
the 2-loop diagrams are then naturally classified as:
\begin{enumerate} 
{\bf \item
Vertex-correction amplitudes} (see Fig.~\ref{fig1}). 

\noindent
They all involve 6 different denominators, the maximal number
of denominators for 2-loop 3-point functions.
According to the kind of vertex corrected, there are the following cases:
\begin{enumerate} 
\item
Ladder diagrams.

\noindent
The ``hard'' $\overline{f}fX$ vertex, where annihilation occurs, is corrected. 
These are the diagrams $(a)$ and $(b)$ in Fig.~\ref{fig1}:
\begin{itemize}
\item
diagram $(a)$ represents the exchange in the $t$ channel
of a pair of $W/Z$ particles between the fermions
and is a ``non-singlet'' contribution to the process;
\item
diagram $(b)$ represents the conversion 
of the initial fermion pair into a boson pair which convert
back into a fermion pair; fermions are emitted in the $t$-channel.
This diagram is a non-singlet contribution which actually
cannot be obtained as correction of the basic one-loop amplitude.
\end{itemize}

\item
Vertex-insertions.

\noindent
One of the 2 vertices involving the external fermions 
is corrected. There are 5 of such diagrams:
\begin{itemize}
\item
diagram $(c)$ containing an abelian correction with a $W$ or a $Z$ 
to the basic one-loop amplitude;
\item
diagram $(d)$ and $(e)$ representing respectively the emission by 
an intermediate boson
of a photon internally or externally to the triangle;
\item
diagram $(f)$ containing the conversion of a photon into an intermediate boson pair,
i.e. the process $\gamma^*\rightarrow W^*W^*$;
\item
diagram $(g)$ containing the splitting $Z^*\rightarrow W^*W^*$; 
this diagram is the only one having 3 massive propagators.
\end{itemize}
The vertex-insertion diagrams are on the same footing as the
ladders but are conventionally named in a different way.
\end{enumerate}
{\bf \item
Self-energy correction amplitudes} (see Fig.~\ref{fig1bis}).

\noindent
There are different cases, according to the kind of line 
corrected and the type of interaction.
Self-energy-corrections may indeed occur on:
\begin{enumerate}
\item 
a fermion line. 

\noindent
This is the case of diagram $(b)$ in Fig.~\ref{fig1bis}, with a bubble insertion. 
This amplitude has 5 different denominators, 
the one representing the corrected line being squared;
\item
a boson line. 

\noindent
This case is more complicated than the previous one because one can have
mixing and quartic interactions. 
Concerning the first point, one can have:
\begin{enumerate}
\item
a ``diagonal'' correction.

\noindent
By this we mean a correction to a $\gamma$, $W$ or $Z$ propagator, involving a
virtual fermion or boson pair. These are the cases $(b),(c),(e),(f)$ and $(h)$ in
Fig.~\ref{fig1bis};
\item
$\gamma-Z^0$ mixing, 

\noindent
By this we mean a contribution to the $\gamma-Z^0$ propagator. 
In this case one has 2 propagators with
the same momenta but with different masses, which can be
disentangled with partial fractioning \cite{UgoRo}.
These are the cases $(d)$ and $(g)$ in Fig.~\ref{fig1bis}.
\end{enumerate}
Let us now consider the second point.
Because of the presence of both cubic and quartic interactions, 
one can have:
\begin{enumerate}
\item
bubble insertion. 

\noindent
These are the cases $(a)-(f)$. There are 5 different denominators.
In the ``diagonal'' case, the denominator representing 
the corrected line is squared.
In the $\gamma-Z^0$ mixing case, the amplitude is a superposition 
of two 5-denominator amplitudes, one containing the photon
propagator and the other the $Z^0$ propagator;
\item
tadpole insertion. 

\noindent
These are the cases $(g)$ and $(h)$.
There are 4 different denominators.
In the ``diagonal'' case, the denominator representing 
the corrected line is squared.
In the $\gamma-Z^0$ mixing case, the amplitude is a superposition 
of two 4-denominator amplitudes, one containing the photon
propagator and the other the $Z^0$ propagator.
The structure of these amplitudes is very simple, as one loop is 
factorized into a momentum-independent expression.
\end{enumerate}
\end{enumerate}
\end{enumerate} 

The exact analytic computation of the 2-loop electro-weak 
form factor is a rather lengthy process, even though the general
method is rather clear.
As explained in detail in \cite{UgoRo}, it involves 2 basic steps:
\begin{enumerate}
\item
the reduction of the original amplitudes considered before 
--- obtained with usual Feynman rules --- to a few independent scalar integrals,
called master integrals (MIs);
\item
the analytic evaluation of all the master integrals generated with the previous 
step with some technique, such as Feynman parameters, dispersion relations, 
small and large momentum expansions or, in our case,
differential equations in the external kinematical invariants. 
\end{enumerate}

\subsection{The reduction to master integrals \label{Master}}

By reduction to master integrals we mean the reduction of all the 
Feynman diagrams to a minimal set of independent scalar integrals.
This process involves, in turn, the following three steps:
\begin{enumerate}
{\bf \item
Projection on invariant form factors.}

The amplitude associated to a given Feynman diagram is, in general, 
a tensor integral of the form:
\be\label{laprima}
{\mathcal F}_{\mu\nu i j\cdots}=
\int d^D k_1 d^D k_2 \frac{{\mathcal N}_{\mu\nu i j\ldots}(p_1,p_2,k_1,k_2)}
{D_{i_1}^{n_1}D_{i_2}^{n_2}D_{i_3}^{n_3}D_{i_4}^{n_4}D_{i_5}^{n_5}D_{i_6}^{n_6}},
\ee
where $\mu\nu i j\ldots$ denotes collectively spinor indices, four-vector 
indices and so on. For a scalar probe, with an interaction for example 
of the form $\overline{f}f X$, the diagram has 2 spinor indices, 
while for a vector probe, with an interaction for instance of 
the form  $\overline{f}\gamma_{\mu}f X^{\mu}$, 
the diagram has also a 4-vector index.
By usual form-factor decomposition of (\ref{laprima}), one can limit 
himself to consider scalar integrals of the form:
\be\label{laseconda}
{\mathcal S}_D=\int d^D k_1 d^D k_2 \frac{ N(S_1,S_2,S_3,S_4,S_5,S_6,S_7) }
{ D_{i_1}^{n_1} D_{i_2}^{n_2}  D_{i_3}^{n_3} D_{i_4}^{n_4} D_{i_5}^{n_5} D_{i_6}^{n_6} }
\ee
where
\bea
&& S_1=k_1^2,~~~~S_2=k_2^2,~~~~S_3=k_1\cdot p_1,~~~~S_4=k_1\cdot p_2,
\\ \nonumber
&& S_5=k_2\cdot p_1,~~~~S_6=k_2\cdot p_2,~~~~S_7=k_1\cdot k_2.
\eea
The function $N(S_1,S_2,S_3,S_4,S_5,S_6,S_7)$ is a polynomial in the kinematical invariants $S_i$.
{\bf \item
Rotation to independent scalar amplitudes.} 

The scalar amplitudes in Eq.~(\ref{laseconda}) are not linearly independent on each other as they
involve for instance 6 denominators and 7 scalar products, i.e. a total of 13 factors, while there are only
7 kinematical invariants depending on the loop momenta $k_1$ and $k_2$.
There are two different methods to eliminate the redundant amplitudes:
\begin{enumerate}
\item
Auxiliary diagram or auxiliary denominator scheme \cite{glover}.

\noindent
We express the scalar products in the numerator of (\ref{laseconda}) in terms 
of the original denominators $D_{i_j}~(j=1\ldots 6)$ and of an auxiliary denominator $D_{i_7}$ by
means of invertible relations of the form:
\be\label{DSP}
S_k=\sum_{j=1}^7 a_{kj} D_{i_j}
\ee
Our task is then shifted to evaluate independent scalar amplitudes
containing formally only denominators
\be
{\mathcal S}_I=\int d^Dk_1d^Dk_2
\frac{1}{D_{i_1}^{n_1}D_{i_2}^{n_2}D_{i_3}^{n_3}D_{i_4}^{n_4}D_{i_5}^{n_5}D_{i_6}^{n_6}D_{i_7}^{n_7}},
\ee
with $n_i\leq 1$ for $i\leq 6$ and $n_7\leq 0$.
Since the numerator $N$ is expressed by means of (\ref{DSP}) 
as a polynomial in the denominators, a term $D_i$ in the denominator can be
canceled by an equal term $D_i$ in the numerator. 
This means that line $i$ is shrunk to a point and the topology number $t$ of the diagram
is decremented by one\footnote{We define the topology number $t$ as the number of different denominators $D_i$ occurring in
a diagram with $n_i\geq 1$, irrespective of the powers of the scalar products \cite{UgoRo}.}. 
A given Feynman diagram then generates a pyramid of sub-diagrams corresponding to all the possible
contractions of $i\geq 1$ internal lines. All the possible resulting diagrams are plotted in 
Figs.~\ref{fig2}, \ref{fig3}, \ref{fig4} and \ref{fig5}.

Since the contracted propagators may be massless as well as massive,
an amplitude with for example 2 massive propagators 
will generally contain, in its decomposition, 
independent subamplitudes with 2,1 and 0 massive propagators.
The amplitudes computed in our previous work \cite{UgoRo}, containing
0 or 1 massive denominators, then, usually appear in the decomposition of the
present amplitudes;

\item
Shift scheme \cite{Lap,Rem3,RoPieRem1}. 

\noindent
No auxiliary denominator is introduced.
A routing of the amplitudes has to be initially assumed with one internal line carrying momentum $k_1$
and another internal line carrying momentum $k_2$. This way one can simplify powers of 
$k_1^2$ and $k_2^2$ in the numerator against denominators, by means of formulas like
$k_i^2/(k_i^2+a) = 1-a/(k_i^2+a)$ with $i=1,2$ and $a=0,m^2$. A set of cancellation rules between
scalar products in the numerator and denominators is established. An example of such
a cancellation is: $k_1\cdot p_1/(k_1^2+2k_1\cdot p_1) = 1/2-k_1^2/2/(k_1^2+2k_1\cdot p_1)$.
We then make all the possible scalar-product denominator cancellations.
After such cancellations, the amplitude may loose the denominators
with momenta $k_1$ and $k_2$, which where initially present by construction.
The latter are then reproduced with proper shifts of the loop
momenta, hence the name of the method. In general, with the shifts, new denominators are generated
and the table of scalar-product denominator cancellation rules has to be updated to include
also the new denominators. We go on through steps of cancellations and shifts till no more
cancellations are feasible. The amplitudes generated in the final step represent the 
independent scalar amplitudes. 
Since the total number of invariants is $7$, an independent amplitude with $t$ denominators
has in general $\overline{t}=7-t$ scalar products in the numerator.
As with the previous scheme, it is clear that sub-amplitudes
with all the possible contractions of the internal lines of the original (linearly dependent) 
amplitude do occur in the decomposition. 

\end{enumerate}

\noindent
The contractions of internal line(s) in an amplitude may lead to the following general simplifications:
\begin{enumerate}
\item
The internal line(s) between 2 external lines are contracted. 

\noindent
This implies that 2 external lines meet in a new effective vertex and
the 3-point function effectively simplifies to a 2-point function.
There are 2 possibilities:
\begin{itemize}
\item
contraction between external fermion lines.

\noindent
Since we are computing a 3-point function with 1 general external 
momentum $q=p_1+p_2$ and 2 light-cone momenta $p_1$ and $p_2$: $p_1^2=p_2^2=0$,
we obtain a 2-point function  with the general momentum 
$q$ flowing through it;
\item
contraction between an external fermion line and the probe line.

\noindent
The simplification is even greater in this case.
The resulting diagram depends only on a single light-cone momentum, 
i.e. on $p_1$ or on $p_2$, but not on both.
A single light-cone momentum is equivalent to a null
momentum, as is clearly seen with Wick rotation.
The resulting diagram is then effectively a vacuum 
amplitude, in which we can set to zero the external
light-cone momentum, i.e. $p_1\rightarrow 0$ or $p_2\rightarrow 0$. 
The above property is the basis of many relations between amplitudes;
\end{itemize}
\item
the internal line where loops overlap is contracted.

\noindent
The amplitude factorizes into the product of two 1-loop amplitudes,
whose computation is trivial.
\end{enumerate}

{\bf \item
Reduction of independent amplitudes to master integrals.}

It is possible to reduce the
independent amplitudes to a small subset of them by means of
identities obtained with integrations by parts \cite{Chet}.
\begin{itemize}
\item
In the auxiliary diagram scheme, these identities are obtained as:
\begin{equation}
0 = \int d^Dk_1 d^Dk_2 \frac{\partial}{\partial k_i^{\mu}}  \left\{
\frac{v^{\mu}}{D_{i_1}^{n_1}D_{i_2}^{n_2}D_{i_3}^{n_3}D_{i_4}^{n_4}D_{i_5}^{n_5}
D_{i_6}^{n_6}D_{i_7}^{n_7}} \right\} ,
\end{equation}
with $i=1,2$ and $v=k_1,k_2,p_1,p_2$. By explicitly performing the derivatives and re-expressing the
generated amplitudes in terms of independent ones as described before, 
one obtains relations among independent amplitudes with shifted indices $n_j\rightarrow n_j\pm 1$.
\item
In the shift scheme, the identities are obtained as:
\begin{equation}
0 = \int d^Dk_1 d^Dk_2 \frac{\partial}{\partial k_i^{\mu}} \left\{
\frac{v^{\mu} S_{i_1}^{l_1}\cdots S_{i_{7-t}}^{l_{7-t}} }{D_{i_1}^{n_1}
\cdots D_{i_t}^{n_t}} \right\} .
\end{equation}
\end{itemize}
Once the integration-by-parts identities and eventual symmetry relations 
\cite{Rem3,RoPieRem1} have been generated,
the next step is to combine them to achieve the complete reductions to the MIs.
At present, there are 3 techniques to solve the ibps:
\begin{enumerate}
\item
symbolic method \cite{Chet}.

\noindent
Historically, this has been the first method used, introduced together with the
ibps identities themselves.
The ibp identities are treated as formal recursion equations in the indices $n_j$ 
of the denominators (and of the scalar products) to take them to some
reference values such as typically $0,1,2,-1$;
\item
numerical-indices method \cite{Lap}.

\noindent
Explicit numerical values are replaced for the indices $n_j$.
In the auxiliary diagram scheme, one takes for the indices of the denominators
both positive and negative values: $n_j=\cdots -2,-1,0,1,2\cdots$, while for the
indices of the auxiliary denominators only negative values: 
$n_{j'}=0,-1,-2\cdots$.
In the shift scheme, one takes positive values for both the indices
of the denominators and the scalar products:
$n_j=0,1,2\cdots$, $l_j=0,1,2\cdots$.
In either scheme, a homogeneous system of linear equations is then
generated, whose unknowns are the amplitudes themselves.
The system is then solved with the method of elimination of variables, 
after having established the order in which amplitudes and equations 
have to be solved. The amplitudes which remain on the r.h.s., after all 
equations have been used, are the MIs for the specific hierarchy assumed;
\item
integral method \cite{Baikov}.

\noindent
An integral representation for the expansion coefficients of the 
independent amplitudes in terms of master integrals is derived.
This technique has not been used up to now, as far as we know, for original
computations but for general studies.
\end{enumerate}

The reduction through the ibps of a given independent amplitude involves MIs with subsets
of denominators of the starting amplitude. In other words, 
the reduction involves MIs in which $0,1,2\cdots$ internal lines
of the original amplitude have been shrunk to a point.
All the MIs appearing in our reductions are given in Fig.~\ref{fig6}.

\end{enumerate}

\bfig
\bc
\subfigure[]{
\begin{fmfgraph*}(25,25)
\fmfleft{i1,i2}
\fmfright{o}
\fmf{photon}{i1,v1}
\fmf{photon}{i2,v2}
\fmf{plain}{v5,o}
\fmf{photon,tension=.3}{v2,v3}
\fmf{photon,tension=.3}{v3,v5}
\fmf{photon,tension=.3}{v1,v4}
\fmf{photon,tension=.3}{v4,v5}
\fmf{plain,tension=0}{v2,v1}
\fmf{plain,tension=0}{v4,v3}
\end{fmfgraph*} }
%
%
\subfigure[]{
\begin{fmfgraph*}(25,25)
\fmfleft{i1,i2}
\fmfright{o}
\fmf{photon}{i1,v1}
\fmf{photon}{i2,v2}
\fmf{plain}{v5,o}
\fmf{plain,tension=.3}{v2,v3}
\fmf{photon,tension=.3}{v3,v5}
\fmf{plain,tension=.3}{v1,v4}
\fmf{photon,tension=.3}{v4,v5}
\fmf{photon,tension=0}{v2,v1}
\fmf{photon,tension=0}{v4,v3}
\end{fmfgraph*} }
%
%
\subfigure[]{
\begin{fmfgraph*}(25,25)
\fmfleft{i1,i2}
\fmfright{o}
\fmfforce{0.2w,0.9h}{v2}
\fmfforce{0.2w,0.1h}{v1}
\fmfforce{0.2w,0.5h}{v3}
\fmfforce{0.8w,0.5h}{v5}
\fmf{photon}{i1,v1}
\fmf{photon}{i2,v2}
\fmf{plain}{v5,o}
\fmf{photon,tension=0}{v2,v5}
\fmf{photon,tension=0}{v3,v4}
\fmf{plain,tension=.4}{v1,v4}
\fmf{photon,tension=.4}{v4,v5}
\fmf{photon,tension=0}{v1,v3}
\fmf{plain,tension=0}{v2,v3}
\end{fmfgraph*} }
%
%
\subfigure[]{
\begin{fmfgraph*}(25,25)
\fmfleft{i1,i2}
\fmfright{o}
\fmfforce{0.2w,0.9h}{v2}
\fmfforce{0.2w,0.1h}{v1}
\fmfforce{0.2w,0.5h}{v3}
\fmfforce{0.8w,0.5h}{v5}
\fmf{photon}{i1,v1}
\fmf{photon}{i2,v2}
\fmf{plain}{v5,o}
\fmf{photon,tension=0}{v2,v5}
\fmf{photon,tension=0}{v3,v4}
\fmf{photon,tension=.4}{v1,v4}
\fmf{photon,tension=.4}{v4,v5}
\fmf{plain,tension=0}{v1,v3}
\fmf{plain,tension=0}{v2,v3}
\end{fmfgraph*} } \\
%
%
\subfigure[]{
\begin{fmfgraph*}(25,25)
\fmfleft{i1,i2}
\fmfright{o}
\fmfforce{0.2w,0.9h}{v2}
\fmfforce{0.2w,0.1h}{v1}
\fmfforce{0.2w,0.5h}{v3}
\fmfforce{0.8w,0.5h}{v5}
\fmf{photon}{i1,v1}
\fmf{photon}{i2,v2}
\fmf{plain}{v5,o}
\fmf{photon,tension=0}{v2,v5}
\fmf{plain,tension=0}{v3,v4}
\fmf{photon,tension=.4}{v1,v4}
\fmf{photon,tension=.4}{v4,v5}
\fmf{photon,tension=0}{v1,v3}
\fmf{plain,tension=0}{v2,v3}
\end{fmfgraph*} }
%
%
\subfigure[]{
\begin{fmfgraph*}(25,25)
\fmfleft{i1,i2}
\fmfright{o}
\fmfforce{0.2w,0.9h}{v2}
\fmfforce{0.2w,0.1h}{v1}
\fmfforce{0.2w,0.5h}{v3}
\fmfforce{0.8w,0.5h}{v5}
\fmf{photon}{i1,v1}
\fmf{photon}{i2,v2}
\fmf{plain}{v5,o}
\fmf{photon,tension=0}{v2,v5}
\fmf{plain,tension=0}{v3,v4}
\fmf{photon,tension=.4}{v1,v4}
\fmf{photon,tension=.4}{v4,v5}
\fmf{plain,tension=0}{v1,v3}
\fmf{photon,tension=0}{v2,v3}
\end{fmfgraph*} }
%
%
\subfigure[]{
\begin{fmfgraph*}(25,25)
\fmfleft{i1,i2}
\fmfright{o}
\fmfforce{0.2w,0.9h}{v2}
\fmfforce{0.2w,0.1h}{v1}
\fmfforce{0.2w,0.5h}{v3}
\fmfforce{0.8w,0.5h}{v5}
\fmf{photon}{i1,v1}
\fmf{photon}{i2,v2}
\fmf{plain}{v5,o}
\fmf{photon,tension=0}{v2,v5}
\fmf{plain,tension=0}{v3,v4}
\fmf{photon,tension=.4}{v1,v4}
\fmf{photon,tension=.4}{v4,v5}
\fmf{plain,tension=0}{v1,v3}
\fmf{plain,tension=0}{v2,v3}
\end{fmfgraph*} }
%
\vspace*{8mm}
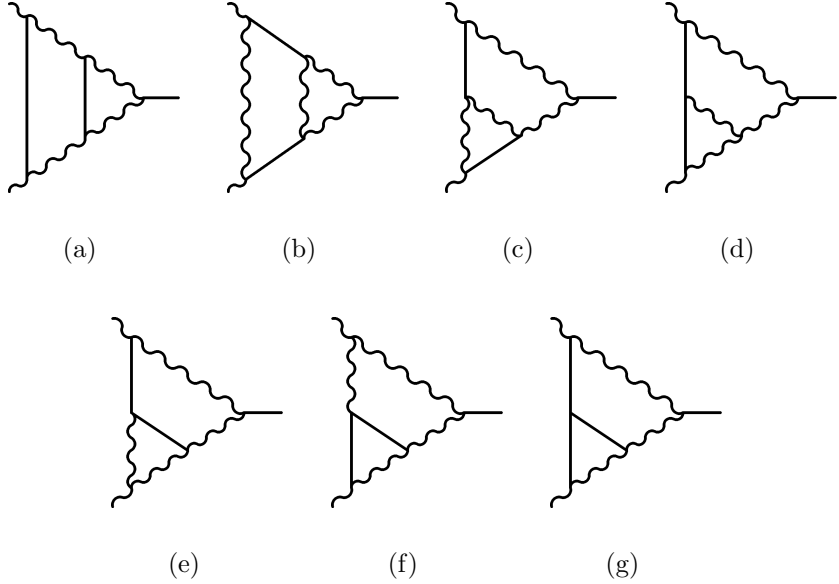
\caption{\label{fig2} The set of 7 independent 6-denominator diagrams, with 2 and 3 massive
propagators.}
\ec
\efig

\bfig
\bc
\subfigure[]{
\begin{fmfgraph*}(20,20)
\fmfleft{i1,i2}
\fmfright{o}
\fmfforce{0.8w,0.5h}{v4}
\fmf{photon}{i1,v1}
\fmf{photon}{i2,v2}
\fmf{plain}{v4,o}
\fmf{photon,tension=.15}{v2,v4}
\fmf{photon,tension=.4}{v1,v3}
\fmf{photon,tension=.2}{v3,v4}
\fmf{plain,tension=0}{v2,v1}
\fmf{plain,tension=0}{v2,v3}
\end{fmfgraph*} }  
%
%
\subfigure[]{
\begin{fmfgraph*}(20,20)
\fmfleft{i1,i2}
\fmfright{o}
\fmfforce{0.8w,0.5h}{v4}
\fmf{photon}{i1,v1}
\fmf{photon}{i2,v2}
\fmf{plain}{v4,o}
\fmf{photon,tension=.4}{v1,v3}
\fmf{photon,tension=.2}{v3,v4}
\fmf{photon,tension=.15}{v2,v4}
\fmf{plain,tension=0}{v2,v1}
\fmf{plain,tension=0,left=.5}{v3,v4}
\end{fmfgraph*} }
%
%
\subfigure[]{
\begin{fmfgraph*}(20,20)
\fmfleft{i1,i2}
\fmfright{o}
\fmfforce{0.8w,0.5h}{v4}
\fmf{photon}{i1,v1}
\fmf{photon}{i2,v2}
\fmf{plain}{v4,o}
\fmf{plain,tension=.4}{v1,v3}\fmf{photon,tension=.2}{v3,v4}
\fmf{plain,tension=.15}{v2,v4}
\fmf{photon,tension=0}{v2,v1}
\fmf{photon,tension=0,left=.5}{v3,v4}
\end{fmfgraph*} }
%
%
\subfigure[]{
\begin{fmfgraph*}(20,20)
\fmfleft{i1,i2}
\fmfright{o}
\fmfforce{0.2w,0.9h}{v2}
\fmfforce{0.2w,0.1h}{v1}
\fmfforce{0.2w,0.5h}{v3}
\fmfforce{0.8w,0.5h}{v4}
\fmf{photon}{i1,v1}
\fmf{photon}{i2,v2}
\fmf{plain}{v4,o}
\fmf{photon,tension=0}{v2,v3}
\fmf{photon,tension=0}{v3,v4}
\fmf{photon,tension=0}{v1,v4}
\fmf{plain,tension=0}{v2,v4}
\fmf{plain,tension=0}{v1,v3}
\end{fmfgraph*} }
%
%
\subfigure[]{
\begin{fmfgraph*}(20,20)
\fmfleft{i1,i2}
\fmfright{o}
\fmfforce{0.2w,0.9h}{v2}
\fmfforce{0.2w,0.1h}{v1}
\fmfforce{0.2w,0.55h}{v3}
\fmfforce{0.2w,0.15h}{v5}
\fmfforce{0.8w,0.5h}{v4}
\fmf{photon}{i1,v1}
\fmf{photon}{i2,v2}
\fmf{plain}{v4,o}
\fmf{plain}{v2,v3}
\fmf{photon,left}{v3,v5}
\fmf{plain,right}{v3,v5}
\fmf{photon}{v1,v4}
\fmf{photon}{v2,v4}
\end{fmfgraph*} } \\
%
%
\subfigure[]{
\begin{fmfgraph*}(20,20)
\fmfleft{i1,i2}
\fmfright{o}
\fmfforce{0.2w,0.9h}{v2}
\fmfforce{0.2w,0.1h}{v1}
\fmfforce{0.2w,0.55h}{v3}
\fmfforce{0.2w,0.15h}{v5}
\fmfforce{0.8w,0.5h}{v4}
\fmf{photon}{i1,v1}
\fmf{photon}{i2,v2}
\fmf{plain}{v4,o}
\fmf{photon}{v2,v3}
\fmf{plain,left}{v3,v5}
\fmf{plain,right}{v3,v5}
\fmf{photon}{v1,v4}
\fmf{photon}{v2,v4}
\end{fmfgraph*} }
%
%
\subfigure[]{
\begin{fmfgraph*}(20,20)
\fmfleft{i1,i2}
\fmfright{o}
\fmfforce{0.2w,0.9h}{v2}
\fmfforce{0.2w,0.1h}{v1}
\fmfforce{0.2w,0.55h}{v3}
\fmfforce{0.2w,0.15h}{v5}
\fmfforce{0.8w,0.5h}{v4}
\fmf{photon}{i1,v1}
\fmf{photon}{i2,v2}
\fmf{plain}{v4,o}
\fmf{plain}{v2,v3}
\fmf{plain,left}{v3,v5}
\fmf{plain,right}{v3,v5}
\fmf{photon}{v1,v4}
\fmf{photon}{v2,v4}
\end{fmfgraph*} } 
%
%
\subfigure[]{
\begin{fmfgraph*}(20,20)
\fmfleft{i1,i2}
\fmfright{o}
\fmfforce{0.2w,0.9h}{v2}
\fmfforce{0.2w,0.1h}{v1}
\fmfforce{0.2w,0.5h}{v3}
\fmfforce{0.8w,0.5h}{v4}
\fmf{photon}{i1,v1}
\fmf{photon}{i2,v2}
\fmf{plain}{v4,o}
\fmf{plain,tension=0}{v1,v3}
\fmf{photon,tension=0}{v3,v4}
\fmf{photon,tension=0}{v2,v4}
\fmf{plain,tension=0}{v2,v3}
\fmf{photon,tension=0}{v1,v4}
\end{fmfgraph*} }
%
%
\subfigure[]{
\begin{fmfgraph*}(20,20)
\fmfleft{i1,i2}
\fmfright{o}
\fmfforce{0.2w,0.9h}{v2}
\fmfforce{0.2w,0.1h}{v1}
\fmfforce{0.2w,0.5h}{v3}
\fmfforce{0.8w,0.5h}{v4}
\fmf{photon}{i1,v1}
\fmf{photon}{i2,v2}
\fmf{plain}{v4,o}
\fmf{photon,tension=0}{v1,v3}
\fmf{plain,tension=0}{v3,v4}
\fmf{photon,tension=0}{v2,v4}
\fmf{plain,tension=0}{v2,v3}
\fmf{photon,tension=0}{v1,v4}
\end{fmfgraph*} }
%
%
\subfigure[]{
\begin{fmfgraph*}(20,20)
\fmfleft{i1,i2}
\fmfright{o}
\fmfforce{0.2w,0.9h}{v2}
\fmfforce{0.2w,0.1h}{v1}
\fmfforce{0.2w,0.5h}{v3}
\fmfforce{0.8w,0.5h}{v4}
\fmf{photon}{i1,v1}
\fmf{photon}{i2,v2}
\fmf{plain}{v4,o}
\fmf{plain,tension=0}{v1,v3}
\fmf{plain,tension=0}{v3,v4}
\fmf{photon,tension=0}{v2,v4}
\fmf{plain,tension=0}{v2,v3}
\fmf{photon,tension=0}{v1,v4}
\end{fmfgraph*} } \\
%
%
\subfigure[]{
\begin{fmfgraph*}(20,20)
\fmfleft{i1,i2}
\fmfright{o}
\fmf{photon}{i1,v1}
\fmf{photon}{i2,v2}
\fmf{plain}{v4,o}
\fmf{plain,tension=.3}{v2,v3}
\fmf{plain,tension=.3}{v1,v3}
\fmf{photon,tension=0}{v2,v1}
\fmf{photon,tension=.2,left}{v3,v4}
\fmf{photon,tension=.2,right}{v3,v4}
\end{fmfgraph*} }
%
%
\subfigure[]{
\begin{fmfgraph*}(20,20)
\fmfleft{i1,i2}
\fmfright{o}
\fmfforce{0.1w,0.1h}{v1}
\fmfforce{0.4w,0.3h}{v2}
\fmfforce{0.1w,0.9h}{v3}
\fmfforce{0.4w,0.7h}{v4}
\fmfforce{0.9w,0.5h}{v5}
\fmf{photon}{i1,v1}
\fmf{phantom}{i2,v3}
\fmf{plain}{v5,o}
\fmf{plain,left}{v1,v2}
\fmf{photon,right}{v1,v2}
\fmf{photon}{v3,v4}
\fmf{plain}{v2,v4}
\fmf{photon}{v4,v5}
\fmf{photon}{v2,v5}
\end{fmfgraph*} } 
%
%
\subfigure[]{
\begin{fmfgraph*}(20,20)
\fmfforce{0.2w,0.5h}{v1}
\fmfforce{0.5w,0.8h}{v2}
\fmfforce{0.5w,0.2h}{v3}
\fmfforce{0.8w,0.5h}{v4}
\fmfleft{i}
\fmfright{o}
\fmf{plain}{i,v1}
\fmf{plain}{v4,o}
\fmf{plain,tension=.2,left=.4}{v1,v2}
\fmf{plain,tension=.2,right=.4}{v1,v3}
\fmf{photon,tension=.2,left=.4}{v2,v4}
\fmf{photon,tension=.2,right=.4}{v3,v4}
\fmf{photon,tension=0}{v2,v3}
\end{fmfgraph*} }
%
%
\subfigure[]{
\begin{fmfgraph*}(20,20)
\fmfforce{0.2w,0.5h}{v1}
\fmfforce{0.5w,0.8h}{v2}
\fmfforce{0.5w,0.2h}{v3}
\fmfforce{0.8w,0.5h}{v4}
\fmfleft{i}
\fmfright{o}
\fmf{photon}{i,v1}
\fmf{photon}{v4,o}
\fmf{plain,tension=.2,left=.4}{v1,v2}
\fmf{photon,tension=.2,right=.4}{v1,v3}
\fmf{plain,tension=.2,left=.4}{v2,v4}
\fmf{photon,tension=.2,right=.4}{v3,v4}
\fmf{photon,tension=0}{v2,v3}
\end{fmfgraph*} }
%
%
\subfigure[]{
\begin{fmfgraph*}(20,20)
\fmfforce{0.2w,0.5h}{v1}
\fmfforce{0.5w,0.8h}{v2}
\fmfforce{0.5w,0.2h}{v3}
\fmfforce{0.8w,0.5h}{v4}
\fmfleft{i}
\fmfright{o}
\fmf{photon}{i,v1}
\fmf{photon}{v4,o}
\fmf{photon,tension=.2,left=.4}{v1,v2}
\fmf{photon,tension=.2,right=.4}{v1,v3}
\fmf{plain,tension=.2,left=.4}{v2,v4}
\fmf{photon,tension=.2,right=.4}{v3,v4}
\fmf{plain,tension=0}{v2,v3}
\end{fmfgraph*} } \\
%
%
\subfigure[]{
\begin{fmfgraph*}(20,20)
\fmfforce{0.2w,0.5h}{v1}
\fmfforce{0.5w,0.8h}{v2}
\fmfforce{0.5w,0.2h}{v3}
\fmfforce{0.8w,0.5h}{v4}
\fmfleft{i}
\fmfright{o}
\fmf{photon}{i,v1}
\fmf{photon}{v4,o}
\fmf{photon,tension=.2,left=.4}{v1,v2}
\fmf{plain,tension=.2,right=.4}{v1,v3}
\fmf{plain,tension=.2,left=.4}{v2,v4}
\fmf{photon,tension=.2,right=.4}{v3,v4}
\fmf{photon,tension=0}{v2,v3}
\end{fmfgraph*} }
%
%
\subfigure[]{
\begin{fmfgraph*}(20,20)
\fmfforce{0.2w,0.5h}{v1}
\fmfforce{0.5w,0.8h}{v2}
\fmfforce{0.5w,0.2h}{v3}
\fmfforce{0.8w,0.5h}{v4}
\fmfleft{i}
\fmfright{o}
\fmf{photon}{i,v1}
\fmf{photon}{v4,o}
\fmf{plain,tension=.2,left=.4}{v1,v2}
\fmf{photon,tension=.2,right=.4}{v1,v3}
\fmf{plain,tension=.2,left=.4}{v2,v4}
\fmf{photon,tension=.2,right=.4}{v3,v4}
\fmf{plain,tension=0}{v2,v3}
\end{fmfgraph*} }
%
%
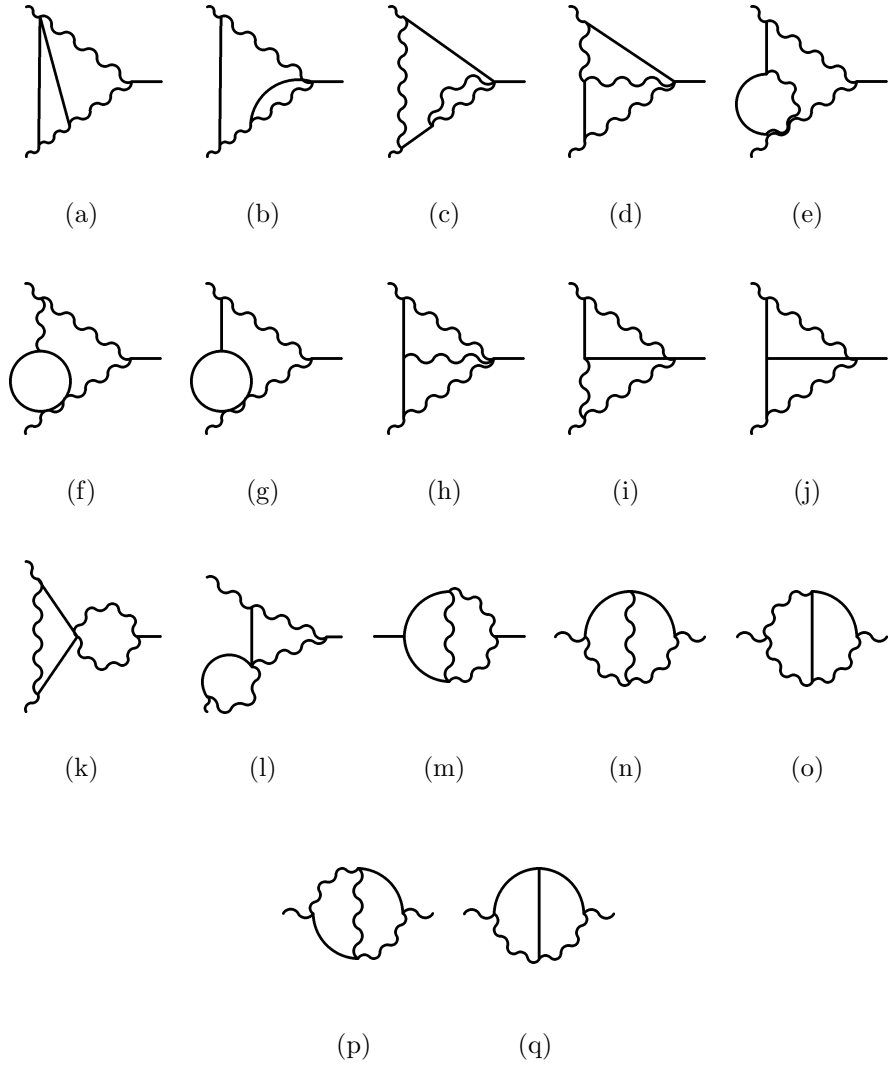
\caption{\label{fig3} The set of 17 independent 5-denominator diagrams, 
with 2 and 3 massive propagators.}
\ec
\efig

\bfig
\bc
%
\subfigure[]{
\begin{fmfgraph*}(20,20)
\fmfleft{i1,i2}
\fmfright{o}
\fmf{photon}{i1,v1}
\fmf{photon}{i2,v2}
\fmf{plain}{v3,o}
\fmf{photon,tension=.3}{v2,v3}
\fmf{photon,tension=.3}{v1,v3}
\fmf{plain,tension=0,right=.5}{v2,v1}
\fmf{plain,tension=0,right=.5}{v1,v2}
\end{fmfgraph*} }
%
%
\subfigure[]{
\begin{fmfgraph*}(20,20)
\fmfleft{i1,i2}
\fmfright{o}
\fmf{photon}{i1,v1}
\fmf{photon}{i2,v2}
\fmf{plain}{v3,o}
\fmf{photon,tension=.3}{v2,v3}
\fmf{photon,tension=.3}{v1,v3}
\fmf{plain,tension=0}{v2,v1}
\fmf{plain,tension=0,right=.5}{v2,v3}
\end{fmfgraph*} } 
%
%
\subfigure[]{
\begin{fmfgraph*}(20,20)
\fmfleft{i1,i2}
\fmfright{o}
\fmf{photon}{i1,v1}
\fmf{photon}{i2,v2}
\fmf{plain}{v3,o}
\fmf{photon,tension=.3}{v2,v3}
\fmf{photon,tension=.3}{v1,v3}
\fmf{plain,tension=0}{v2,v1}
\fmf{plain,right=45}{v3,v3}
\end{fmfgraph*} }
%
%
\subfigure[]{
\begin{fmfgraph*}(20,20)
\fmfleft{i}
\fmfright{o}
\fmfforce{0.2w,0.5h}{v1}
\fmfforce{0.5w,0.2h}{v2}
\fmfforce{0.8w,0.5h}{v3}
\fmf{plain}{i,v1}
\fmf{plain}{v3,o}
\fmf{plain,left}{v1,v3}
\fmf{plain,right=.4}{v1,v2}
\fmf{photon,right=.4}{v2,v3}
\fmf{photon,left=.6}{v2,v3}
\end{fmfgraph*} } 
%
%
\subfigure[]{
\begin{fmfgraph*}(20,20)
\fmfleft{i}
\fmfright{o}
\fmf{plain}{i,v1}
\fmf{plain}{v3,o}
\fmf{plain,tension=.2,left}{v1,v2}
\fmf{plain,tension=.2,right}{v1,v2}
\fmf{photon,tension=.2,left}{v2,v3}
\fmf{photon,tension=.2,right}{v2,v3}
\end{fmfgraph*} } \\
%
%
\subfigure[]{
\begin{fmfgraph*}(20,20)
\fmfleft{i}
\fmfright{o}
\fmfforce{0.2w,0.5h}{v1}
\fmfforce{0.5w,0.2h}{v2}
\fmfforce{0.8w,0.5h}{v3}
\fmf{photon}{i,v1}
\fmf{photon}{v3,o}
\fmf{plain,left}{v1,v3}
\fmf{photon,right=.4}{v1,v2}
\fmf{photon,right=.4}{v2,v3}
\fmf{plain,left=.6}{v2,v3}
\end{fmfgraph*} }  
%
%
\subfigure[]{
\begin{fmfgraph*}(20,20)
\fmfbottom{v5}
\fmftop{v4}
\fmfleft{i}
\fmfright{o}
\fmf{photon}{i,v1}
\fmf{photon}{v3,o}
\fmf{phantom}{v5,v2} 
\fmf{plain}{v2,v4} 
\fmf{plain,tension=.2,left}{v1,v2}
\fmf{photon,tension=.2,right}{v1,v2}
\fmf{photon,tension=.2,left}{v2,v3}
\fmf{plain,tension=.2,right}{v2,v3}
\end{fmfgraph*} } 
%
%
\subfigure[]{
\begin{fmfgraph*}(20,20)
\fmfbottom{v5}
\fmftop{v4}
\fmfleft{i}
\fmfright{o}
\fmf{photon}{i,v1}
\fmf{photon}{v3,o}
\fmf{phantom}{v5,v2} 
\fmf{phantom}{v2,v4} 
\fmf{plain,tension=.2,left}{v1,v2}
\fmf{photon,tension=.2,right}{v1,v2}
\fmf{plain,tension=.2,left}{v2,v3}
\fmf{photon,tension=.2,right}{v2,v3}
\end{fmfgraph*} }   
%
%
\subfigure[]{
\begin{fmfgraph*}(20,20)
\fmfleft{i}
\fmfright{o}
\fmfforce{0.2w,0.5h}{v1}
\fmfforce{0.5w,0.8h}{v2}
\fmfforce{0.8w,0.5h}{v3}
\fmf{photon}{i,v1}
\fmf{photon}{v3,o}
\fmf{photon,right}{v1,v3}
\fmf{plain,left=.4}{v1,v2}
\fmf{plain,left=.4}{v2,v3}
\fmf{photon,right=.6}{v2,v3}
\end{fmfgraph*} } 
%
%
\subfigure[]{
\begin{fmfgraph*}(20,20)
\fmfleft{i}
\fmfright{o}
\fmfforce{0.2w,0.5h}{v1}
\fmfforce{0.5w,0.8h}{v2}
\fmfforce{0.8w,0.5h}{v3}
\fmf{photon}{i,v1}
\fmf{photon}{v3,o}
\fmf{photon,right}{v1,v3}
\fmf{photon,left=.4}{v1,v2}
\fmf{plain,left=.4}{v2,v3}
\fmf{plain,right=.6}{v2,v3}
\end{fmfgraph*} } \\
%
%
\subfigure[]{
\begin{fmfgraph*}(20,20)
\fmfleft{i}
\fmfright{o}
\fmfforce{0.2w,0.5h}{v1}
\fmfforce{0.5w,0.8h}{v2}
\fmfforce{0.8w,0.5h}{v3}
\fmf{photon}{i,v1}
\fmf{photon}{v3,o}
\fmf{photon,right}{v1,v3}
\fmf{plain,left=.4}{v1,v2}
\fmf{plain,left=.4}{v2,v3}
\fmf{plain,right=.6}{v2,v3}
\end{fmfgraph*} } 
%
%
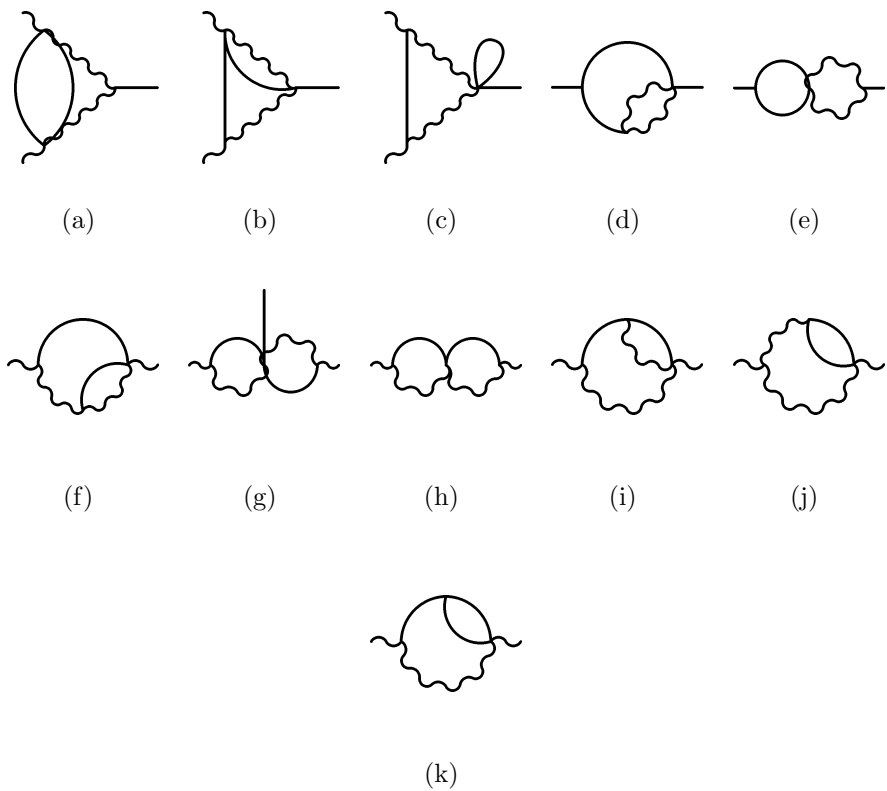
\caption{\label{fig4} The set of 11 independent 4-denominator diagrams,
with 2 and 3 massive denominators.}
\ec
\efig

\bfig
\bc
\subfigure[]{
\begin{fmfgraph*}(20,20)
\fmfleft{i}
\fmfright{o}
\fmf{photon}{i,v1}
\fmf{photon}{v2,o}
\fmf{plain,tension=.15,left}{v1,v2}
\fmf{plain,tension=.15}{v1,v2}
\fmf{photon,tension=.15,right}{v1,v2}
\end{fmfgraph*} }
%
%
\subfigure[]{
\begin{fmfgraph*}(20,20)
\fmfleft{i}
\fmfright{o}
\fmf{photon}{i,v1}
\fmf{photon}{v2,o}
\fmf{plain,tension=.22,left}{v1,v2}
\fmf{photon,tension=.22,right}{v1,v2}
\fmf{plain,right=45}{v2,v2}
\end{fmfgraph*} }
%
%
\subfigure[]{
\begin{fmfgraph*}(20,20)
\fmfleft{i}
\fmfright{o}
\fmfforce{0.5w,0.1h}{v1}
\fmfforce{0.25w,0.62h}{v3}
\fmfforce{0.5w,0.9h}{v7}
\fmfforce{0.74w,0.62h}{v11}
\fmf{plain,left=.1}{v1,v3}
\fmf{plain,left=.5}{v3,v7}
\fmf{photon,left=.5}{v7,v11}
\fmf{photon,left=.1}{v11,v1}
\fmf{plain}{v1,v7}
\end{fmfgraph*}}
%
%
\subfigure[]{
\begin{fmfgraph*}(20,20)
\fmfleft{i}
\fmfright{o}
\fmfforce{0.5w,0.1h}{v1}
\fmfforce{0.25w,0.62h}{v3}
\fmfforce{0.5w,0.9h}{v7}
\fmfforce{0.74w,0.62h}{v11}
\fmf{plain,left=.1}{v1,v3}
\fmf{plain,left=.5}{v3,v7}
\fmf{plain,left=.5}{v7,v11}
\fmf{plain,left=.1}{v11,v1}
\fmf{plain}{v1,v7}
\end{fmfgraph*}} \\
%
%
\subfigure[]{
\begin{fmfgraph*}(20,20)
\fmfleft{i}
\fmfright{o}
\fmf{phantom}{i,v1}
\fmf{phantom}{v1,o}
\fmf{plain,left}{v1,v1}
\fmf{plain,right}{v1,v1}
\end{fmfgraph*}} 
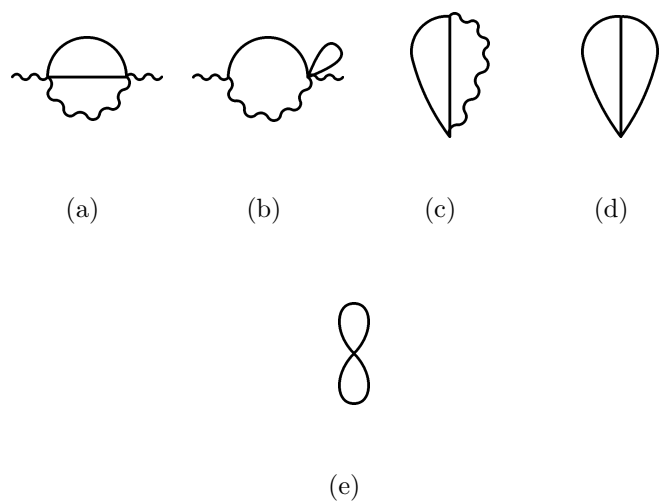
\caption{\label{fig5} The set of 4 independent 3-denominator 
diagrams together with the single 2-denominator diagram, with 2 and 3 massive 
propagators.}
\ec
\efig

\bfig
\bc
\subfigure[]{
\begin{fmfgraph*}(20,20)
\fmfleft{i1,i2}
\fmfright{o}
\fmf{photon}{i1,v1}
\fmf{photon}{i2,v2}
\fmf{plain}{v5,o}
\fmf{photon,tension=.3}{v2,v3}
\fmf{photon,tension=.3}{v3,v5}
\fmf{photon,tension=.3}{v1,v4}
\fmf{photon,tension=.3}{v4,v5}
\fmf{plain,tension=0}{v2,v1}
\fmf{plain,tension=0}{v4,v3}
\end{fmfgraph*} }
%
%
\hspace{1mm}
\subfigure[]{
\begin{fmfgraph*}(20,20)
\fmfleft{i1,i2}
\fmfright{o}
\fmf{photon}{i1,v1}
\fmf{photon}{i2,v2}
\fmf{plain}{v5,o}
\fmf{plain,tension=.3}{v2,v3}
\fmf{photon,tension=.3}{v3,v5}
\fmf{plain,tension=.3}{v1,v4}
\fmf{photon,tension=.3}{v4,v5}
\fmf{photon,tension=0}{v2,v1}
\fmf{photon,tension=0}{v4,v3}
\end{fmfgraph*} }
%
%
\hspace{1mm}
\subfigure[]{
\begin{fmfgraph*}(20,20)
\fmfleft{i1,i2}
\fmfright{o}
\fmfforce{0.2w,0.9h}{v2}
\fmfforce{0.2w,0.1h}{v1}
\fmfforce{0.2w,0.5h}{v3}
\fmfforce{0.8w,0.5h}{v5}
\fmf{photon}{i1,v1}
\fmf{photon}{i2,v2}
\fmf{plain}{v5,o}
\fmf{photon,tension=0}{v2,v5}
\fmf{photon,tension=0}{v3,v4}
\fmf{plain,tension=.4}{v1,v4}
\fmf{photon,tension=.4}{v4,v5}
\fmf{photon,tension=0}{v1,v3}
\fmf{plain,tension=0}{v2,v3}
\end{fmfgraph*} }
%
%
\hspace{1mm}
\subfigure[]{
\begin{fmfgraph*}(20,20)
\fmfleft{i1,i2}
\fmfright{o}
\fmfforce{0.2w,0.9h}{v2}
\fmfforce{0.2w,0.1h}{v1}
\fmfforce{0.2w,0.5h}{v3}
\fmfforce{0.8w,0.5h}{v5}
\fmf{photon}{i1,v1}
\fmf{photon}{i2,v2}
\fmf{plain}{v5,o}
\fmf{photon,tension=0}{v2,v5}
\fmf{plain,tension=0}{v3,v4}
\fmf{photon,tension=.4}{v1,v4}
\fmf{photon,tension=.4}{v4,v5}
\fmf{plain,tension=0}{v1,v3}
\fmf{plain,tension=0}{v2,v3}
\end{fmfgraph*} }
%
%
%
\hspace{1mm}
\subfigure[]{
\begin{fmfgraph*}(20,20)
\fmfleft{i1,i2}
\fmfright{o}
\fmfforce{0.8w,0.5h}{v4}
\fmf{photon}{i1,v1}
\fmf{photon}{i2,v2}
\fmf{plain}{v4,o}
\fmf{photon,tension=.15}{v2,v4}
\fmf{photon,tension=.4}{v1,v3}
\fmf{photon,tension=.2}{v3,v4}
\fmf{plain,tension=0}{v2,v1}
\fmf{plain,tension=0}{v2,v3}
\end{fmfgraph*} }  \\
%
%
\subfigure[]{
\begin{fmfgraph*}(20,20)
\fmfleft{i1,i2}
\fmfright{o}
\fmfforce{0.8w,0.5h}{v4}
\fmf{photon}{i1,v1}
\fmf{photon}{i2,v2}
\fmf{plain}{v4,o}
\fmf{plain,tension=.4}{v1,v3}
\fmf{photon,tension=.2}{v3,v4}
\fmf{plain,tension=.15}{v2,v4}
\fmf{photon,tension=0}{v2,v1}
\fmf{photon,tension=0,left=.5}{v3,v4}
\end{fmfgraph*} }
%
%
\hspace{3mm}
\subfigure[]{
\begin{fmfgraph*}(20,20)
\fmfleft{i1,i2}
\fmfright{o}
\fmfforce{0.2w,0.9h}{v2}
\fmfforce{0.2w,0.1h}{v1}
\fmfforce{0.2w,0.5h}{v3}
\fmfforce{0.8w,0.5h}{v4}
\fmf{photon}{i1,v1}
\fmf{photon}{i2,v2}
\fmf{plain}{v4,o}
\fmf{photon,tension=0}{v2,v3}
\fmf{photon,tension=0}{v3,v4}
\fmf{photon,tension=0}{v1,v4}
\fmf{plain,tension=0}{v2,v4}
\fmf{plain,tension=0}{v1,v3}
\end{fmfgraph*} }
%
%
\hspace{3mm}
\subfigure[]{
\begin{fmfgraph*}(20,20)
\fmfleft{i1,i2}
\fmfright{o}
\fmfforce{0.2w,0.9h}{v2}
\fmfforce{0.2w,0.1h}{v1}
\fmfforce{0.2w,0.5h}{v3}
\fmfforce{0.8w,0.5h}{v4}
\fmf{photon}{i1,v1}
\fmf{photon}{i2,v2}
\fmf{plain}{v4,o}
\fmflabel{$(p_{1} \cdot k_{2})$}{o}
\fmf{photon,tension=0}{v2,v3}
\fmf{photon,tension=0}{v3,v4}
\fmf{photon,tension=0}{v1,v4}
\fmf{plain,tension=0}{v2,v4}
\fmf{plain,tension=0}{v1,v3}
\end{fmfgraph*} }
%
%
\hspace{15mm}
\subfigure[]{
\begin{fmfgraph*}(20,20)
\fmfleft{i1,i2}
\fmfright{o}
\fmfforce{0.2w,0.9h}{v2}
\fmfforce{0.2w,0.1h}{v1}
\fmfforce{0.2w,0.5h}{v3}
\fmfforce{0.8w,0.5h}{v4}
\fmf{photon}{i1,v1}
\fmf{photon}{i2,v2}
\fmf{plain}{v4,o}
\fmf{plain,tension=0}{v1,v3}
\fmf{photon,tension=0}{v3,v4}
\fmf{photon,tension=0}{v2,v4}
\fmf{plain,tension=0}{v2,v3}
\fmf{photon,tension=0}{v1,v4}
\end{fmfgraph*} }  \\
%
%
\hspace{-18mm}
\subfigure[]{
\begin{fmfgraph*}(20,20)
\fmfleft{i1,i2}
\fmfright{o}
\fmfforce{0.2w,0.9h}{v2}
\fmfforce{0.2w,0.1h}{v1}
\fmfforce{0.2w,0.5h}{v3}
\fmfforce{0.8w,0.5h}{v4}
\fmf{photon}{i1,v1}
\fmf{photon}{i2,v2}
\fmf{plain}{v4,o}
\fmflabel{$(p_{2} \cdot k_{1})$}{o}
\fmf{plain,tension=0}{v1,v3}
\fmf{photon,tension=0}{v3,v4}
\fmf{photon,tension=0}{v2,v4}
\fmf{plain,tension=0}{v2,v3}
\fmf{photon,tension=0}{v1,v4}
\end{fmfgraph*} }
%
%
\hspace{15mm}
\subfigure[]{
\begin{fmfgraph*}(20,20)
\fmfleft{i1,i2}
\fmfright{o}
\fmfforce{0.2w,0.9h}{v2}
\fmfforce{0.2w,0.1h}{v1}
\fmfforce{0.2w,0.55h}{v3}
\fmfforce{0.2w,0.15h}{v5}
\fmfforce{0.8w,0.5h}{v4}
\fmf{photon}{i1,v1}
\fmf{photon}{i2,v2}
\fmf{plain}{v4,o}
\fmf{plain}{v2,v3}
\fmf{plain,left}{v3,v5}
\fmf{plain,right}{v3,v5}
\fmf{photon}{v1,v4}
\fmf{photon}{v2,v4}
\end{fmfgraph*} }
%
%
\hspace{-2mm}
\subfigure[]{
\begin{fmfgraph*}(20,20)
\fmfleft{i1,i2}
\fmfright{o}
\fmfforce{0.2w,0.9h}{v2}
\fmfforce{0.2w,0.1h}{v1}
\fmfforce{0.2w,0.5h}{v3}
\fmfforce{0.8w,0.5h}{v4}
\fmf{photon}{i1,v1}
\fmf{photon}{i2,v2}
\fmf{plain}{v4,o}
\fmf{plain,tension=0}{v1,v3}
\fmf{plain,tension=0}{v3,v4}
\fmf{photon,tension=0}{v2,v4}
\fmf{plain,tension=0}{v2,v3}
\fmf{photon,tension=0}{v1,v4}
\end{fmfgraph*} }
%
%
\hspace{-2mm}
\subfigure[]{
\begin{fmfgraph*}(20,20)
\fmfleft{i1,i2}
\fmfright{o}
\fmfforce{0.2w,0.9h}{v2}
\fmfforce{0.2w,0.1h}{v1}
\fmfforce{0.2w,0.5h}{v3}
\fmfforce{0.8w,0.5h}{v4}
\fmf{photon}{i1,v1}
\fmf{photon}{i2,v2}
\fmf{plain}{v4,o}
\fmflabel{$(p_{2} \cdot k_{1})$}{o}
\fmf{plain,tension=0}{v1,v3}
\fmf{plain,tension=0}{v3,v4}
\fmf{photon,tension=0}{v2,v4}
\fmf{plain,tension=0}{v2,v3}
\fmf{photon,tension=0}{v1,v4}
\end{fmfgraph*} } \\
%
%
\subfigure[]{
\begin{fmfgraph*}(20,20)
\fmfleft{i1,i2}
\fmfforce{.4w,.5h}{d1}
\fmfright{o}
\fmfforce{0.2w,0.9h}{v2}
\fmfforce{0.2w,0.1h}{v1}
\fmfforce{0.2w,0.5h}{v3}
\fmfforce{0.8w,0.5h}{v4}
\fmf{photon}{i1,v1}
\fmf{photon}{i2,v2}
\fmf{plain}{v4,o}
\fmf{plain,tension=0}{v1,v3}
\fmf{plain,tension=0}{v3,v4}
\fmf{photon,tension=0}{v2,v4}
\fmf{plain,tension=0}{v2,v3}
\fmf{photon,tension=0}{v1,v4}
\fmfv{decor.shape=circle,decor.filled=full,decor.size=.1w}{d1}
\end{fmfgraph*} }
%
%
\hspace{1mm}
\subfigure[]{
\begin{fmfgraph*}(20,20)
\fmfleft{i1,i2}
\fmfright{o}
\fmf{photon}{i1,v1}
\fmf{photon}{i2,v2}
\fmf{plain}{v4,o}
\fmf{plain,tension=.3}{v2,v3}
\fmf{plain,tension=.3}{v1,v3}
\fmf{photon,tension=0}{v2,v1}
\fmf{photon,tension=.2,left}{v3,v4}
\fmf{photon,tension=.2,right}{v3,v4}
\end{fmfgraph*} }
%
%
\hspace{1mm}
\subfigure[]{
\begin{fmfgraph*}(20,20)
\fmfforce{0.2w,0.5h}{v1}
\fmfforce{0.5w,0.8h}{v2}
\fmfforce{0.5w,0.2h}{v3}
\fmfforce{0.8w,0.5h}{v4}
\fmfleft{i}
\fmfright{o}
\fmf{plain}{i,v1}
\fmf{plain}{v4,o}
\fmf{plain,tension=.2,left=.4}{v1,v2}
\fmf{plain,tension=.2,right=.4}{v1,v3}
\fmf{photon,tension=.2,left=.4}{v2,v4}
\fmf{photon,tension=.2,right=.4}{v3,v4}
\fmf{photon,tension=0}{v2,v3}
\end{fmfgraph*} }
%
%
%
\hspace{1mm}
\subfigure[]{
\begin{fmfgraph*}(20,20)
\fmfleft{i1,i2}
\fmfright{o}
\fmf{photon}{i1,v1}
\fmf{photon}{i2,v2}
\fmf{plain}{v3,o}
\fmf{photon,tension=.3}{v2,v3}
\fmf{photon,tension=.3}{v1,v3}
\fmf{plain,tension=0,right=.5}{v2,v1}
\fmf{plain,tension=0,right=.5}{v1,v2}
\end{fmfgraph*} }
%
%
\hspace{1mm}
\subfigure[]{
\begin{fmfgraph*}(20,20)
\fmfleft{i1,i2}
\fmfforce{.5w,.65h}{d1}
\fmfright{o}
\fmf{photon}{i1,v1}
\fmf{photon}{i2,v2}
\fmf{plain}{v3,o}
\fmf{photon,tension=.3}{v2,v3}
\fmf{photon,tension=.3}{v1,v3}
\fmf{plain,tension=0,right=.5}{v2,v1}
\fmf{plain,tension=0,right=.5}{v1,v2}
\fmfv{decor.shape=circle,decor.filled=full,decor.size=.1w}{d1}
\end{fmfgraph*} } \\
%
%
\hspace{1mm}
\subfigure[]{
\begin{fmfgraph*}(20,20)
\fmfleft{i1,i2}
\fmfright{o}
\fmf{photon}{i1,v1}
\fmf{photon}{i2,v2}
\fmf{plain}{v3,o}
\fmf{photon,tension=.3}{v2,v3}
\fmf{photon,tension=.3}{v1,v3}
\fmf{plain,tension=0}{v2,v1}
\fmf{plain,tension=0,right=.5}{v2,v3}
\end{fmfgraph*} }
%
%
\subfigure[]{
\begin{fmfgraph*}(20,20)
\fmfleft{i1,i2}
\fmfright{o}
\fmf{photon}{i1,v1}
\fmf{photon}{i2,v2}
\fmf{plain}{v3,o}
\fmflabel{$(p_{2} \cdot k_{2})$}{o}
\fmf{photon,tension=.3}{v2,v3}
\fmf{photon,tension=.3}{v1,v3}
\fmf{plain,tension=0}{v2,v1}
\fmf{plain,tension=0,right=.5}{v2,v3}
\end{fmfgraph*} }
%
%
\hspace{12mm}
\subfigure[]{
\begin{fmfgraph*}(20,20)
\fmfleft{i1,i2}
\fmfright{o}
\fmf{photon}{i1,v1}
\fmf{photon}{i2,v2}
\fmf{plain}{v3,o}
\fmf{photon,tension=.3}{v2,v3}
\fmf{photon,tension=.3}{v1,v3}
\fmf{plain,tension=0}{v2,v1}
\fmf{plain,right=45}{v3,v3}
\end{fmfgraph*} }
%
%
\subfigure[]{
\begin{fmfgraph*}(20,20)
\fmfleft{i}
\fmfright{o}
\fmfforce{0.2w,0.5h}{v1}
\fmfforce{0.5w,0.2h}{v2}
\fmfforce{0.8w,0.5h}{v3}
\fmf{plain}{i,v1}
\fmf{plain}{v3,o}
\fmf{plain,left}{v1,v3}
\fmf{plain,right=.4}{v1,v2}
\fmf{photon,right=.4}{v2,v3}
\fmf{photon,left=.6}{v2,v3}
\end{fmfgraph*} }   \\
%
%
\subfigure[]{
\begin{fmfgraph*}(20,20)
\fmfleft{i}
\fmfright{o}
\fmf{plain}{i,v1}
\fmf{plain}{v3,o}
\fmf{plain,tension=.2,left}{v1,v2}
\fmf{plain,tension=.2,right}{v1,v2}
\fmf{photon,tension=.2,left}{v2,v3}
\fmf{photon,tension=.2,right}{v2,v3}
\end{fmfgraph*} }
%
%
\subfigure[]{
\begin{fmfgraph*}(20,20)
\fmfleft{i}
\fmfright{o}
\fmfforce{0.5w,0.1h}{v1}
\fmfforce{0.25w,0.62h}{v3}
\fmfforce{0.5w,0.9h}{v7}
\fmfforce{0.74w,0.62h}{v11}
\fmf{plain,left=.1}{v1,v3}
\fmf{plain,left=.5}{v3,v7}
\fmf{plain,left=.5}{v7,v11}
\fmf{plain,left=.1}{v11,v1}
\fmf{plain}{v1,v7}
\end{fmfgraph*}}
%
%
\subfigure[]{
\begin{fmfgraph*}(20,20)
\fmfleft{i}
\fmfright{o}
\fmf{phantom}{i,v1}
\fmf{phantom}{v1,o}
\fmf{plain,left}{v1,v1}
\fmf{plain,right}{v1,v1}
\end{fmfgraph*} }
%
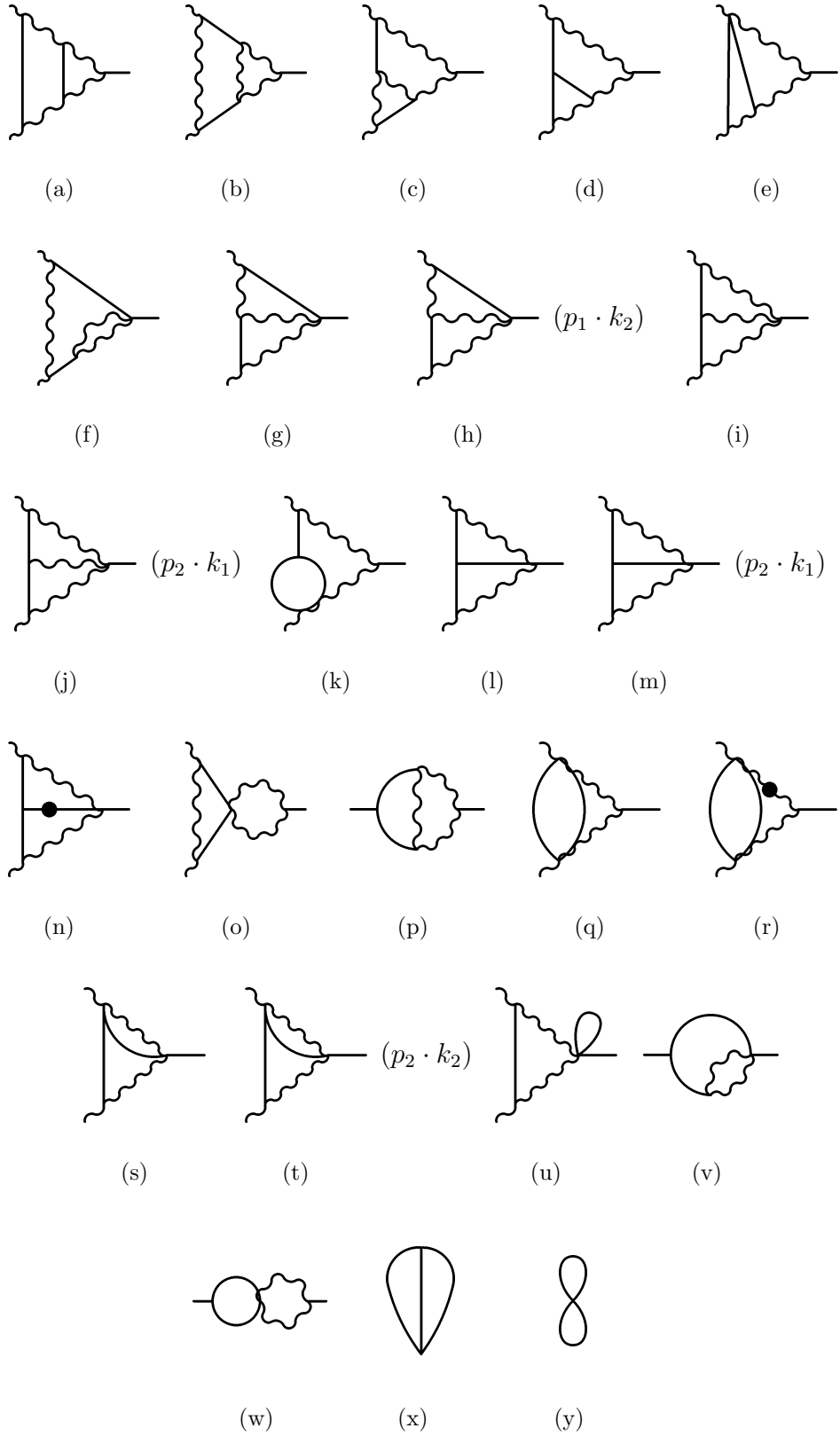
\caption{\label{fig6} The set of 25 MIs with 2 and 3 massive 
denominators. Out of them, 4 are the product of
1-loop MIs. The dot on a line indicates a square of the corresponding
denominator.}
\ec
\efig
%
%
%
\subsection{The evaluation of the master integrals: the differential equation method}

In the previous Section we outlined how to reduce the 2-loop Feynman diagrams
of the electroweak form factor to the master integrals.
Let us now sketch how to explicitly evaluate the master integrals with the
differential equation method \cite{Kotikov1,Kotikov2,Kotikov3,Rem1,Rem2}.
We may identify 2 basic steps:
\begin{enumerate}
{\bf \item
Generation of the differential equations.}

\noindent
This step is rather ``automatic'' in the sense that it does not
offer specific difficulties.  
It involves, in turn, the following steps:
\begin{enumerate}
\item
We take the derivatives of the MIs with respect
to $x$ at fixed $a$ by differentiating the integrand
with respect to the external momenta $p_1$ or $p_2$;
the derivatives of the MIs are then expressed
as a superposition of linearly-dependent scalar amplitudes;
\item
We rotate the dependent scalar amplitudes to independent ones,
either in the auxiliary diagram scheme or in the shift scheme;
\item
We reduce the independent amplitudes to MIs by using the ibps identities
and other eventual symmetry relations. This way the system closes on the MIs
themselves.
Linear systems of first-order differential equations with variable coefficients are then
generated, whose unknowns are the MIs themselves. 
\end{enumerate}
{\bf \item
Solution of the differential equations.}

\noindent
In order to have a unique solution --- as it should ---
initial conditions can often be obtained by studying the behaviour of the MIs
close to thresholds or pseudothresholds.
The system of differential equations is then solved by recursion in $\epsilon$.
Let us remark that this step is not ``automatic'', in the sense that
obtaining a solution is in some cases a matter of intuition and
experience.
\end{enumerate}
Let us make a few comments about specific characteristics of the present computation.
The most complicated cases involve topologies with more than 1 MI:
\begin{itemize}
\item
$t=4$: we encounter 2 vertex topologies having 2 MIs;
\item
$t=5$: we encounter 2 vertex topologies having 2 MIs and
1 topology having 3 MIs\footnote{
The $t=6$ crossed ladder topology also has 3 MIs.}
\footnote{We have also used the symbolic method to reduce the topology
with 3 MIs, with similar results.}.  
\end{itemize}

\noindent
A convenient basis for topologies with 2 MIs consists
of the following amplitudes:
\begin{enumerate}
\item
the ``basic'' amplitude, i.e. the amplitude with unitary numerator 
and with all the denominators having unitary indices;
\item
the amplitude with unitary numerator and
with one of the denominators squared, or with an independent scalar product 
left on the numerator.
\end{enumerate}
The resulting system of 2 differential
equations is often triangular in $D=4$, allowing for a simple
solution.
For the only case of 3 MIs, we found that a convenient basis 
comprises the following amplitudes:
\begin{enumerate}
\item
the ``basic'' amplitude defined above;
\item
the amplitude with unitary numerator and one denominator squared,
as in the previous case; 
\item
the amplitude with an irreducible scalar product in the numerator
and the denominators all with unitary indices.
\end{enumerate}
With this choice, the system of 3 ordinary differential
equations becomes triangular in $D=4$, allowing for a
simple recursive solution in 
$\epsilon$.

\section{Harmonic Polylogarithms of one variable \label{HPLs}}

As discussed in the introduction, amplitudes with 2 and 3 massive 
propagators may have thresholds in $s=4m^2$ and pseudothresholds
in $s=-4m^2$. As a consequence, a generalization of the usual
harmonic polylogarithms of one variable is needed.
In the next Section we summarize the salient features
of the ordinary harmonic polylogarithms (HPLs) that we want to preserve with 
the generalization, while in Section \ref{Def} we discuss such a generalization.

\subsection{Ordinary Harmonic Polylogarithms}

The basic idea of the HPLs is that of representing
a given integral introducing a minimal set
of transcendental functions by making only linear
transformations, such as partial fractioning 
and integration by parts. 
Furthermore, the transcendental functions are defined once and for all.
Let us begin with a simple example.
The integration of the power function with an integer exponent $n$,
\bea
\label{primocaso}
\int_1^x {x'}^n dx' &=& \frac{x^{n+1}}{n+1}-\frac{1}{n+1}~~~{\rm if}~~~n\neq -1
\\ \nonumber
            &=& \log x~~~~~~~~~~~~~~~~{\rm if}~~~n = -1 \, ,
\eea
is again a power function, except for the case $n=-1$ for which we have the logarithm.
Integration then introduces 1 transcendental function, plus the original set of elementary
functions.
Larger sets of transcendental functions are needed to represent double integrals,
such as for instance
\be
\int_0^x \frac{dx'}{1-x'} \log x' = \int_0^x \frac{dx'}{1-x'} \int_{1}^{x'} \frac{dx''}{x''} = 
-\log x\,\log (1 - x) - \rm{Li_2}(x),
\ee
which involves a new transcendental functions, the well-known dilogarithm:
\be
{\rm Li}_2(x)\equiv 
-\int_0^x \frac{dx'}{x'} \int_0^{x'}\frac{dx''}{1-x''} = 
\sum_{n=1}^{\infty}~\frac{x^n}{n^2}~~~{\rm for}~|x|\leq 1.
\ee
However, many double integrals do not involve new transcendental functions with respect to
the ones needed to represent single integrals such as (\ref{primocaso}).
Let us indeed consider:
\be
\int_{0}^{x} \frac{dx'}{(1-x')^2} \log x' = 
\int_{0}^{x} \frac{dx'}{(1-x')^2} \int_{1}^{x'} \frac{dx''}{x''} = 
\log(1-x) + \left( \frac{1}{1-x} -1\right) \log x. 
\ee
The reduction above is done with an integration by parts: we integrate $1/(1-x)^2$, whose
integral is elementary, and differentiate $\log x$, obtaining a simpler, elementary function.
We then make a partial fractioning of 
\be
\frac{1}{x(1-x)} = \frac{1}{x} + \frac{1}{(1-x)} \, ,
\ee
i.e. another linear transformation. 
In general, whenever we have to integrate the product of a transcendental function times an elementary function
whose integral is also elementary, we do a by-parts integration in order
to derive, i.e. to simplify, the transcendental function. 
This applies in particular to the integral of the transcendental function itself:
\be
\int_0^x dx' \log x' = x \log x - x
\ee
and in general to all the integrals of the form:
\be
\int_0^x \frac{dx'}{(1-x')^n} \log x'
\ee
with $n\neq 1$.
The examples given above generalize to multiple repeated integrations, of the form:
\be
\int^x \frac{dx_l}{(x_l+a_l)^{n_l}} \cdots \cdots
\int^{x_3} \frac{dx_2}{(x_2+a_2)^{n_2}}   \int^{x_2} \frac{dx_1}{(x_1+a_1)^{n_1}},
\ee
with $a_1\cdots a_l$ some constants.
One has to introduce transcendental functions given by the repeated 
integrations of the inverse linear functions $1/(x+a_i)$:
\be
\int^x \frac{dx_l}{x_l+a_l} \cdots \cdots
\int^{x_3} \frac{dx_2}{x_2+a_2}   \int^{x_2} \frac{dx_1}{x_1+a_1}.
\ee
For many applications, such as our previous computation of 2-loop electroweak amplitudes with 1 
massive propagator \cite{UgoRo},
it is sufficient to introduce the following set of basis functions:
\bea
g(0,x)  & = & \frac{1}{x}  \, , \\
g(1,x)  & = & \frac{1}{1-x} \, , \\
g(-1,x) & = & \frac{1}{1+x}  \, . 
\eea
Note that $g(0,x)$ has a non-integrable singularity for $x\rightarrow 0$, while the other functions
are finite in the same limit. 
The HPLs of weight 1 are defined as integrals of the basis functions:
\bea
H(0,x) & = &  \int_{1}^{x} \frac{dt}{t} = \log{x} \, ,  \\
H(1,x) & = &  \int_{0}^{x} \frac{dt}{1-t} = - \log{(1-x)} \, , \\
H(-1,x) & = & \int_{0}^{x} \frac{dt}{1+t} = \log{(1+x)} \, , 
\eea
Note the slight asymmetry related to the lower bound of integration for $H(0,x)$.

Let us now define the general HPL $H(\vec{w};x)$, 
where $\vec{w}$ is a vector with $w$ components, consisting of a sequence of ``1'', ``0'', and ``-1''.
The weight of a HPL is the number of its indices, $w$, coinciding with the number of 
the repeated integrations.
The HPLs of weight $w+1$ have the following integral recursive definition:
\be
\label{defHPL}
H(a,\vec{w};x) = \int_0^x f(a;x') H(\vec{w},x'), 
\ee
except for the case with all the indices zero:
\be
H(\vec{0}_w;x) = \frac{1}{w!}\log^{w} x,
\ee
or, if a recursive definition is preferred:
\be
H(0,\vec{0}_w;x) = \int_1^x f(0;x') H(\vec{0}_w,x') = \frac{1}{(w+1)!}\log^{w+1} x,
\ee
where $\vec{0}_w=(0,0,\cdots,0)$ is a vector containing $w$ zeroes.
\subsection{Generalized Harmonic Polylogarithms \label{Def}}
The basis functions for the GHPLs involve various kinds of extensions
of the basis functions defining the HPLs.

\noindent
A trivial extension concerns functions involving different real constants:
\bea
g(-4,x) & = & \frac{1}{4+x}  \, , 
\label{meno4} \\
g(4,x)  & = & \frac{1}{4-x}.
\label{piu4} 
\eea
The above functions are related to amplitudes with threshold/pseudothreshold
at $s=\pm 4m^2$ respectively, just as the old functions $g(\mp 1;x)$ represent
amplitudes with threshold/pseudothresold in $s=\pm m^2$.
We will see in a moment however that the above extension in not sufficient.

To represent $3-$point functions with 3 massive propagators 
one also needs to introduce basis functions involving a {\it complex}
constant:\footnote{
An alternative representation uses only real functions at the price
of introducing squares of the integration variable:
$1/(x^2-x+1)$ and $x/(x^2-x+1)$.}
\bea
g(c,x) & = & \frac{1}{ x-\frac{1}{2}-i \frac{\sqrt{3}}{2} } \, , 
\label{ci} \\
g(\overline{c},x) & = & \frac{1}{ x-\frac{1}{2}+i \frac{\sqrt{3}}{2} }  
\label{cibarra} \, .
\eea
Note that $\frac{1}{2}\pm i\frac{\sqrt{3}}{2} = \exp(\pm i\pi/3)$ are the non trivial cubic roots of ``-1''
and are the inverse of each other.

\noindent
The non trivial extension, however, involves:
\begin{itemize}
\item 
radicals of the form
\bea
g(-r,x) & = & \frac{1}{\sqrt{x(4+x)}}  \, , 
\label{menoerre} \\
g(r,x)  & = & \frac{1}{\sqrt{x(4-x)}} \, ; 
\label{piuerre} 
\eea
These functions also describe amplitudes with thresholds and pseudothresholds in $s=\pm 4 m^2$ respectively.
It is indeed well-known that amplitudes involving 2 particles with the same mass $m\ne 0$ contain square roots
of similar form as a phase-space reduction effect; 
\item
products of radical with inverse linear functions:\footnote{
Alternative basis functions having the same asymptotic limits $\sim 1/x$ for
$x\rightarrow\infty$ of the previous ones can be taken for instance as
$\sqrt{(4\pm x)/x}/(1\pm x)$.}
\bea
g(-1-r,x) & = & \frac{1}{\sqrt{x(4+x)}~(1+x)}  \, , 
\label{meno1menoR} \\
g(1-r,x) & = & \frac{1}{\sqrt{x(4+x)}~(1-x)}  \, , 
\label{1menoerre} \\
g(-1+r,x) & = & \frac{1}{\sqrt{x(4-x)}~(1+x)}  \, , 
\label{meno1piuR} \\
g(1+r,x) & = & \frac{1}{\sqrt{x(4-x)}~(1-x)}  \, .
\label{1piuR} 
\eea
\end{itemize}
In total, we have added 10 functions to the old basis.
All the new basis functions above are finite in $x=0$ or have at most
an integrable singularity $\sim 1/\sqrt{x}$ for $x\rightarrow 0$.

The GHPLs of weight 1 are given by integrals of the basis functions 
and can be written, in general, in terms of logarithms of complex functions of $x$:
\bea
H(-4,x) & = & \int_{0}^{x} \frac{dt}{4+t} = \log{(4+x)} - 2 \log{2} \, , \\
H(4,x) & = & \int_{0}^{x} \frac{dt}{4-t} = - \log{(4-x)} + 2 \log{2} \, , \\
H(c,x) & = & \int_{0}^{x} 
\frac{dt}{ t-\frac{1}{2}-i \frac{\sqrt{3}}{2} } \, , \nn\\
& & 
        = \log\left(x-1/2- i \sqrt{3}/2\right) - 
             \log\left( - 1/2-i \sqrt{3}/2\right),~~~~~ \\
H(\overline{c},x) & = & \int_{0}^{x} 
\frac{dt}{ t-\frac{1}{2}+i \frac{\sqrt{3}}{2} } \, , \nn\\
& &  
= \log\left(x-1/2+i \sqrt{3}/2\right) - 
  \log\left( - 1/2+i \sqrt{3}/2\right),~~~~~ \\
H(-r,x) & = & \int_{0}^{x} \frac{dt}{\sqrt{t(4+t)}} = 
2~{\rm arcsinh}\left(\frac{\sqrt{x}}{2}\right) 
\, , \\
        & = & 2 \log{(\sqrt{x+4}+\sqrt{x})} - 2 \log{2} 
\, , \\
H(r,x) & = & \int_{0}^{x} \frac{dt}{\sqrt{t(4-t)}} = 2~{\rm arcsin }\left(\frac{\sqrt{x}}{2}\right) 
\, , \\
        & = & - i \log{ \left\{ \frac{ \sqrt{4-x}+i\sqrt{x} }{ \sqrt{4-x}-i\sqrt{x} }
	\right\} }
\, , \\
H(-1-r,x) & = & \int_{0}^{x} \frac{dt}{ \sqrt{t(4+t)}(1+t) } 
   = \frac{2}{ \sqrt{3} }~{\rm arctan}\left( \sqrt{ \frac{3 x}{4+x}  }\right) 
\, , \\
   &=& \frac{i}{ \sqrt{3} }\log{ \left\{ 
   \frac{\sqrt{4+x} - i\sqrt{ 3 x } } {\sqrt{4+x} + i\sqrt{ 3 x } } \right\} }
 \, , \\
H(1-r,x) & = & \int_{0}^{x} \frac{dt}{\sqrt{t(4+t)}(1-t)} 
= \frac{1}{\sqrt{5}} \log{ \left\{ \frac{ \sqrt{4+x}+\sqrt{5x} }{ \sqrt{4+x}-\sqrt{5x} }
	\right\} },~~~~~~ \\
H(-1+r,x) & = & \int _{0}^{x}\frac{dt}{\sqrt{t(4-t)}(1+t)} 
= \frac{2}{ \sqrt{5} }~{\rm arctan}\left( \sqrt{ \frac{5 x}{4-x}  }\right)
\, , \\
& = & \frac{i}{\sqrt{5}}\log{ \left\{ 
\frac{ \sqrt{4-x}-i\sqrt{5 x} } { \sqrt{4-x}+i\sqrt{5 x} } \right\} } ,~~~~~~\\
H(1+r,x) & = & \int_{0}^{x} \frac{dt}{\sqrt{t(4-t)}(1-t)}  
= \frac{1}{\sqrt{3}} \log{ \left\{ \frac{ \sqrt{4-x}+\sqrt{3x} }{ \sqrt{4-x}-\sqrt{3x} }
	\right\} } \, .
\eea
Note that all the integrals above have zero as the lower limit of integration.

The GHPLs of general weight $w>1$ are defined exactly in the same way as the HPLs, 
according to Eq.~(\ref{defHPL}), where now the components of $\vec{w}$ can also
take the values $\pm 4,c,\overline{c},\pm r,\pm 1 \pm r$. 

Let us now discuss how the integrals appearing in the evaluation of the MIs
can be reduced to the GHPLs plus, of course, elementary functions.
As far as the indices $\pm 4,c$ and $\overline{c}$ are concerned, there is basically 
nothing new with respect to the HPLs: the only difference is that individual terms in 
the expressions for the MIs are in general complex and only their sum is, at it 
should, real. 

When radicals, i.e. semi-integer powers are involved, the situation is more complicated.
The integral of a semi-integer power,
\be
\int (x+a)^{n-1/2} dx = \frac{(x+a)^{n+1/2}}{n+1/2}~~~~~(n~{\rm is~an~integer}),
\ee
is always a semi-integer power, i.e. an elementary function. There is no need, then, 
to introduce basis functions of this kind.
On the other hand, integrals of products of radicals, such as
\be
\label{twoindex}
I(\alpha,\beta)=\int (x+a)^{\alpha} (x+b)^{\beta} dx,
\ee
with $\alpha$ and $\beta$ half integers,
involve in general new transcendental 
functions\footnote{
On the contrary, products of integer powers do not require the introduction of new basis
functions because partial fractioning completely disentangles the factors:
\be
\frac{1}{(x+a)^n}\frac{1}{(x+b)^k}=\frac{A_n}{(x+a)^n}+\cdots+\frac{A_1}{x+a}+
                                   \frac{B_n}{(x+b)^n}+\cdots+ \frac{B_1}{x+b}.
\ee}.
By taking $a=0$ and $b=\pm 4$ (or $a=\pm 4$ and $b=0$), we cover with $I(\alpha,\beta)$ 
the cases related to the basis functions with indices $\pm r$.
The idea is to shift the indices to reference values,
such as $\alpha=-1/2$ and $\beta=-1/2$ in our (arbitrary) choice of the basis functions, 
by using recursive relations.

Let us consider the general case in which {\it both} indices differ from their 
target values: $\alpha\neq -1/2$ {\it and} $\beta\neq -1/2$.
The integrand of (\ref{twoindex}) is conveniently rewritten as
\be
(x+a)^{\alpha} (x+b)^{\beta} = \frac{1}{\sqrt{(x+a)(x+b)}} (x+a)^n (x+b)^l,
\ee
with $n=\alpha+1/2$ and $l=\beta+1/2$ general integers.

We first make an algebraic reduction on the integrand to take {\it one} 
of the {\it two} indices $n$ and $l$
to its target value zero. There are two possibilities:
\begin{itemize}
\item
If 1 of the 2 indices is positive, let's say $n>0$, we expand $(x+a)^n$ in powers of
$x+b$ with the binomial formula:
\be
(x+a)^n = \sum_{s=0}^{n} {n\choose s} (a-b)^{n-s}~(x+b)^s, 
\ee
so that the integrand takes the form
\be
\label{algebr1}
(x+a)^{\alpha} (x+b)^{\beta} = 
\sum_{s=0}^{n} {n\choose s} (a-b)^{n-s}~(x+a)^{-1/2} 
(x+b)^{s+l-1/2}~~~~~({\rm if}~n>0),~~~~
\ee
in which the index $\alpha$ reached its target value.
\item
If {\it both} indices are negative, $n<0$ and $l<0$, 
we do ordinary partial fractioning:
\begin{eqnarray}
   \frac{1}{(x+a)^{|n|}} \frac{1}{(x+b)^{|l|}} 
&=& \frac{A_{|n|}}{(x+a)^{|n|}} + \frac{A_{|n|-1}}{(x+a)^{|n|-1}}+\cdots +\frac{A_1}{x+a}~~~~
\\ \nonumber  
&+& \frac{B_{|l|}}{(x+b)^{|l|}} + \frac{B_{|l|-1}}{(x+b)^{|l|-1}}+\cdots +\frac{B_1}{x+b},
\end{eqnarray}
so that
\begin{eqnarray}
\label{algebr2}
\nonumber
 (x+a)^{\alpha} (x+b)^{\beta}  &=&       
        A_{|n|}   (x+a)^{-1/2-|n|} (x+b)^{-1/2} \nn\\
& + &   A_{|n|-1} (x+a)^{1/2-|n|} (x+b)^{-1/2} \nn\\
& + & \cdots~ + ~ A_{1}~(x+a)^{-3/2}~(x+b)^{-1/2} \nn\\
& + & B_{|l|}   (x+a)^{-1/2} (x+b)^{-1/2-|l|} \nn\\
& + & B_{|l|-1} (x+a)^{-1/2} (x+b)^{1/2-|l|} \nn\\
& + & \cdots~ + ~ B_{1}~(x+a)^{-1/2}~(x+b)^{-3/2} \nn\\
&  & 
({\rm if}~n<0~{\rm and}~l<0).
\end{eqnarray}
In any of the above terms on the l.h.s. one of the two indices reached its target value.
\end{itemize}

To take also the second index to its target value, we use the integral identity:
\be
\label{easyrec}
I(\alpha,\beta) =
\frac{1}{\alpha +\beta +1} (x+a)^{\alpha +1} (x+b)^{\beta}
-\frac{\beta(a-b)}{\alpha +\beta +1} I(\alpha,\beta-1).
\ee
The above relation is obtained by doing an integration by parts on $I(\alpha,\beta)$:
we integrate $(x+a)^{\alpha}$ and differentiate $(x+b)^{\beta}$ with respect to $x$;
we then simplify the ratio $(x+a)/(x+b)$ as $1+(a-b)/(x+b)$.

The first term on the r.h.s. is a finite one and is therefore known.
The above equation can then be used to lower the index $\beta$ by one unit.
For example $I(\alpha,3/2)$ can be transformed to a linear combination
of $I(\alpha,1/2)$ and $I(\alpha,-1/2)$, while $I(\alpha,1/2)$ can be reduced to $I(\alpha,-1/2)$.
In general, by recursively using equation (\ref{easyrec}), one can reduce any 
$\beta<-1/2$ to $\beta=-1/2$\footnote{
Since we can take $\alpha=-1/2$ and the equation is used for $\beta>-1/2$, it holds that $\alpha+\beta>-1$, 
so the singularity in $\alpha+\beta=-1$ is always avoided.}.

By solving Eq.~(\ref{easyrec}) with respect to the last term on the r.h.s, $I(\alpha,\beta-1)$, 
and shifting $\beta\rightarrow \beta+1$, one obtains an identity to raise
the index $\beta$ by one unit:\footnote{
Note the the singularity for $\beta=-1$ is never reached since $\beta$ takes half-integer values only.} 
\be\label{easyrec2}
I(\alpha,\beta) = \frac{(x+a)^{\alpha +1} (x+b)^{\beta+1}}{(\beta +1)(a-b)} 
- \frac{\alpha +\beta +2} {(\beta +1)(a-b)} I(\alpha,\beta+1).
\ee
By using Eqs.~(\ref{easyrec}) and (\ref{easyrec2}) 
we can then take the index $\beta$ to any desired reference value, such as the value $-1/2$ of our basis.
Then, we succeeded in reducing an integral of the form (\ref{twoindex}) to $H(\pm r;x)$ plus terms
containing elementary functions only.

In general, we encounter integrals containing both radicals and GHPLs, of the form:
\be
\label{secondred}
J(\alpha,\beta)=\int (x+a)^{\alpha} (x+b)^{\beta} H(v,\vec{w};x) dx,
\ee
where $H(v,\vec{w};x)$ is a GHPL whose first index $v$ has been separated out for a 
reason that will become clear soon. Since $\vec{w}$ has $w$ components 
$H(v,\vec{w};x)$ has weight $w+1$.
As with the previous integral, we can transform one of the two indices $\alpha$ and $\beta$,
let's say $\alpha$, to its target value $\alpha=-1/2$.
The identity to reduce the second index is:
\bea
\label{redJ}
J(\alpha,\beta) &=&
\frac{1}{\alpha +\beta +1} (x+a)^{\alpha +1} (x+b)^{\beta} H(v,\vec{w};x)
\\ \nonumber
&-&\frac{1}{\alpha +\beta +1} \int (x+a)^{\alpha +1} (x+b)^{\beta} g(v;x) H(\vec{w};x) dx
\\ \nonumber
&-&\frac{\beta(a-b)}{\alpha +\beta +1} J(\alpha,\beta-1).
\eea
The second term on the r.h.s. involves the integration of a GHPL of smaller weight.
We then consider the last term as the only relevant one in the recursion. 
As before, the above equation can be directly used to lower the index $\beta$ by one unit.
Note that the indices inside the GHPL are not touched by the reduction.
Analogously to the previous case, by solving Eq.~(\ref{redJ}) with respect to $J(\alpha,\beta-1)$ 
and sending $\beta\rightarrow\beta+1$, one obtains an identity to raise $\beta$ by one unit.

In some cases, it is also necessary to integrate expressions involving 3 factors,
of the form:
\be
\label{ultint}
L(\alpha,\beta;k)=\int~dx~(x+a)^{\alpha}~(x+b)^{\beta}~(x+c)^{-k},
\ee
with $\alpha$ and $\beta$ half integers and $k$ a general integer\footnote{
We have put a minus sign in front of $k$ just for practical convenience, 
in order to simplify the forthcoming results
a little bit.}.
For our computation, the relevant cases are $a=0$, $b=\pm 4$ and $c=\pm 1$ 
(or $a=\pm 4$, $b=0$ and $c=\pm 1$).
Let us consider the various possibilities for the $k$ index.
For $k<0$ one can reduce the integral (\ref{ultint}) to the simpler form 
of $I(\alpha,\beta)$ in (\ref{twoindex}) by means of the binomial expansion 
of $(x+c)^{|k|}$ in powers of $x+a$ or of $x+b$:
\be
(x+c)^{|k|} = \sum_{s=0}^{|k|} {|k| \choose s} (c-a)^{|k|-s} (x+a)^{s} = 
\sum_{s=0}^{|k|} {|k| \choose s} (c-b)^{|k|-s} (x+b)^{s}.
\ee
Therefore we just need to consider the case $k>0$ from now on.

By means of the algebraic relations in (\ref{algebr1}) and (\ref{algebr2}), 
we can shift one of the half-integer indices $\alpha$ and $\beta$
to the target value $-1/2$; let us assume for instance that $\alpha=-1/2$.
We can then assume the integral of the form:
\be
L^{'}(\beta;k)~=~L(-1/2,\beta;k)~=~\int \frac{dx}{ \sqrt{(x+a) (x+b)} }  (x+b)^{l} (x+c)^{-k},
\ee
with $l=\beta+1/2$ a general integer.

With an algebraic reduction analogous to the one in Eqs.~(\ref{algebr1}) and 
(\ref{algebr2}), we can reduce the product $(x+b)^{l} (x+c)^{-k}$
to a linear combination of terms involving powers of either $(x+b)$ or $(x+c)$, 
but not the product. The terms not containing $(x+c)$ belong to the simpler class 
of the integrals $I(\alpha,\beta)$ defined in Eq.~(\ref{twoindex}), whose reduction 
has already been discussed. The terms not containing $(x+b)$ have the indices 
$\alpha$ and $\beta$ both equal to their target value $-1/2$. 
Our task is then the evaluation of integrals of the form:
\be
\tilde{L}(k)~=~L(-1/2,-1/2;k)~=~\int  \frac{dx}{ \sqrt{(x+a) (x+b)} } (x+c)^{-k},
\ee
with $k$ integer and strictly positive, $k \in N$.
The required identity is:
\begin{eqnarray}
\label{genred}
\tilde{L}(k) = &-&\frac{1}{k-1}~(x+a)^{-1/2} (x+b)^{-1/2} (x+c)^{-k+1} 
\\ \nonumber
&-&\frac{1}{2(k-1)}~\frac{1}{(c-a)^{k-1}} \int dx~(x+a)^{-3/2}~(x+b)^{-1/2}
\\ \nonumber
&-&\frac{1}{2(k-1)}~\frac{1}{(c-b)^{k-1}} \int dx~(x+a)^{-1/2}~(x+b)^{-3/2}
\\ \nonumber
&+&\frac{1}{2(k-1)}\sum_{l=1}^{k-1}\left[ \frac{1}{(c-a)^{k-l}}+\frac{1}{(c-b)^{k-l}} \right] \tilde{L}(l).
\end{eqnarray}
The second and the third terms on the r.h.s. do not contain any power of $(x+c)$, 
can be reduced with the previous identities and are therefore known terms.
The last term contains integrals $\tilde{L}(l)$ of the same form as the one on the 
l.h.s.: $\tilde{L}(k)$.
Eq.~(\ref{genred}) is then to be used to relate $\tilde{L}$ integrals, 
treating the other terms as known functions. 
By using Eq.~(\ref{genred}) for instance for $k=3$ one can reduce $\tilde{L}(3)$ 
to a superposition of $\tilde{L}(2)$ and $\tilde{L}(1)$, while using it for
$k=2$ one can reduce $\tilde{L}(2)$ to $\tilde{L}(1)$.
In general, by recursively using Eq.~(\ref{genred}), one can reduce any $k>1$ to $k=1$,
i.e. to the basic 
integral\footnote{
Note that the singularity of the coefficients in Eq.~(\ref{genred}) for $k=1$ 
forbids any further reduction.}
\be
\tilde{L}(1) = \int dx 
\frac{1}{ \sqrt{(x+a)(x+b)}~(x+c) },
\ee
which defines the $H(\pm r\pm 1;x)$'s.

The derivation of Eq.~(\ref{genred}) is analogous to the ones of the previous identities:
one has to integrate $(x+c)^{-k}$ and to differentiate
$(x+a)^{-1/2} (x+b)^{-1/2}$ with respect to $x$. 
The resulting expressions have to be simplified using
$(x+c)/(x+a) = 1+(c-a)/(x+a)$ and an analogous equation with $b$ replacing $a$.
One also needs the following result:
\be
\label{induction}
\frac{1}{(x+c)^k}~\frac{1}{x+a} = 
-\sum_{l=0}^{k-1}  \frac{1}{(c-a)^{l+1}}~\frac{1}{(x+c)^{k-l}} 
+\frac{1}{(c-a)^k}~\frac{1}{x+a}
\ee
and an analogous equation with $b$ replacing $a$.

Finally, the last class of integrals we need to consider is:
\be
\label{ultultint}
F(\alpha,\beta;k) = \int~dx~(x+a)^{\alpha}~(x+b)^{\beta}~(x+c)^{-k}~H(v,\vec{w};x).
\ee
As with the previous case, we can reduce ourselves to the case 
$\alpha=-1/2$, $\beta=-1/2$ and $k>0$, i.e. to:
\be
\label{ultultint2}
\tilde{F}(k)~=~F(-1/2,-1/2;k)~=~\int~\frac{dx}{ \sqrt{(x+a)~(x+b)}~(x+c)^{k} }~H(v,\vec{w};x).
\ee
The identity reads:
\begin{eqnarray}
\label{genred2}
\tilde{F}(k) = &-&\frac{1}{k-1}~(x+a)^{-1/2} (x+b)^{-1/2} (x+c)^{-k+1} H(v,\vec{w};x)
\\ \nonumber
&-&\frac{1}{2(k-1)}~\frac{1}{(c-a)^{k-1}} \int dx~(x+a)^{-3/2}~(x+b)^{-1/2}~H(v,\vec{w};x)
\\ \nonumber
&-&\frac{1}{2(k-1)}~\frac{1}{(c-b)^{k-1}} \int dx~(x+a)^{-1/2}~(x+b)^{-3/2}~H(v,\vec{w};x)
\\ \nonumber
&+&\frac{1}{k-1}\int dx~(x+a)^{-1/2}~(x+b)^{-1/2}~(x+c)^{-k+1} g(v;x) H(\vec{w};x)
\\ \nonumber
&+&\frac{1}{2(k-1)}\sum_{l=1}^{k-1}\left[ \frac{1}{(c-a)^{k-l}}+\frac{1}{(c-b)^{k-l}} \right]\tilde{F}(l).
\end{eqnarray}
The fourth term on the r.h.s. involves the integration of an elementary function 
times a GHPL of weight $w$, while the l.h.s involves the integration of a GHPL of weight $w+1$. 
This term then has a smaller recursive weight than $\tilde{F}$ 
and has to be considered as a known integral.
For the rest, analogous considerations to the one of the previous identity hold.
By recursively using Eq.~(\ref{genred2}), one can reduce any $k>1$ to $k=1$,
i.e. to the basic integral
\be
\tilde{F}(1) = \int dx 
\frac{1}{ \sqrt{(x+a)(x+b)}~(x+c) } H(v,\vec{w};x) ,
\ee
which defines the GHPLs of weight $w+2$.

\subsection{Closure under the transformation $x \rightarrow 1/x$ --- 
a further extension}
The extended basis function set introduced in the previous Section preserves 
the following fundamental properties of the HPLs:
\begin{itemize}
\item
the uniqueness of representation as repeated integration noted before 
\cite{Polylog};
\item
the fulfilling of an algebra. This means that the product of two GHPLs of weights
$w_1$ and $w_2$ can be written as a sum of GHPLs of weight
$w_1 + w_2$. 
\end{itemize}
The ordinary HPLs, however, have also the property of ``closure'' under the
inverse transformation $x \rightarrow 1/x$, i.e. anyone of the functions $g(\pm 1;x),g(0;x)$
transforms into a linear combination of the same functions under $x \rightarrow 1/x$. 
This property, which is useful for the large momentum expansion of the MIs, is not
preserved by the generalization discussed in the previous Section.
One can impose this property on the GHPLs at the price of a second basis
extension. Before doing that, however, let us discuss in detail how
the expansion of the ordinary HPLs for large values of the argument $x$ is performed. 
The followings steps are taken in order:
\begin{enumerate}
\item we set $x=1/y$ in the HPL under consideration:
\be
H(\vec{w};x)=H(\vec{w};1/y) \, ;
\ee
\item we use the identities which allow to reduce  $H(\vec{w};1/y)$ to a
combination of $H(\vec{w}';y)$'s;
\item we expand $H(\vec{w}';y)$ for $y \rightarrow 0$, i.e. for small value of the
argument, as explained in some detail in Section \ref{PiccoliP};
\item we perform the inverse substitution $y=1/x$ in the final result.
\end{enumerate}

As an example consider the simple weight-1 HPL 
\be
H(-1;x) = \int_{0}^{x} \frac{dt}{1+t} = \log{(1+x)} \, .
\ee

We have:
\bea
H(-1;x) & = & H(-1;1/y) 
\\
&=& \int_{0}^{1} \frac{dt}{1+t} + \int_{1}^{1/y} 
\frac{dt}{1+t} \, , 
\label{arriv1} \\
& = & H(-1;1) - \int_{1}^{y} dt' \left[ \frac{1}{t'} - \frac{1}{1+t'}
\right] \, , 
\label{arriv2} \\
& = & - H(0;y) + H(-1;y) \, ,
\label{arriv3}
\eea
where, moving from Eq.~(\ref{arriv1}) to Eq.~(\ref{arriv2}), we divided
the integral into the sum of two integrals (the first between 0 and 1
and the second between 1 and $1/y$) and we replaced $t$ by $1/t'$.

Expanding the r.h.s. of Eq.~(\ref{arriv3}) in series of $y$ and
re-expressing $y$ as $1/x$ we finally have:
\be
H(-1;x) \stackrel{x \rightarrow \infty}{=} \log{x} + \frac{1}{x} - \frac{1}{2x^2} 
+ {\mathcal O} \left( \frac{1}{x^3} \right) \, .
\ee

For the weight-2 HPL 
\be
H(0,-1;x) = \int_{0}^{x} \frac{dt}{t} H(-1;t) \, ,
\ee
we have:
\bea
H(0,-1;x) & = & H(0,-1;1/y) 
\\
&=& H(0,-1;1) + \int_{1}^{1/y} 
\frac{dt}{t} H(-1;t) \, , 
\label{arriv4} \\
& = & H(0,-1;1) - \int_{1}^{y} \frac{dt'}{t'} H(-1;1/t')  \, , 
\label{arriv5} \\
& = & H(0,-1;1) - \int_{1}^{y} \frac{dt'}{t'} \left[ - H(0;t') 
+ H(-1;t') \right] \, ,
\label{arriv6} \\
& = & 2 H(0,-1;1) + H(0,0;y) - H(0,-1;y) \, ,
\label{arriv7}
\eea
where, moving from Eq.~(\ref{arriv5}) to Eq.~(\ref{arriv6}) we used
Eq.~(\ref{arriv3}).

Expanding the r.h.s. of Eq.~(\ref{arriv7}) in series of $y$ and
re-expressing $y$ as $1/x$ we have finally:
\be
H(0,-1;x) \stackrel{x \rightarrow \infty}{=} 2 H(0,-1;1)
+ \frac{1}{2} \log^{2}{x} - \frac{1}{x} + \frac{1}{4x^2} 
+ {\mathcal O} \left( \frac{1}{x^3} \right) \, .
\ee

Let us remark that performing the substitution $x=1/y$ the basis
functions $g(n,1/y)$ are expressed in terms of the basis functions
belonging to the same set, $g(n',y)$: this is the closure under the
transformation $x \rightarrow 1/x$. Moreover, in the final result
the constants $H(\vec{w};1)$ do appear. The latter have been systematically
evaluated and tabulated in \cite{vermaseren} up to weight $4$ included.

Following exactly the same steps in the case of
the GHPLs, we find that the set of basis functions given in the previous
Section is too small to preserve the closure under the transformation 
$x \rightarrow 1/x$. We can understand it, for example, trying to expand
the GHPL
\be
H(4;x) = \int_{0}^{x} \frac{dt}{4-t} = - \log{(4-x)} - 2 \log{2} \, .
\ee
We have:
\bea
H(4;1/y) & = & H(4;1) + \int_{1}^{1/y} \frac{dt}{4-t} \, , \\
& = & H(4;1) - \frac{1}{4} \int_{1}^{y} \frac{dt'}{t'} \frac{1}{t'-
\frac{1}{4}} \, , \\
& = & H(4;1) + \int_{1}^{y} dt' \left[ \frac{1}{t'} - \frac{1}{t'-
\frac{1}{4}} \right] \, , \\
& = & H(4;1) + H(0;y) - \int_{1}^{y} dt' \frac{dt'}{t'-
\frac{1}{4}} \, .
\label{arriv8}
\eea
The integral in Eq.~(\ref{arriv8}) does not belong to the set of GHPLs
of the previous Section. 
We then add to the basis functions
defined in Eqs.~(\ref{meno4}--\ref{1piuR}) the following functions:
\bea
g(\pm 1/4;x) & = & \frac{1}{\frac{1}{4} \mp x} 
\, ,\label{prim} \\
g(\pm 1 \pm r/4;x) & = & \frac{1}{\sqrt{x \mp \frac{1}{4}} (1 \mp x)} 
\, , \\
g(r_0/4;x) & = & \frac{1-2i \sqrt{x-\frac{1}{4}}}{x \sqrt{x-\frac{1}{4}}}
\, , \\
g(-r_0/4;x) & = & \frac{1-2 \sqrt{x+\frac{1}{4}}}{x \sqrt{x+\frac{1}{4}}} 
\, .
\label{ult}
\eea
To avoid a pole in $x=0$, we have subtracted $g(0;x)=1/x$ with a proper
coefficient, the residue in $x=0$ of the new functions. 
In taking the limit $x\rightarrow 0$
inside square roots one has to remember that $x\rightarrow x-i\epsilon$,
with $\epsilon$ a small positive number.
The related GHPLs of weight 1 are defined as usual:
\be
H(w;x) = \int_{0}^{x} g(w;t) dt \, ,
\ee
since all the functions (\ref{prim}--\ref{ult}) are integrable in $x=0$.

As an example, the reader may verify that:
\bea
H(4,-r;x) & = & H(4,-r;1/y) 
\\
&=& - H(4;1) H(r/4;1) + \frac{1}{2} H(-1,-r_0/4;1)
\nn\\
& & + H(1,r/4;1) - \frac{1}{2} H(-r_0/4,4;1) \nn\\
& & + \left[ H(r/4;1) + \frac{1}{2} H(-r_0/4;1)
\right] \Bigl[ H(0;y) + H(4;y) \Bigr] \nn\\
& & - \frac{1}{2} H(4,-r_0/4;y) - \frac{1}{2} H(-1,-r_0/4;y) 
\, .
\eea
Expanding the GHPLs for $y \rightarrow 0$ and substituting in the final
expression $y=1/x$ we finally obtain:
\bea
H(4,-r;x) & \stackrel{x \to \infty}{=} & - H(4;1) H(r/4;1) 
+ \frac{1}{2} H(-1,-r_0/4;1) \nn\\
& & + H(1,r/4;1) - \frac{1}{2} H(-r_0/4,4;1) \nn\\
& & - \left[ H(r/4;1) + \frac{1}{2} H(-r_0/4;1) \right] \, \log{x} 
+ {\mathcal O} \left( \frac{1}{x} \right) \, .
\label{asyexp}
\eea

As in the case of the HPLs, the final expression contains the 
constants $H(\vec{w};1)$, which have to be evaluated.
The $x=1$ GHPLs of weight one, $H(a;1)$, can be expressed in terms of
known transcendental constants in an elementary way.
In Eq.~(\ref{asyexp}), the coefficient of the leading
logarithm is, for example:
\be
H(r;1) + \frac{1}{2} H(-r_0/4;1) = \frac{\pi}{3}
- 2 \bigl[ \log{(1+\sqrt{5})} - \log{2} \bigr] \, .
\ee
Similar reductions are also possible in other cases
but, in general, it is non trivial to reduce the $H(\vec{w};1)$'s 
with a general weight $\vec{w}$ to 
a minimal set containing known transcendental constants 
such as $\zeta(n)$ and eventual new transcendental constants. 

The reduction of the $H(\vec{w};1)$'s to a minimal set of transcendental
constants is beyond the goal of the present paper  \cite{inpreparation};
we restrict ourselves to the evaluation of the asymptotic expansions
of the diagrams involving only ordinary HPLs.

\section{Results for the master integrals \label{Results}}

In this Section we present the results of our computation of the MIs 
involving up to 6 denominators included, which constitute a  
necessary input for the calculation of the 2-loop vertex diagrams 
in Figs.~\ref{fig1} and \ref{fig1bis}. They are expanded in a Laurent 
series in 
\begin{equation}
\epsilon = 2-D/2, 
\end{equation}
up to the required order in $\epsilon$. The coefficients of the series 
are expressed in terms of GHPL's (see Section~\ref{HPLs}) of the 
variable $x$, defined as :
\be
x = \frac{-s}{a} \, ,
\ee
where $s=-q^2$ is the c.m. energy 
squared\footnote{We define the scalar product of two 4-vectors as:
$a\cdot b=-a_0b_0+\vec{a}\cdot\vec{b}$.}. 
It holds $q = p_{1}+p_{2}$ and we defined $ a = m^2 $.
We denote by $\mu$ the mass scale of the Dimensional Regularization (DR)
--- the so-called unit of mass. 
We work in Minkowski space and we normalize the loop measures as:
\be
{\mathfrak D}^D k = \frac{d^{D}k}{i\pi^{\frac{D}{2}} \Gamma 
\left( 3 - \frac{D}{2} \right) } = \frac{d^{4-2\epsilon}k}{i
\pi^{2-\epsilon} \Gamma 
\left( 1+\epsilon \right) }  .
\ee
This definition makes the expression of the 1-loop tadpole 
--- the simplest of all loop diagrams --- particularly simple 
\cite{UgoRo}.
The denominators appearing in the master integrals are listed below:

\bea
{\mathcal D}_{1} & = & k_{1}^{2} \, , \\
{\mathcal D}_{2} & = & k_{2}^{2} \, , \\
{\mathcal D}_{3} & = & (k_{1}+k_{2})^{2} \, , \\
{\mathcal D}_{4} & = & (p_{1}-k_{1})^{2} \, , \\
{\mathcal D}_{5} & = & (p_{2}+k_{1})^{2} \, , \\
{\mathcal D}_{6} & = & (p_{2}-k_{2})^{2} \, , \\
{\mathcal D}_{7} & = & (p_{1}-k_{1}+k_{2})^{2} \, , \\
{\mathcal D}_{8} & = & (p_{2}+k_{1}-k_{2})^{2} \, , \\
{\mathcal D}_{9} & = & (p_{1}+p_{2}-k_{1})^{2} \, , \\
{\mathcal D}_{10} & = & (p_{1}+p_{2}-k_{2})^{2} \, , \\
{\mathcal D}_{11} & = & (p_{1}+p_{2}-k_{1}-k_{2})^{2} \, , \\
{\mathcal D}_{12} & = & k_{1}^{2} + a \, , \\
{\mathcal D}_{13} & = & k_{2}^{2} + a \, , \\
{\mathcal D}_{14} & = & (k_{1}+k_{2})^{2} + a \, , \\
{\mathcal D}_{15} & = & (p_{1}-k_{1})^{2} + a \, , \\
{\mathcal D}_{16} & = & (p_{2}+k_{1})^{2} + a \, , \\
{\mathcal D}_{17} & = & (p_{2}-k_{2})^{2} + a \, , \\
{\mathcal D}_{18} & = & (p_{1}-k_{1}+k_{2})^{2} + a \, , \\
{\mathcal D}_{19} & = & (p_{2}+k_{1}-k_{2})^{2} + a \, , \\
{\mathcal D}_{20} & = & (p_{1}+p_{2}-k_{1})^{2} + a \, , \\
{\mathcal D}_{21} & = & (p_{1}+p_{2}-k_{2})^{2} + a \, , \\
{\mathcal D}_{22} & = & (p_{1}+p_{2}-k_{1}-k_{2})^{2} + a
\, .
\eea
The MIs are listed according to the increasing number of the denominators,
which corresponds more or less to the level of complexity.

\subsection{Topology $t=3$ \label{3den}}

\bea
\parbox{20mm}{\begin{fmfgraph*}(15,15)
\fmfleft{i}
\fmfright{o}
\fmfforce{0.5w,0.1h}{v1}
\fmfforce{0.25w,0.62h}{v3}
\fmfforce{0.5w,0.9h}{v7}
\fmfforce{0.74w,0.62h}{v11}
\fmf{plain,left=.1}{v1,v3}
\fmf{plain,left=.5}{v3,v7}
\fmf{plain,left=.5}{v7,v11}
\fmf{plain,left=.1}{v11,v1}
\fmf{plain}{v1,v7}
\end{fmfgraph*}} & = & \mu^{2(4-D)} 
\int {\mathfrak D}^D k_1 {\mathfrak D}^D k_2
\frac{1}{{\mathcal D}_{12} {\mathcal D}_{13} {\mathcal D}_{14} } \\
& = & \left( \frac{\mu^{2}}{a} \right) ^{2 \epsilon} 
\sum_{i=-2}^{2} \epsilon^{i} F^{(1)}_{i} + {\mathcal O} \left( 
\epsilon^{3} \right) , 
\eea
where:
\bea
\frac{F^{(1)}_{-2}}{a} & = & - \frac{3}{2} \, , \\
\frac{F^{(1)}_{-1}}{a} & = & - \frac{9}{2} \, , \\
\frac{F^{(1)}_{0}}{a} & = & - \frac{21}{2} - \sqrt{3} H(r,0;1) \, , \\
\frac{F^{(1)}_{1}}{a} & = & - \frac{45}{2} - \sqrt{3} \Bigl[ 3 H(r,0;1)
+ H(r,0,0;1) + H(4,r,0;1) \Bigr]
\, , \\
\frac{F^{(2)}_{2}}{a} & = &  - \frac{93}{2} 
- \sqrt{3} \Bigl[ 7 H(r,0;1) + 3 H(r,0,0;1) + 3 H(4,r,0;1) 
\nn\\
& &~~~~~+ H(r,0,0,0;1) + H(4,r,0,0;1) + H(4,4,r,0;1) 
\Bigr]
\, .
\eea
This diagram has been originally evaluated in ref.~\cite{Davy1}.

Even though the above MI is a vacuum amplitude, its computation
is non trivial because of the presence of 3 massive propagators.
We have evaluated it with the following method. 
We consider a vacuum sunrise with 2 
internal lines with equal mass $m$ and the third internal line
with the different mass (squared) ${m'}^2 = z m^2$. We then differentiate
the vacuum sunrise with respect to $z$ and rewrite the result 
in terms of MIs by using the ibps identities. The resulting 
differential equation represents the evolution in one of the masses and is 
solved as in usual cases. We set at the end $z=1$ to obtain the equal-mass case.

Let us make a few remarks about the above result:
\begin{itemize}
\item 
The finite part of the MI, i.e. the $O(\epsilon^0)$, involves 1 transcendental
constant: $H(r,0;1)$, related to the Clausen function;
\item
the $O(\epsilon)$ part involves 2 new 
independent transcendental constants: $H(r,0,0;1)$ and $H(4,r,0;1)$; 
\item
the $O(\epsilon^2)$ part involves 3 new transcendental constants:
$H(r,0,0,0;1)$, $H(4,r,0,0;1)$ and $H(4,4,r,0;1)$.
\end{itemize}
To simplify the above expressions, we have used the following identities
to move the ``0'' indices to the right:
\bea
H(0,r;1)&=&-H(r,0;1),
\nn \\
H(0,r,0;1)&=&-2H(r,0,0;1),
\nn \\
H(0,r,0,0;1)&=&-3H(r,0,0,0;1),
\nn \\
H(0,0,r,0;1)&=& 3H(r,0,0,0;1),
\nn \\
H(0,4,r,0;1)&=&-H(4,0,r,0;1)-2 H(4,r,0,0;1).
\eea
The above relations are obtained by transforming products of $H$'s
into linear combinations of $H$'s, as for example in:
\be
0=H(0;1)H(r;1)=H(0,r;1)+H(r,0;1).
\ee

\subsection{Topology $t=4$ \label{4den}}

\bea
\parbox{20mm}{\begin{fmfgraph*}(15,15)
\fmfleft{i}
\fmfright{o}
\fmfforce{0.2w,0.5h}{v1}
\fmfforce{0.5w,0.2h}{v2}
\fmfforce{0.8w,0.5h}{v3}
\fmf{plain}{i,v1}
\fmf{plain}{v3,o}
\fmf{plain,left}{v1,v3}
\fmf{plain,right=.4}{v1,v2}
\fmf{photon,right=.4}{v2,v3}
\fmf{photon,left=.6}{v2,v3}
\end{fmfgraph*}} & = & \mu^{2(4-D)} 
\int {\mathfrak D}^D k_1 {\mathfrak D}^D k_2
\frac{1}{{\mathcal D}_{2} 
         {\mathcal D}_{11} 
	 {\mathcal D}_{12} 
	 {\mathcal D}_{20} } \\
& = & \left( \frac{\mu^{2}}{a} \right) ^{2 \epsilon} 
\sum_{i=-2}^{2} \epsilon^{i} F^{(2)}_{i} + {\mathcal O} \left( 
\epsilon^{3} \right) , 
\eea
where:
\bea
F^{(2)}_{-2} & = & \frac{1}{2} \, , \\
F^{(2)}_{-1} & = & \frac{5}{2}
          - \frac{x+4}{\sqrt{x(x+4)}} 
            H( - r;x)\, , \\
F^{(2)}_{0} & = & \frac{19}{2}
          + \zeta(2)
          - \frac{1}{2} H(0,-1;x)
       - \frac{x+4}{\sqrt{x(x+4)}}   \Biggl[
            4 H( - r;x)
          - \frac{3}{2} H( - r,-1;x) \nn\\
& & 
          - H(-4, - r;x)
          \Biggr]
       - \Biggl[ \frac{1}{x} + 1 \Biggr]  
           H(-1;x)
          \, , \\
F^{(2)}_{1} & = & 
            \frac{65}{2}
          + 5 \zeta(2)
          - \zeta(3) - 3 H(0,-1;x)
          + 2 H(0,-1,-1;x)
          - \frac{1}{2} H(0,0,-1;x)  \nn\\
& & 
	  + \Biggl[ \frac{1}{x} + 1 \Biggr]   \Bigl\{
          - 7 H(-1;x)
          + 4 H(-1,-1;x)
          - H(0,-1;x)
          \Bigr\} \nn\\
& & 
       + \frac{x+4}{\sqrt{x(x+4)}}  \Biggl\{
          -  \! 2(6 \! +  \! \zeta(2) ) H( - r;x) \! 
          +  \! 6 H( - r,-1;x) \! 
          -  \! 6 H( - r, \! -1, \! -1;x) \nn\\
& & 
          + 3 H( - r,0,-1;x)
          + 4 H(-4, - r;x)
          - \frac{3}{2} H(-4, - r,-1;x) \nn\\
& & 
          - H(-4,-4, - r;x)
          \Biggr\}
\, , \\
F^{(2)}_{2} & = &  
            \frac{211}{2}
          + 19 \zeta(2)
          + \frac{9}{5} \zeta^2(2)
          - 5 \zeta(3)
          - \Bigl[ 13+\zeta(2) \Bigr] H(0,-1;x)  \nn\\
& &         
          - 8 H(0,-1,-1,-1;x)
          + 3 H(0,-1,0,-1;x)
          + 12 H(0,-1,-1;x)   \nn\\
& &         
	  - 3 H(0,0,-1;x)
          + 2 H(0,0,-1,-1;x)
          - \frac{1}{2} H(0,0,0,-1;x) \nn\\
& & 
       + \! \frac{x+4}{\sqrt{x(x+4)}} \Biggl\{
            \! 2(\zeta(3)\! - \! 4 \zeta(2) \! - \! 16 ) H( - r;x) \!
          + \! (6 \! + \! \zeta(2)) \bigl[ 3 H( - r,-1;x) \nn\\
& & 
	  + 2 H(-4, - r;x) \bigr]
          - 24 H( - r,-1,-1;x)
          + 24 H( - r,-1,-1,-1;x) \nn\\
& & 
          - 9 H( - r,-1,0,-1;x)
          + 12 H( - r,0,-1;x)
          - 12 H( - r,0,-1,-1;x) \nn\\
& & 
          + 3 H( - r,0,0,-1;x)
          - 6 H(-4, - r,-1;x)
          + 6 H(-4, - r,-1,-1;x) \nn\\
& & 
          - 3 H(-4, - r,0,-1;x)
          - 4 H(-4,-4, - r;x)
          + \frac{3}{2} H(-4,-4, - r,-1;x) \nn\\
& & 
          + H(-4,-4,-4, - r;x)
          \Biggr\}
       - \Biggl[ \frac{1}{x} + 1 \Biggr] \Bigl\{
            (33 + 2 \zeta(2) ) H(-1;x) \nn\\
& & 
          - 28 H(-1,-1;x)
          + 16 H(-1,-1,-1;x)
          - 6 H(-1,0,-1;x) \nn\\
& & 
          + 7 H(0,-1;x)
          - 4 H(0,-1,-1;x)
          + H(0,0,-1;x)
          \Bigr\}
	  \, .
\eea
The above 2-point function is the simplest amplitude having thresholds both in
$s=m^2$ and $s=4m^2$. It does not have any pseudo-threshold.
The indices appearing in the GHPLs are indeed
only $0,-1,-4$ and $-r$. Note that the index $-r$ appears eventually only once in the $H$'s.
The related terms have coefficients always containing radicals, in order to reproduce the right
causality structure.

\bea 
\parbox{20mm}{\begin{fmfgraph*}(15,15)
\fmfleft{i1,i2}
\fmfright{o}
\fmf{photon}{i1,v1}
\fmf{photon}{i2,v2}
\fmf{plain}{v3,o}
\fmf{photon,tension=.3}{v2,v3}
\fmf{photon,tension=.3}{v1,v3}
\fmf{plain,tension=0}{v2,v1}
\fmf{plain,tension=0,right=.5}{v2,v3}
\end{fmfgraph*}} & = & \mu^{2(4-D)} 
\int {\mathfrak D}^D k_1 {\mathfrak D}^D k_2
\frac{1}{{\mathcal D}_{5} 
         {\mathcal D}_{7} 
	 {\mathcal D}_{12} 
	 {\mathcal D}_{13}} \\
& = & \left( \frac{\mu^{2}}{a} \right) ^{2 \epsilon} 
\sum_{i=-2}^{1} \epsilon^{i} F^{(3)}_{i} + {\mathcal O} \left( 
\epsilon^{3} \right) , 
\eea
where:
\bea
F^{(3)}_{-2} & = & \frac{1}{2} \, , \\
F^{(3)}_{-1} & = & \frac{3}{2} \, , \\
F^{(3)}_{0} & = & \frac{5}{2} 
          + H(-1,x)
          - H(0,-1,x)
       + \frac{1}{x} \Bigl[
            H(-1,x)
          - \zeta(2) H(1,x) \nn\\
& & 
          + 2 H(1,0,-1,x)
          \Bigr]
\, , \\
F^{(3)}_{1} & = & 
          - \frac{1}{2} \! 
          - \zeta(2) \! 
          +  \! 7 H(-1,x) \! 
          - \zeta(2) H(1,x) \! 
          - 4 H(-1,-1,x) \! 
          - 2 H(0,-1,x) \nn\\
& & 
          + 4 H(0,-1,-1,x)
          + H(0,0,-1,x)
          + 2 H(1,0,-1,x)
       + \frac{1}{x} \Bigl[
            7 H(-1,x) \nn\\
& & 
          - \bigl( \zeta(2) \! 
          - \zeta(3) \bigr) H(1,x) \! 
          - 4 H(-1,-1,x) \! 
          +  \! H(0,-1,x)
          - \zeta(2) H(0,1,x) \nn\\
& & 
          + 2 H(0,1,0,-1,x)
          + 2 H(1,0,-1,x)
          - 8 H(1,0,-1,-1,x)
          \Bigr] \, .
\eea

\bea
\parbox{35mm}{\begin{fmfgraph*}(15,15)
\fmfleft{i1,i2}
\fmfright{o}
\fmf{photon}{i1,v1}
\fmf{photon}{i2,v2}
\fmf{plain}{v3,o}
\fmflabel{$(p_{2} \cdot k_{2})$}{o}
\fmf{photon,tension=.3}{v2,v3}
\fmf{photon,tension=.3}{v1,v3}
\fmf{plain,tension=0}{v2,v1}
\fmf{plain,tension=0,right=.5}{v2,v3}
\end{fmfgraph*}} & = & \mu^{2(4-D)} 
\int {\mathfrak D}^D k_1 {\mathfrak D}^D k_2
\frac{p_2 \cdot k_2}{{\mathcal D}_{5} 
                     {\mathcal D}_{7} 
	             {\mathcal D}_{12} 
	             {\mathcal D}_{13}} \\
& = & \left( \frac{\mu^{2}}{a} \right) ^{2 \epsilon} 
\sum_{i=-2}^{2} \epsilon^{i} F^{(4)}_{i} + {\mathcal O} \left( 
\epsilon^{3} \right) , 
\eea
where:
\bea
\frac{F^{(4)}_{-2}}{a} & = & - \frac{1}{8} x \, , \\
\frac{F^{(4)}_{-1}}{a} & = & - \frac{5}{16} x \, , \\
\frac{F^{(4)}_{0}}{a} & = & 
            \frac{1}{8} \! 
          - \! \frac{1}{2} H(-1,x) \! 
          - \! \frac{1}{8x} H(-1,x) \! 
	  - \! x \Biggl[
            \frac{7}{32} \! 
          +  \! \frac{3}{8} H(-1,x) \! 
          -  \! \frac{1}{4} H(0,-1,x) \Biggr] , \\
\frac{F^{(4)}_{1}}{a} & = & 
            \frac{9}{16} \! 
          +  \! \frac{1}{4} \zeta(2) \! 
          - 3 H(-1,x) \! 
          +  \! 2 H(-1,-1,x) \! 
          - H(0,-1,x) \! 
	 - \frac{1}{x} \Biggl[
            \frac{7}{16} H(-1,x) \nn\\
& & 
          + \frac{1}{4} \zeta(2) H(1,x)
          - \frac{1}{2} H(-1,-1,x)
          + \frac{1}{8} H(0,-1,x)
          - \frac{1}{2} H(1,0,-1,x)
             \Biggr] \nn\\
& & 
	 + x \Biggl[
            \frac{123}{64}
          + \frac{3}{8} \zeta(2)
          - \frac{41}{16} H(-1,x)
          + \frac{1}{4} \zeta(2) H(1,x)
          + \frac{3}{2} H(-1,-1,x) \nn\\
& & 
          +  \! \frac{1}{4} H(0, \! -1,x) \! 
          -  \! H(0, \! -1, \! -1,x) \! 
          -  \! \frac{1}{4} H(0,0, \! -1,x) \! 
          -  \! \frac{1}{2} H(1,0, \! -1,x) \! 
             \Biggr]
\, , \\
\frac{F^{(4)}_{2}}{a} & = &  
            \frac{39}{32}
          + \zeta(2)
          - \frac{1}{4} \zeta(3)
          - \frac{25}{2} H(-1,x)
          - \zeta(2) H(-1,x)
          - \frac{1}{2} \zeta(2) H(1,x) \nn\\
& & 
          +  \! 12 H(-1, \! -1,x) \! 
          -  \! \frac{21}{4} H(0, \! -1,x) \! 
          -  \! 8 H(-1, \! -1, \! -1,x) \! 
          +  \! 3 H(-1,0, \! -1,x) \nn\\
& & 
          + 4 H(0,-1,-1,x)
          - \frac{1}{2} H(0,0,-1,x)
          + H(1,0,-1,x)
	 - \frac{1}{x} \Biggl[
            \frac{21}{32} H(-1,x) \nn\\
& & 
          + \frac{1}{4} \zeta(2) H(-1,x)
          + \frac{1}{2} \zeta(2) H(1,x)
          - \frac{1}{4} \zeta(3) H(1,x)
          - \frac{7}{4} H(-1,-1,x) \nn\\
& & 
          +  \! \frac{7}{16} H(0,-1,x) \!  \! 
          +  \! \frac{1}{4} \zeta(2) H(0,1,x) \! 
          +  \! 2 H(-1, \! -1, \! -1,x) \! 
          -  \! \frac{3}{4} H(-1,0, \! -1,x) \nn\\
& & 
          - \frac{1}{2} H(0, \! -1, \! -1,x) \! 
          +  \! \frac{1}{8} H(0,0, \! -1,x) \! 
          -  \! H(1,0, \! -1,x) \! 
          -  \! \frac{1}{2} H(0,1,0, \! -1,x) \nn\\
& & 
          + 2 H(1,0,-1,-1,x)
             \Biggr]
	 + x \Biggl[
            \frac{1681}{128}
          + \frac{41}{16} \zeta(2)
          - \frac{3}{8} \zeta(3)
          - \frac{379}{32} H(-1,x) \nn\\
& & 
          - \frac{3}{4} \zeta(2) H(-1,x)
          + \zeta(2) H(1,x)
          - \frac{1}{4} \zeta(3) H(1,x)
          + \frac{41}{4} H(-1,-1,x) \nn\\
& & 
          -  \! \frac{17}{8} H(0, \! -1,x) \! 
          +  \! \frac{1}{2} \zeta(2) H(0, \! -1,x) \! 
          -  \! \frac{1}{4} \zeta(2) H(0,1,x) \! 
          -  \! 6 H(-1, \! -1, \! -1,x) \nn\\
& & 
          +  \! \frac{9}{4} H(-1,0, \! -1,x) \! 
          -  \! H(0, \! -1, \! -1,x) \! 
          -  \! \frac{7}{4} H(0,0, \! -1,x) \! 
          -  \! 2 H(1,0,-1,x) \nn\\
& & 
          + 4 H(0,-1,-1,-1,x)
          - \frac{3}{2} H(0,-1,0,-1,x)
          + H(0,0,-1,-1,x) \nn\\
& & 
          + \frac{3}{4} H(0,0,0,-1,x)
          + \frac{1}{2} H(0,1,0,-1,x)
          + 2 H(1,0,-1,-1,x)
             \Biggr]
\, .
\eea
The above 2 MIs, which contain 2 massive denominators,
can be expressed in terms of ordinary HPLs. The reason is that the 2 massive lines,
roughly speaking, are in different channels: one is in the $s$ channel while the other 
is in the $t$ channel.
The amplitudes do not have thresholds/pseudothresholds in $s=\pm 4m^2$, but only in
$s=\pm m^2$. Both the indices ``$1$'' and ``$-1$'' do indeed appear inside the HPLs.
The presence of 2 massive denominators is then a necessary but not a sufficient
condition in order to have thresholds or pseudothresholds in $s=\pm 4m^2$.

\bea
\parbox{20mm}{\begin{fmfgraph*}(15,15)
\fmfleft{i1,i2}
\fmfright{o}
\fmf{photon}{i1,v1}
\fmf{photon}{i2,v2}
\fmf{plain}{v3,o}
\fmf{photon,tension=.3}{v2,v3}
\fmf{photon,tension=.3}{v1,v3}
\fmf{plain,tension=0,right=.5}{v2,v1}
\fmf{plain,tension=0,right=.5}{v1,v2}
\end{fmfgraph*}} & = & \mu^{2(4-D)} 
\int {\mathfrak D}^D k_1 {\mathfrak D}^D k_2
\frac{1}{{\mathcal D}_{7} 
         {\mathcal D}_{8} 
	 {\mathcal D}_{12} 
	 {\mathcal D}_{13}} \\
& = & \left( \frac{\mu^{2}}{a} \right) ^{2 \epsilon} 
\sum_{i=-2}^{1} \epsilon^{i} F^{(5)}_{i} + {\mathcal O} \left( 
\epsilon^{3} \right) , 
\eea
where:
\bea
F^{(5)}_{-2} & = & \frac{1}{2} \, , \\
F^{(5)}_{-1} & = & \frac{5}{2} - H(0,x) \, , \\
F^{(5)}_{0} & = & \frac{19}{2}
          - \zeta(2)
          - 5 H(0,x)
          + H(0,0,x)
        + \frac{4-x}{\sqrt{x(4-x)}} H(r,0,x) \nn\\
& &
          + \frac{2}{x} H(r,r,0,x)
\, , \\
F^{(5)}_{1} & = & 
            \frac{65}{2}
          - 5 \zeta(2)
          - 2 \zeta(3)
          - 19 H(0,x)
          + \zeta(2) H(0,x)
          - H(r,r,0,x) \nn\\
& &
          + 5 H(0,0,x)
          - H(0,0,0,x)
        + \frac{4-x}{\sqrt{x(4-x)}} \Bigl[
            \zeta(2) H(r,x)
          + 5 H(r,0,x) \nn\\
& &
          - H(r,0,0,x)
          - H(0,r,0,x)
          + 2 H(4,r,0,x) \Bigr]
        + \frac{2}{x}  \Bigl[
            \zeta(2) H(r,r,x) \nn\\
& &
          + 3 H(r,r,0,x)
          - H(r,r,0,0,x)
          - H(r,0,r,0,x)
          + 2 H(r,4,r,0,x) \nn\\
& &
          + H(0,r,r,0,x)
           \Bigr]
\, .
\eea
The double and the simple poles in the MI above have ultraviolet origin.
The amplitude has indeed an ultraviolet sub-divergence related 
to the integration of the bubble together with an over-all UV divergence.

\bea
\parbox{20mm}{\begin{fmfgraph*}(15,15)
\fmfleft{i1,i2}
\fmfforce{.55w,.65h}{d1}
\fmfright{o}
\fmf{photon}{i1,v1}
\fmf{photon}{i2,v2}
\fmf{plain}{v3,o}
\fmf{photon,tension=.3}{v2,v3}
\fmf{photon,tension=.3}{v1,v3}
\fmf{plain,tension=0,right=.5}{v2,v1}
\fmf{plain,tension=0,right=.5}{v1,v2}
\fmfv{decor.shape=circle,decor.filled=full,decor.size=.1w}{d1}
\end{fmfgraph*}} & = & \mu^{2(4-D)} 
\int {\mathfrak D}^D k_1 {\mathfrak D}^D k_2
\frac{1}{{\mathcal D}_{7}^{2} 
         {\mathcal D}_{8} 
	 {\mathcal D}_{12} 
	 {\mathcal D}_{13}} \\
& = & \left( \frac{\mu^{2}}{a} \right) ^{2 \epsilon} 
\sum_{i=-2}^{2} \epsilon^{i} F^{(6)}_{i} + {\mathcal O} \left( 
\epsilon^{3} \right) , 
\eea
where:
\bea
\frac{F^{(6)}_{-2}}{a} & = & - \frac{1}{x} \, , \\
\frac{F^{(6)}_{-1}}{a} & = & \frac{1}{x} H(0,x) \, , \\
\frac{F^{(6)}_{0}}{a} & = & - \frac{1}{x} \Bigl[
            4
          - \zeta(2)
          - 2 H(0,x)
          + H(0,0,x) \Bigr]
	- \frac{4-x}{x \sqrt{x (4-x)}} H(r,0,x)
 \, , \\
\frac{F^{(6)}_{1}}{a} & = & \frac{1}{x} \Bigl[ 
            2 \zeta(2)
          + 2 \zeta(3)
          + 4 H(0,x)
          - \zeta(2) H(0,x)
          - 2 H(0,0,x)
          + H(r,r,0,x) \nn\\
& & 
          +  \! H(0,0,0,x)
              \Bigr] \! 
	- \frac{4-x}{x \sqrt{x (4-x)}} \Bigl[ 
            \zeta(2) H(r,x) \! 
          +  \! 2 H(r,0,x)
          - H(r,0,0,x) \nn\\
& & 
          - H(0,r,0,x)
          + 2 H(4,r,0,x)
              \Bigr]
\, , \\
\frac{F^{(6)}_{2}}{a} & = &  - \frac{1}{x} \Bigl[ 
            16
          - 4 \zeta(2)
          - \frac{9}{10} \zeta^2(2)
          - 4 \zeta(3)
          - 2 \bigl( 
	    4 
          - \zeta(2) 
          - \zeta(3) \bigr) H(0,x) \nn\\
& & 
          + \bigl( 4 
          - \zeta(2) \bigr) H(0,0,x)
          - \zeta(2) H(r,r,x)
          - 2 H(r,r,0,x)
          + H(r,r,0,0,x) \nn\\
& & 
          + H(r,0,r,0,x)
          - 2 H(r,4,r,0,x)
          - H(0,r,r,0,x)
          - 2 H(0,0,0,x) \nn\\
& & 
          + H(0,0,0,0,x)
              \Bigr] \! 
	-  \! \frac{4-x}{x\sqrt{x (4-x)}} \Bigl[ 
            2 \zeta(2) H(r,x) \! 
          +  \! 2 \zeta(3) H(r,x) \! 
          +  \! 4 H(r,0,x) \nn\\
& & 
          - \zeta(2) H(r,0,x)
          - \zeta(2) H(0,r,x)
          + 2 \zeta(2) H(4,r,x)
          + 3 H(r,r,r,0,x) \nn\\
& & 
          - 2 H(r,0,0,x)
          + 4 H(4,r,0,x)
          - 2 H(0,r,0,x)
          + H(r,0,0,0,x) \nn\\
& & 
          + H(0,r,0,0,x)
          + H(0,0,r,0,x)
          - 2 H(0,4,r,0,x)
          - 2 H(4,r,0,0,x) \nn\\
& & 
          - 2 H(4,0,r,0,x)
          + 4 H(4,4,r,0,x)
              \Bigr]
\, .
\eea
The double pole in $\epsilon$ in the MI above is the product of a simple UV pole
coming from the nested bubble and of an IR pole coming from the massless line
squared.

The above 2 MIs have a pseudothreshold in $s=-4m^2$ related to the
exchange of 2 massive particles in the $t$ channel and consequently
only GHPLs with indices $0,4$, and $r$ do appear in the $\epsilon$ expansion.
These MIs have also been computed in \cite{RoPieRem2} by means of a 
transformation well-known in QED eliminating the square roots:
\be
x= \frac{(1+z)^2}{z}.
\ee
In general, this change of variable is very convenient for amplitudes not having the 
pseudothreshold in $s=-m^2$.

\subsection{Topology $t=5$ \label{5den}}

\bea
\parbox{20mm}{\begin{fmfgraph*}(15,15)
\fmfforce{0.2w,0.5h}{v1}
\fmfforce{0.5w,0.8h}{v2}
\fmfforce{0.5w,0.2h}{v3}
\fmfforce{0.8w,0.5h}{v4}
\fmfleft{i}
\fmfright{o}
\fmf{plain}{i,v1}
\fmf{plain}{v4,o}
\fmf{plain,tension=.2,left=.4}{v1,v2}
\fmf{plain,tension=.2,right=.4}{v1,v3}
\fmf{photon,tension=.2,left=.4}{v2,v4}
\fmf{photon,tension=.2,right=.4}{v3,v4}
\fmf{photon,tension=0}{v2,v3}
\end{fmfgraph*}} & = & \mu^{2(4-D)} 
\int {\mathfrak D}^D k_1 {\mathfrak D}^D k_2
\frac{1}{{\mathcal D}_{2} 
         {\mathcal D}_{3} 
	 {\mathcal D}_{11} 
	 {\mathcal D}_{12} 
	 {\mathcal D}_{20}} \\
& = & \left( \frac{\mu^{2}}{a} \right) ^{2 \epsilon} 
\sum_{i=0}^{1} \epsilon^{i} F^{(7)}_{i} + {\mathcal O} \left( 
\epsilon^{3} \right) , 
\eea
where:
\bea
aF^{(7)}_{0} & = & \frac{1}{x} \Bigl[ 
            H(0,0,-1,x)
          + 3 H( - r, - r,-1,x)
          - 2 H( - r, - r,0,x) \nn\\
& & 
          - 2 H( - r,0, - r,x)
		   \Bigr] 
\, , \\
aF^{(7)}_{1} & = & \frac{1}{x} \Bigl[ 
            2 H(0,0,-1,x)
          - 6 \zeta(2) H( - r, - r,x)
          + 6 H( - r, - r,-1,x) \nn\\
& & 
          - 4 H( - r, - r,0,x)
          - 4 H( - r,0, - r,x)
          - 4 H(0,0,-1,-1,x) \nn\\
& & 
          + H(0,0,0,-1,x)
          - 12 H( - r, - r,-1,-1,x)
          + 6 H( - r, - r,0,-1,x) \nn\\
& & 
          + 2 H( - r, - r,0,0,x)
          - 3 H( - r,-4, - r,-1,x)
          + 2 H( - r,-4, - r,0,x) \nn\\
& & 
          + 2 H( - r,-4,0, - r,x)
          + 2 H( - r,0, - r,0,x)
          + 2 H( - r,0,-4, - r,x) \nn\\
& & 
          + 2 H( - r,0,0, - r,x)
		   \Bigr]
\, .
\eea
The above 2-point function has thresholds in $s=0,m^2$ as well as in 
$s=4m^2$ and no pseudothresholds. The index $-r$ always appears twice 
in the $H$'s and the related coefficients contain no radicals. The 
finite part $O(\epsilon^0)$ of $F^{(7)}$ has been computed by the 
authors of \cite{fleischer} by means of a resummed small momentum 
expansion. With the help of the Mathematica \cite{Mathe} we have 
computed the first 20 terms of the small momentum expansion of our 
result and compared with their result, finding complete agreement.

\bea
\parbox{20mm}{\begin{fmfgraph*}(15,15)
\fmfleft{i1,i2}
\fmfright{o}
\fmfforce{0.2w,0.9h}{v2}
\fmfforce{0.2w,0.1h}{v1}
\fmfforce{0.2w,0.5h}{v3}
\fmfforce{0.8w,0.5h}{v4}
\fmf{photon}{i1,v1}
\fmf{photon}{i2,v2}
\fmf{plain}{v4,o}
\fmf{plain,tension=0}{v1,v3}
\fmf{plain,tension=0}{v3,v4}
\fmf{photon,tension=0}{v2,v4}
\fmf{plain,tension=0}{v2,v3}
\fmf{photon,tension=0}{v1,v4}
\end{fmfgraph*}} & = & \mu^{2(4-D)} 
\int {\mathfrak D}^D k_1 {\mathfrak D}^D k_2
\frac{1}{{\mathcal D}_{4} 
         {\mathcal D}_{6} 
	 {\mathcal D}_{12} 
	 {\mathcal D}_{13} 
	 {\mathcal D}_{14} } \\
& = & \left( \frac{\mu^{2}}{a} \right) ^{2 \epsilon} 
 F^{(8)}_{0} + {\mathcal O} \left( 
\epsilon^{3} \right) , 
\eea
where:
\bea
aF^{(8)}_{0} & = & 
       - \frac{1}{x}   \Bigl[ 
            \zeta(2) \Bigl( H(0,c,x)
          + 2 H(0,1,x)
          + H(0, \overline{c},x) \Bigr)
          - 3 H(0, \overline{c},0,-1,x) \nn\\
& & 
          - 3 H(0,c,0,-1,x)
          - 4 H(0,1,0,-1,x)
          \Bigr]
       + \frac{i}{x}  H(r,0,1) \Bigl[
            H(0,c,x) \nn\\
& & 
          - H(0, \overline{c},x)
          \Bigr]
 \, .
\eea

\bea
\parbox{35mm}{\begin{fmfgraph*}(15,15)
\fmfleft{i1,i2}
\fmfright{o}
\fmfforce{0.2w,0.9h}{v2}
\fmfforce{0.2w,0.1h}{v1}
\fmfforce{0.2w,0.5h}{v3}
\fmfforce{0.8w,0.5h}{v4}
\fmf{photon}{i1,v1}
\fmf{photon}{i2,v2}
\fmf{plain}{v4,o}
\fmflabel{$(p_{2} \cdot k_{1})$}{o}
\fmf{plain,tension=0}{v1,v3}
\fmf{plain,tension=0}{v3,v4}
\fmf{photon,tension=0}{v2,v4}
\fmf{plain,tension=0}{v2,v3}
\fmf{photon,tension=0}{v1,v4}
\end{fmfgraph*}} & = & \mu^{2(4-D)} 
\int {\mathfrak D}^D k_1 {\mathfrak D}^D k_2
\frac{p_2 \cdot k_1}{{\mathcal D}_{4} 
                     {\mathcal D}_{6} 
		     {\mathcal D}_{12} 
		     {\mathcal D}_{13} 
		     {\mathcal D}_{14} } \\
& = & \left( \frac{\mu^{2}}{a} \right) ^{2 \epsilon} 
\sum_{i=0}^{1} \epsilon^{i} F^{(9)}_{i} + {\mathcal O} \left( 
\epsilon^{3} \right) , 
\eea
where:
\bea
F^{(9)}_{0} & = & 
          - 1
          + H(-1,x)
          - \zeta(2) H(1,x)
	  - \frac{1}{2} \zeta(2) \bigl[ 
	     H( \overline{c},x) 
	  +  H(c,x) \bigr]  \nn\\
& & 
          - H(0,-1,x)
          + 2 H(1,0,-1,x)
          + \frac{3}{2} \bigl[ 
	     H( \overline{c},0,-1,x)
           + H(c,0,-1,x) \bigr] \nn\\
& & 
       + \frac{1}{x} \Biggl\{ 
            \frac{1}{4} \Bigl( \sqrt{3} H(r,0,1) 
	    + \zeta(2) \Bigr)
	    \Bigl[
            H(c,x)
          + H( \overline{c},x)
          \Bigr]
          - \frac{3}{4} H( \overline{c},0,-1,x) \nn\\
& & 
          - \frac{3}{4} H(c,0,-1,x)
          + H(-1,x)
          + \zeta(2) H(1,x)
          - 2 H(1,0,-1,x) \Biggr\} \nn\\
& & 
       + \frac{i}{4} \Biggl\{ 
	    H(r,0,1) \Biggl( 2 - \frac{1}{x} \Biggr) \Bigl[
	    H(c,x)
	  - H( \overline{c},x) \Bigr]
	  + \frac{\sqrt{3}}{x}  \Bigl[
            \zeta(2) \Bigl( H(c,x) \nn\\
& & 
          - H( \overline{c},x) \Bigr)
          - 3 H(c,0,-1,x)
          + 3 H( \overline{c},0,-1,x)
          \Bigr]  \Biggr\}
\, , \\
F^{(9)}_{1} & = & 
          - 7
          - \zeta(2)
          + 6 H(-1,x)
          - 2 \zeta(2) H(1,x)
          + \zeta(3) H(1,x) \nn\\
& & 
          - \frac{1}{2} \Bigl( \zeta(2) 
	  - \zeta(3) \Bigr) \Bigl[ H(c,x)
	  + H( \overline{c},x) \Bigr]
          - 4 H(-1,-1,x) \nn\\
& & 
          - H(0,-1,x)
          - 3 \zeta(2) H(0,1,x)
          - \frac{3}{2} \zeta(2) \Bigl[ H(0,c ,x)
          + H(0,\overline{c},x) \Bigr] \nn\\
& & 
          + \zeta(2) \Bigl[ H(c,c,x)  \! 
          + H( \overline{c}, \! \overline{c},x) \Bigr] \! 
          + \! \frac{1}{2} \zeta(2) \Bigl[ H( \overline{c},c,x)\! 
          +  H(c, \! \overline{c},x) \Bigr] \nn\\
& & 
          + 2 \zeta(2) \Bigl[ H( \overline{c},1,x)
          + H(c,1,x)\Bigr] 
          + 4 H(0,-1,-1,x)
          + H(0,0,-1,x) \nn\\
& & 
          + 4 H(1,0,-1,x)
          + \frac{3}{2} \Bigl[ H( \overline{c},0,-1,x)
          + H(c,0,-1,x) \Bigr] \nn\\
& & 
          - 3 \Bigl[ H(c,c,0,-1,x)
          + H( \overline{c}, \overline{c},0,-1,x) \Bigr]
          - \frac{3}{2} \Bigl[ H(c, \overline{c},0,-1,x) \nn\\
& & 
          + H( \overline{c},c,0,-1,x) \Bigr]
          - 6 \Bigl[ H( \overline{c},0,-1,-1,x)
          + H(c,0,-1,-1,x) \Bigr] \nn\\
& & 
          - \frac{1}{2} \Bigl[ H( \overline{c},0,0,-1,x)
          + H(c,0,0,-1,x) \Bigr]
          - 4 \Bigl[ H( \overline{c},1,0,-1,x) \nn\\
& & 
          + H(c,1,0,-1,x) \Bigr]
          + \frac{9}{2} \Bigl[ H(0, \overline{c},0,-1,x)
          + H(0,c,0,-1,x) \Bigr] \nn\\
& & 
          + 6 H(0,1,0,-1,x)
          - 8 H(1,0,-1,-1,x)
       + \frac{1}{4x} \sqrt{3}  \Bigl\{
            \Bigl[ H(r,0,1)  \nn\\
& & 
          - H(r,0,0,1) \Bigr] 
	    \Bigl[ H(c,x) + H( \overline{c},x) \Bigr]
          + H(r,0,1) \Bigl[ H(c, \overline{c},x) \nn\\
& & 
          + H( \overline{c},c,x)
          - 2 H(c,c,x)
          - 2 H( \overline{c}, \overline{c},x)
          + H(0, \overline{c},x)
          + H(0,c,x) \Bigr] \nn\\
& & 
          - H(0,r,0,1) \Bigl[ H(c,x)
          + H(-c,x)
          + H(4,r,0,1) \Bigl[ H(c,x)
          + H( \overline{c},x) \Bigr]
          \Bigr\} \nn\\
& & 
       + \frac{i}{2} \Biggl\{
         \Bigl[ H(r,0,1) 
	  + H(r,0,0,1)
	  + H(4,r,0,1) \Bigr] \Bigl[
	    H(c,x)
          - H( \overline{c},x) \Bigr] \nn\\
& & 
	  + H(r,0,1) \Bigl[
            H(c, \overline{c},x)
          - H( \overline{c},c,x)
          - 2 H(c,c,x)
          + 2 H( \overline{c}, \overline{c},x) \nn\\
& & 
          + 3 H(0,c,x)
          - 3 H(0, \overline{c},x) \Bigr]
	 - \frac{1}{2x} \Bigl[H(r,0,1) 
	  + H(r,0,0,1) \nn\\
& & 
	  + H(4,r,0,1) \Bigr] \Bigl[
	    H(c,x)
          - H( \overline{c},x) \Bigr]
	  + H(r,0,1) \Bigl[
          H(c, \overline{c},x) \nn\\
& & 
          - H( \overline{c},c,x)
          - 2 H(c,c,x)
          + 2 H( \overline{c}, \overline{c},x)
          + H(0,c,x)
          - H(0, \overline{c},x) \Bigr] \nn\\
& & 
	 + \frac{\sqrt{3}}{2x} \Bigl[
            \Bigl( \zeta(2) 
          - \zeta(3) \Bigr) \Bigl( H(c,x)
          - H( \overline{c},x) \Bigr)
          + \zeta(2) H(0,c,x) \nn\\
& & 
          - \zeta(2) H(0, \overline{c},x)
          - 4 \zeta(2) H(c,1,x)
          + 4 \zeta(2) H( \overline{c},1,x)
          - 2 \zeta(2) H(c,c,x) \nn\\
& & 
          + 2 \zeta(2) H( \overline{c}, \overline{c},x)
          - \zeta(2) H(c, \overline{c},x)
          + \zeta(2) H( \overline{c},c,x)
          - 3 H(c,0,-1,x) \nn\\
& & 
          + 3 H( \overline{c},0,-1,x)
          + 8 H(c,1,0,-1,x)
          - 8 H( \overline{c},1,0,-1,x) \nn\\
& & 
          - 3 H( \overline{c},c,0,-1,x)
          + 3 H(c, \overline{c},0,-1,x)
          + 3 H(c,c,0,-1,x) \nn\\
& & 
          - 3 H( \overline{c}, \overline{c},0,-1,x)
          + 6 H(c,0,-1,-1,x)
          - 6 H( \overline{c},0,-1,-1,x) \nn\\
& & 
          +  H(c,0,0,-1,x)
          -  H( \overline{c},0,0,-1,x)
          - 3 H(0,c,0,-1,x) \nn\\
& & 
          + 3 H(0, \overline{c},0,-1,x) \Bigr] 
          \Biggr\} \, .
\eea

\bea
\parbox{20mm}{\begin{fmfgraph*}(15,15)
\fmfleft{i1,i2}
\fmfforce{.4w,.5h}{d1}
\fmfright{o}
\fmfforce{0.2w,0.9h}{v2}
\fmfforce{0.2w,0.1h}{v1}
\fmfforce{0.2w,0.5h}{v3}
\fmfforce{0.8w,0.5h}{v4}
\fmf{photon}{i1,v1}
\fmf{photon}{i2,v2}
\fmf{plain}{v4,o}
\fmf{plain,tension=0}{v1,v3}
\fmf{plain,tension=0}{v3,v4}
\fmf{photon,tension=0}{v2,v4}
\fmf{plain,tension=0}{v2,v3}
\fmf{photon,tension=0}{v1,v4}
\fmfv{decor.shape=circle,decor.filled=full,decor.size=.1w}{d1}
\end{fmfgraph*}} & = & \mu^{2(4-D)} 
\int {\mathfrak D}^D k_1 {\mathfrak D}^D k_2
\frac{1}{{\mathcal D}_{4} 
         {\mathcal D}_{6} 
	 {\mathcal D}_{12} 
	 {\mathcal D}_{13} 
	 {\mathcal D}_{14}^{2} } \\
& = & \left( \frac{\mu^{2}}{a} \right) ^{2 \epsilon} 
\sum_{i=0}^{1} \epsilon^{i} F^{(10)}_{i} + {\mathcal O} \left( 
\epsilon^{3} \right) , 
\eea
where:
\bea
a^2 F^{(10)}_{0} & = & 
         - \frac{1}{x}  \Biggl\{
            \frac{\sqrt{3}}{3} H(r,0,1) \Bigl[ H(c,x) 
	                              + H( \overline{c},x) \Bigr]
       + i \frac{\sqrt{3}}{3} \Bigl[
           \zeta(2) \Bigl( H(c,x)
                       - H( \overline{c},x) \Bigr) \nn\\
& & 
          - 3 H(c,0,-1,x)
          + 3 H( \overline{c},0,-1,x)
          \Bigr] \Biggr\}
\, , \\
a^2 F^{(10)}_{1} & = &    - \frac{1}{x}    \Biggl\{ 
         \frac{\sqrt{3}}{3}  \Bigl[ 
	    H(r,0,0,1)
	  + H(4,r,0,1)  \Bigr]  \Bigl[ H(c,x)
                       + H( \overline{c},x) \Bigr]  \nn\\
& & 
          + H(r,0,1) \Bigl[ 
	    H(c, \overline{c},x)
          + H( \overline{c},c,x)
          - 2 H(c,c,x)
          - 2 H( \overline{c}, \overline{c},x) \nn\\
& & 
          + H(0,c,x)
          + H(0, \overline{c},x) \Bigr]
	- i \frac{\sqrt{3}}{3}  \Bigl[ 
             \zeta(3) \Bigl( H(c,x)
                       - H( \overline{c},x) \Bigr) \nn\\
& & 
          - \zeta(2) \Bigl( H(0,c,x)
          - H(0, \overline{c},x)
          - 4 H(c,1,x)
          + 4 H( \overline{c},1,x)
          - 2 H(c,c,x) \nn\\
& & 
          + 2 H( \overline{c}, \overline{c},x)
          - H(c, \overline{c},x)
          + H( \overline{c},c,x) \Bigr)
          - 2 H(c,c,0,-1,x) \nn\\
& & 
          + 2 H( \overline{c}, \overline{c},0,-1,x)
          - H(c, \overline{c},0,-1,x)
          + H( \overline{c},c,0,-1,x) \nn\\
& & 
          - 4 H(c,0,-1,-1,x)
          + 4 H( \overline{c},0,-1,-1,x)
          - \frac{1}{3} H(c,0,0,-1,x) \nn\\
& & 
          + \frac{1}{3} H( \overline{c},0,0,-1,x)
          - \frac{8}{3} H(c,1,0,-1,x)
          + \frac{8}{3} H( \overline{c},1,0,-1,x) \nn\\
& & 
          + H(0,c,0,-1,x)
          - H(0, \overline{c},0,-1,x) \Bigr]
	   \Biggr\}
\, .
\eea
The above 3 MIs are real as complex $H(\cdots c\cdots;x)$'s always 
appear in the combinations: 
$H(\cdots c\cdots;x)+H(\cdots\overline{c}\cdots;x)$ and 
$i[H(\cdots c\cdots;x)-H(\cdots\overline{c}\cdots;x)]$.
As discussed in the previous Section, the above topology is the only one having 3 MIs. 
The amplitudes have a threshold in $s=m^2$ in agreement with Cutkowsky rule
as well as a pseudo thresholds in $s=-m^2$.

\bea
\parbox{20mm}{\begin{fmfgraph*}(15,15)
\fmfleft{i1,i2}
\fmfright{o}
\fmfforce{0.2w,0.9h}{v2}
\fmfforce{0.2w,0.1h}{v1}
\fmfforce{0.2w,0.55h}{v3}
\fmfforce{0.2w,0.15h}{v5}
\fmfforce{0.8w,0.5h}{v4}
\fmf{photon}{i1,v1}
\fmf{photon}{i2,v2}
\fmf{plain}{v4,o}
\fmf{plain}{v2,v3}
\fmf{plain,left}{v3,v5}
\fmf{plain,right}{v3,v5}
\fmf{photon}{v1,v4}
\fmf{photon}{v2,v4}
\end{fmfgraph*}} & = & \mu^{2(4-D)} 
\int {\mathfrak D}^D k_1 {\mathfrak D}^D k_2
\frac{1}{{\mathcal D}_{4} 
         {\mathcal D}_{5} 
	 {\mathcal D}_{12} 
	 {\mathcal D}_{13} 
	 {\mathcal D}_{14} } \\
& = & \left( \frac{\mu^{2}}{a} \right) ^{2 \epsilon} 
\sum_{i=-1}^{1} \epsilon^{i} F^{(11)}_{i} + {\mathcal O} \left( 
\epsilon^{3} \right) , 
\eea
where:
\bea
aF^{(11)}_{-1} & = & - \frac{1}{x} H(1,0,x) \, , \\
aF^{(11)}_{0} & = & - \frac{1}{x} \Bigl\{ 
            \zeta(2) H(1,x)
          + 2 H(1,0,x)
          - H(r,r,0,x)
          + H(0,1,0,x) \nn\\
& & 
          - H(1,0,0,x)
          + H(1,1,0,x)
          - 3 H( 1+r,r,0,x) \nn\\
& & 
          + \sqrt{3} H(r,0,1) H(1,x)
       \Bigr\} \, , \\
aF^{(11)}_{1} & = & \frac{1}{x} \Bigl\{
          - 2 \zeta(2) H(1,x)
          - 2 \zeta(3) H(1,x)
          - \zeta(2) H(0,1,x)
          - 4 H(1,0,x) \nn\\
& & 
          + \zeta(2) H(1,0,x)
          - \zeta(2) H(1,1,x)
          + \zeta(2) H(r,r,x)
          + 3 \zeta(2) H( 1+r,r,x) \nn\\
& & 
          - 2 H(0,1,0,x)
          + 2 H(1,0,0,x)
          - 2 H(1,1,0,x)
          + 2 H(r,r,0,x) \nn\\
& & 
          + 6 H( 1+r,r,0,x)
          - H(0,0,1,0,x)
          + H(0,1,0,0,x)
          - H(0,1,1,0,x) \nn\\
& & 
          - H(1,0,0,0,x)
          - H(1,0,1,0,x)
          + H(1,1,0,0,x)
          - H(1,1,1,0,x) \nn\\
& & 
          - H(r,r,0,0,x)
          - H(r,0,r,0,x)
          + 2 H(r,4,r,0,x)
          + H(0,r,r,0,x) \nn\\
& & 
          - 3 H( 1+r,r,0,0,x)
          - 3 H( 1+r,0,r,0,x)
          + 6 H( 1+r,4,r,0,x) \nn\\
& & 
          + 3 H(0, 1+r,r,0,x)
          + 3 H(1, 1+r,r,0,x)
       - \sqrt{3} \Bigl[
            \Bigl( 2 H(r,0,1)  \nn\\
& & 
          + H(r,0,0,1)
          + H(4,r,0,1) \Bigr) H(1,x)
	  + H(r,0,1) \Bigl( H(0,1,x)  \nn\\
& & 
          + H(1,1,x) \Bigr)
          \Bigr] \Bigr\} \, .
\eea
The above amplitude has pseudothresholds in $s=-m^2$ and $s=-4m^2$ and 
is the only one containing GHPLs with the index $1+r$. The latter is then
related to the (virtual) transition of a particle with mass $m\neq 0$
into a pair of particles with the same mass, i.e. a bubble with 2 equal mass lines. 
The index $r$ appears in the GHPLs only 0 or 2 times.

\bea
\parbox{20mm}{\begin{fmfgraph*}(15,15)
\fmfleft{i1,i2}
\fmfright{o}
\fmfforce{0.2w,0.9h}{v2}
\fmfforce{0.2w,0.1h}{v1}
\fmfforce{0.2w,0.5h}{v3}
\fmfforce{0.8w,0.5h}{v4}
\fmf{photon}{i1,v1}
\fmf{photon}{i2,v2}
\fmf{plain}{v4,o}
\fmf{plain,tension=0}{v1,v3}
\fmf{photon,tension=0}{v3,v4}
\fmf{photon,tension=0}{v2,v4}
\fmf{plain,tension=0}{v2,v3}
\fmf{photon,tension=0}{v1,v4}
\end{fmfgraph*}} & = & \mu^{2(4-D)} 
\int {\mathfrak D}^D k_1 {\mathfrak D}^D k_2
\frac{1}{{\mathcal D}_{3} 
         {\mathcal D}_{4} 
	 {\mathcal D}_{6} 
	 {\mathcal D}_{12} 
	 {\mathcal D}_{13} } \\
& = & \left( \frac{\mu^{2}}{a} \right) ^{2 \epsilon} 
F^{(12)}_{0} + {\mathcal O} \left( 
\epsilon \right) , 
\eea
where:
\bea
aF^{(12)}_{0} & = & \frac{2}{x} \Bigl[
            \zeta(2) H(0,1,x)
          + H(0,1,1,0,x)
          \Bigr]
\, .
\eea

\bea
\parbox{35mm}{\begin{fmfgraph*}(15,15)
\fmfleft{i1,i2}
\fmfright{o}
\fmfforce{0.2w,0.9h}{v2}
\fmfforce{0.2w,0.1h}{v1}
\fmfforce{0.2w,0.5h}{v3}
\fmfforce{0.8w,0.5h}{v4}
\fmf{photon}{i1,v1}
\fmf{photon}{i2,v2}
\fmf{plain}{v4,o}
\fmflabel{$(p_{2} \cdot k_{1})$}{o}
\fmf{plain,tension=0}{v1,v3}
\fmf{photon,tension=0}{v3,v4}
\fmf{photon,tension=0}{v2,v4}
\fmf{plain,tension=0}{v2,v3}
\fmf{photon,tension=0}{v1,v4}
\end{fmfgraph*}} & = & \mu^{2(4-D)} 
\int {\mathfrak D}^D k_1 {\mathfrak D}^D k_2
\frac{p_2 \cdot k_1}{{\mathcal D}_{3} 
         {\mathcal D}_{4} 
	 {\mathcal D}_{6} 
	 {\mathcal D}_{12} 
	 {\mathcal D}_{13} } \\
& = & \left( \frac{\mu^{2}}{a} \right) ^{2 \epsilon} 
\sum_{i=0}^{1} \epsilon^{i} F^{(13)}_{i} + {\mathcal O} \left( 
\epsilon^{2} \right) , 
\eea
where:
\bea
\frac{F^{(13)}_{0}}{a} & = & 
       - 1
          + \zeta(2)
          + H(0,x)
          + \zeta(2) H(1,x)
          + H(1,0,x)
          + H(1,1,0,x) \nn\\
& &
       + \frac{1}{x}  \Bigl[
          - \zeta(2) H(1,x)
          - H(1,0,x)
          - H(1,1,0,x)
          \Bigr]
\, , \\
\frac{F^{(13)}_{1}}{a} & = & 
          - 7
          + 4 \zeta(2)
          - \zeta(3)
          + 6 H(0,x)
          + 3 \zeta(2) H(1,x)
          - \zeta(3) H(1,x) \nn\\
& &
          - 2 H(0,0,x)
          + 3 \zeta(2) H(0,1,x)
          + 3 H(1,0,x)
          + 5 \zeta(2) H(1,1,x) \nn\\
& &
          + H(0,1,0,x)
          - 2 H(1,0,0,x)
          + 3 H(1,1,0,x)
          + H(1,0,1,0,x) \nn\\
& &
          - 2 H(1,1,0,0,x)
          + 5 H(1,1,1,0,x)
          + 3 H(0,1,1,0,x)
       + \frac{1}{x} \Bigl[
            \zeta(3) H(1,x) \nn\\
& &
          - 3 \zeta(2) H(1,x)
          - \zeta(2) H(0,1,x)
          - 3 H(1,0,x)
          - 5 \zeta(2) H(1,1,x) \nn\\
& &
          - H(0,1,0,x)
          + 2 H(1,0,0,x)
          - 3 H(1,1,0,x)
          - H(0,1,1,0,x) \nn\\
& &
          - H(1,0,1,0,x)
          + 2 H(1,1,0,0,x)
          - 5 H(1,1,1,0,x)
          \Bigr]
\, .
\eea
The above 2 MIs contain HPLs with indices ``0'' and ``1'' only.
They represent the emission of a photon
by a charged vector boson in the $t$ channel, 
and therefore have only a pseudothreshold in $s=-m^2$.
They are IR (as well as UV) finite because the photon is emitted internally to the basic 
1-loop triangle.

\bea
\parbox{20mm}{\begin{fmfgraph*}(15,15)
\fmfleft{i1,i2}
\fmfright{o}
\fmfforce{0.2w,0.9h}{v2}
\fmfforce{0.2w,0.1h}{v1}
\fmfforce{0.2w,0.5h}{v3}
\fmfforce{0.8w,0.5h}{v4}
\fmf{photon}{i1,v1}
\fmf{photon}{i2,v2}
\fmf{plain}{v4,o}
\fmf{photon,tension=0}{v2,v3}
\fmf{photon,tension=0}{v3,v4}
\fmf{photon,tension=0}{v1,v4}
\fmf{plain,tension=0}{v2,v4}
\fmf{plain,tension=0}{v1,v3}
\end{fmfgraph*}} & = & \mu^{2(4-D)} 
\int {\mathfrak D}^D k_1 {\mathfrak D}^D k_2
\frac{1}{{\mathcal D}_{1} 
         {\mathcal D}_{3} 
	 {\mathcal D}_{6} 
	 {\mathcal D}_{13} 
	 {\mathcal D}_{15}} \\
& = & \left( \frac{\mu^{2}}{a} \right) ^{2 \epsilon} 
F^{(14)}_{0} + {\mathcal O} \left( 
\epsilon \right) , 
\eea
where:
\bea
aF^{(14)}_{0} & = & 
         \frac{1}{x}  \Bigl[
            \zeta(2) H(0,-1,x)
          + \zeta(2) H(0,1,x)
          + H(0,-1,0,-1,x) \nn\\
& & 
          - 2 H(0,1,0,-1,x)
          \Bigr]
\, .
\eea

\bea
\parbox{35mm}{\begin{fmfgraph*}(15,15)
\fmfleft{i1,i2}
\fmfright{o}
\fmfforce{0.2w,0.9h}{v2}
\fmfforce{0.2w,0.1h}{v1}
\fmfforce{0.2w,0.5h}{v3}
\fmfforce{0.8w,0.5h}{v4}
\fmf{photon}{i1,v1}
\fmf{photon}{i2,v2}
\fmf{plain}{v4,o}
\fmflabel{$(p_{1} \cdot k_{2})$}{o}
\fmf{photon,tension=0}{v2,v3}
\fmf{photon,tension=0}{v3,v4}
\fmf{photon,tension=0}{v1,v4}
\fmf{plain,tension=0}{v2,v4}
\fmf{plain,tension=0}{v1,v3}
\end{fmfgraph*}} & = & \mu^{2(4-D)} 
\int {\mathfrak D}^D k_1 {\mathfrak D}^D k_2
\frac{p_1 \cdot k_2}{{\mathcal D}_{1} 
         {\mathcal D}_{3} 
	 {\mathcal D}_{6} 
	 {\mathcal D}_{13} 
	 {\mathcal D}_{15} } \\
& = & \left( \frac{\mu^{2}}{a} \right) ^{2 \epsilon} 
\sum_{i=0}^{1} \epsilon^{i} F^{(15)}_{i} + {\mathcal O} \left( 
\epsilon^{2} \right) , 
\eea
where:
\bea
F^{(15)}_{0} & = & 
          - \frac{3}{2}
          + \frac{1}{2} \Bigl[ 3
          + \zeta(2) \Bigr] H(-1,x)
          + \frac{1}{2} \zeta(2) H(1,x)
          - \frac{3}{2} H(0,-1,x) \nn\\
& & 
          + \frac{1}{2} H(-1,0,-1,x)
          - H(1,0,-1,x)
       + \frac{1}{x}  \Biggl\{
            \frac{1}{2} \Bigl[ 3
          + \zeta(2) \Bigr] H(-1,x) \nn\\
& & 
          - \frac{1}{2} \zeta(2) H(1,x)
          + \frac{1}{2} H(-1,0,-1,x)
          + H(1,0,-1,x)
          \Biggr\}
\, , \\
F^{(15)}_{1} & = & 
          - 9
          - \zeta(2)
          + \frac{1}{2} \Biggl[ 15
          - \zeta(3) \Biggr] H(-1,x)
          - \Biggl[ \zeta(2)
          + \frac{1}{2} \zeta(3) \Biggr] H(1,x) \nn\\
& & 
          - \Bigl[ 6 
          + \zeta(2) \Bigr] H(-1,-1,x)
          + \frac{3}{2} \zeta(2) \Bigl[ H(0,-1,x) + H(0,1,x) \Bigr] \nn\\
& & 
          + 6 H(0,-1,-1,x)
          + \frac{1}{2} H(0,0,-1,x)
          + 2 H(1,0,-1,x) \nn\\
& & 
          - H(-1,-1,0,-1,x)
          - 2 H(-1,0,-1,-1,x)
          - \frac{1}{2} H(-1,0,0,-1,x) \nn\\
& & 
          + \frac{3}{2} H(0,-1,0,-1,x)
          - 3 H(0,1,0,-1,x)
          + 4 H(1,0,-1,-1,x) \nn\\
& & 
       + \frac{1}{x} \Biggl\{
            \frac{1}{2} \Bigl[ 15  
          -  \zeta(3) \Bigr] H(-1,x) 
          +  \Biggl[ \zeta(2) 
          +  \frac{1}{2} \zeta(3) \Biggr] H(1,x)  \nn\\
& & 
          -  \Bigl[ 6  
          +  \zeta(2) \Bigr] H(-1, -1,x)
          + \frac{1}{2} \Bigl[ 3
          + \zeta(2) \Bigr] H(0,-1,x)
          - \frac{1}{2} \zeta(2) H(0,1,x)  \nn\\
& & 
          - 2 H(1,0,-1,x)
          - H(-1,-1,0,-1,x)
          - 2 H(-1,0,-1,-1,x)  \nn\\
& & 
          - \frac{1}{2} H(-1,0,0,-1,x)
          + \frac{1}{2} H(0,-1,0,-1,x)
          + H(0,1,0,-1,x)  \nn\\
& & 
          - 4 H(1,0,-1,-1,x)
          \Biggr\}
\, .
\eea

\bea
\parbox{20mm}{\begin{fmfgraph*}(15,15)
\fmfleft{i1,i2}
\fmfright{o}
\fmfforce{0.8w,0.5h}{v4}
\fmf{photon}{i1,v1}
\fmf{photon}{i2,v2}
\fmf{plain}{v4,o}
\fmf{plain,tension=.4}{v1,v3}
\fmf{photon,tension=.2}{v3,v4}
\fmf{plain,tension=.15}{v2,v4}
\fmf{photon,tension=0}{v2,v1}
\fmf{photon,tension=0,left=.5}{v3,v4}
\end{fmfgraph*}} & = & \mu^{2(4-D)} 
\int {\mathfrak D}^D k_1 {\mathfrak D}^D k_2
\frac{1}{{\mathcal D}_{1} 
         {\mathcal D}_{2} 
	 {\mathcal D}_{8} 
	 {\mathcal D}_{15} 
	 {\mathcal D}_{16} } \\
& = & \left( \frac{\mu^{2}}{a} \right) ^{2 \epsilon} 
\sum_{i=-1}^{1} \epsilon^{i} F^{(16)}_{i} + {\mathcal O} \left( 
\epsilon^{2} \right) , 
\eea
where:
\bea
aF^{(16)}_{-1} & = & \frac{2}{x} H( - r, - r,x) \, , \\
aF^{(16)}_{0} & = & \frac{1}{x}  \Bigl[
            4 H( - r, - r,x)
          - 3 H( - r, - r,-1,x)
          - 2 H( - r,-4, - r,x) \nn\\
& & 
          + 2 H(0, - r, - r,x)
          \Bigr] \, , \\
aF^{(16)}_{1} & = & \frac{1}{x} \Bigl[
            4 \Bigl( 2 
          + \zeta(2) \Bigr) H( - r, - r,x)
          - 6 H( - r, - r,-1,x)
          + 4 H(0, - r, - r,x) \nn\\
& & 
          - 4 H( - r,-4, - r,x)
          + 12 H( - r, - r,-1,-1,x)
          - 6 H( - r, - r,0,-1,x) \nn\\
& & 
          + 3 H( - r,-4, - r,-1,x)
          + 2 H( - r,-4,-4, - r,x) \nn\\
& & 
          - 3 H(0, - r, - r,-1,x)
          - 2 H(0, - r,-4, - r,x) \nn\\
& & 
          + 2 H(0,0, - r, - r,x)
          \Bigr] \, .
\eea
The above MI has a simple UV pole coming from the sub-divergence in the 
bubble.

\bea
\parbox{20mm}{\begin{fmfgraph*}(15,15)
\fmfleft{i1,i2}
\fmfright{o}
\fmfforce{0.8w,0.5h}{v4}
\fmf{photon}{i1,v1}
\fmf{photon}{i2,v2}
\fmf{plain}{v4,o}
\fmf{photon,tension=.15}{v2,v4}
\fmf{photon,tension=.4}{v1,v3}
\fmf{photon,tension=.2}{v3,v4}
\fmf{plain,tension=0}{v2,v1}
\fmf{plain,tension=0}{v2,v3}
\end{fmfgraph*}} & = & \mu^{2(4-D)} 
\int {\mathfrak D}^D k_1 {\mathfrak D}^D k_2
\frac{1}{{\mathcal D}_{5} 
         {\mathcal D}_{7} 
	 {\mathcal D}_{8} 
	 {\mathcal D}_{12} 
	 {\mathcal D}_{13}} \\
& = & \left( \frac{\mu^{2}}{a} \right) ^{2 \epsilon} 
F^{(17)}_{0} + {\mathcal O} \left( 
\epsilon \right) , 
\eea
where:
\bea
aF^{(17)}_{0} & = & - \frac{1}{x} \Bigl[
            \zeta(2) H(0,1,x)
	  - 2 H(0,1,0,-1,x)
	  + 2 H(0,r,r,0,x) \Bigr]
\, .
\eea

\subsection{Topology $t=6$ \label{6den}}

\bea
\parbox{20mm}{\begin{fmfgraph*}(15,15)
\fmfleft{i1,i2}
\fmfright{o}
\fmfforce{0.2w,0.9h}{v2}
\fmfforce{0.2w,0.1h}{v1}
\fmfforce{0.2w,0.5h}{v3}
\fmfforce{0.8w,0.5h}{v5}
\fmf{photon}{i1,v1}
\fmf{photon}{i2,v2}
\fmf{plain}{v5,o}
\fmf{photon,tension=0}{v2,v5}
\fmf{plain,tension=0}{v3,v4}
\fmf{photon,tension=.4}{v1,v4}
\fmf{photon,tension=.4}{v4,v5}
\fmf{plain,tension=0}{v1,v3}
\fmf{plain,tension=0}{v2,v3}
\end{fmfgraph*}} & = & \mu^{2(4-D)} 
\int {\mathfrak D}^D k_1 {\mathfrak D}^D k_2
\frac{1}{{\mathcal D}_{4} 
         {\mathcal D}_{5} 
	 {\mathcal D}_{6} 
	 {\mathcal D}_{12} 
	 {\mathcal D}_{13} 
	 {\mathcal D}_{14}} 
\label{6MI1} \\
& = & \left( \frac{\mu^{2}}{a} \right) ^{2 \epsilon} 
F^{(18)}_{0} + {\mathcal O} \left( 
\epsilon^{2} \right) , 
\eea
where:
\bea
a^2 F^{(18)}_{0} & = & 
          \frac{1}{x}  \Bigl[
            \zeta(2) \Bigl( H(0,1,x)
          + H(1,1,x)
          + H(1,c,x)
          + H(1, \overline{c},x) \nn\\
& & 
          + H(0,c,x)
          + H(0, \overline{c},x) \Bigr)
          - 2 H(1,1,0,-1,x)
          - 2 H(0,1,0,-1,x) \nn\\
& & 
          - 3 H(0,c,0,-1,x)
          - 3 H(0, \overline{c},0,-1,x)
          - 2 H(1,r,r,0,x) \nn\\
& & 
          - 2 H(0,r,r,0,x)
          - 3 H(1,c,0,-1,x)
          - 3 H(1, \overline{c},0,-1,x)
          \Bigr] \nn\\
& & 
       - i \frac{1}{x}   H(r,0,1) \Bigl[
            H(0,c,x)
          - H(0, \overline{c},x)
          + H(1,c,x) \nn\\
& & 
          - H(1, \overline{c},x)
          \Bigr]
\, .
\eea
The index $r$ appears in the GHPLs only 0 or 2 times, so the 
coefficients of the related terms do not contain radicals.

\bea
\parbox{20mm}{\begin{fmfgraph*}(15,15)
\fmfleft{i1,i2}
\fmfright{o}
\fmfforce{0.2w,0.9h}{v2}
\fmfforce{0.2w,0.1h}{v1}
\fmfforce{0.2w,0.5h}{v3}
\fmfforce{0.8w,0.5h}{v5}
\fmf{photon}{i1,v1}
\fmf{photon}{i2,v2}
\fmf{plain}{v5,o}
\fmf{photon,tension=0}{v2,v5}
\fmf{photon,tension=0}{v3,v4}
\fmf{plain,tension=.4}{v1,v4}
\fmf{photon,tension=.4}{v4,v5}
\fmf{photon,tension=0}{v1,v3}
\fmf{plain,tension=0}{v2,v3}
\end{fmfgraph*}} & = & \mu^{2(4-D)} 
\int {\mathfrak D}^D k_1 {\mathfrak D}^D k_2
\frac{1}{{\mathcal D}_{2} 
         {\mathcal D}_{3} 
	 {\mathcal D}_{4} 
	 {\mathcal D}_{5} 
	 {\mathcal D}_{12} 
	 {\mathcal D}_{17} } 
\label{6MI2} \\
& = & \left( \frac{\mu^{2}}{a} \right) ^{2 \epsilon} 
F^{(19)}_{0} + {\mathcal O} \left( 
\epsilon \right) , 
\eea
where:
\bea
\hspace{-5mm}
a^2 F^{(19)}_{0} & = & \frac{1}{x}  \Bigl\{
           \zeta(2) \Bigl[ H(0,-1,x)
          - H(0,1,x)
          - H(1,1,x)
          + H(1,-1,x) \Bigr] \nn\\
\hspace{-5mm}
& & 
          - 2 H(0,-1,0,-1,x)
          + H(0,-1,0,0,x)
          + 2 H(0,1,0,-1,x) \nn\\
\hspace{-5mm}
& & 
          - 2 H(1,-1,0,-1,x)
          + H(1,-1,0,0,x)
          + 2 H(1,1,0,-1,x)
          \Bigr\} \, .
\eea
Because of analogous considerations to the previous ones,
the above MI is expressed in terms of ordinary HPLs.

\bea
\parbox{20mm}{\begin{fmfgraph*}(15,15)
\fmfleft{i1,i2}
\fmfright{o}
\fmf{photon}{i1,v1}
\fmf{photon}{i2,v2}
\fmf{plain}{v5,o}
\fmf{plain,tension=.3}{v2,v3}
\fmf{photon,tension=.3}{v3,v5}
\fmf{plain,tension=.3}{v1,v4}
\fmf{photon,tension=.3}{v4,v5}
\fmf{photon,tension=0}{v2,v1}
\fmf{photon,tension=0}{v4,v3}
\end{fmfgraph*}} & = & \mu^{2(4-D)} 
\int {\mathfrak D}^D k_1 {\mathfrak D}^D k_2
\frac{1}{{\mathcal D}_{1} 
         {\mathcal D}_{2} 
	 {\mathcal D}_{7} 
	 {\mathcal D}_{8} 
	 {\mathcal D}_{15} 
	 {\mathcal D}_{16} } 
\label{6MI3} \\
& = & \left( \frac{\mu^{2}}{a} \right) ^{2 \epsilon} 
F^{(20)}_{0} + {\mathcal O} \left( 
\epsilon \right) , 
\eea
where:
\bea
a^2 F^{(20)}_{0} & = & \frac{1}{x\sqrt{x(x+4)}}  \Biggl\{
            12 \zeta(2) H( - r,-1,x) 
          - 6 H( - r,-1,0,-1,x) \nn\\
& & 
          + 6 H( - r,-1,0,0,x)
          - 12 H( - r, - r, - r,-1,x) \nn\\
& & 
          + 8 H( - r, - r, - r,0,x)
          + 8 H( - r, - r,0, - r,x) \nn\\
& & 
          + 4 H( - r,0, - r, - r,x)
          - 2 H( - r,0,0,-1,x) \Biggr\}
\, .
\eea

\bea
\parbox{20mm}{\begin{fmfgraph*}(15,15)
\fmfleft{i1,i2}
\fmfright{o}
\fmf{photon}{i1,v1}
\fmf{photon}{i2,v2}
\fmf{plain}{v5,o}
\fmf{photon,tension=.3}{v2,v3}
\fmf{photon,tension=.3}{v3,v5}
\fmf{photon,tension=.3}{v1,v4}
\fmf{photon,tension=.3}{v4,v5}
\fmf{plain,tension=0}{v2,v1}
\fmf{plain,tension=0}{v4,v3}
\end{fmfgraph*}} & = & \mu^{2(4-D)} 
\int {\mathfrak D}^D k_1 {\mathfrak D}^D k_2
\frac{1}{{\mathcal D}_{4} 
         {\mathcal D}_{5} 
	 {\mathcal D}_{7} 
	 {\mathcal D}_{8} 
	 {\mathcal D}_{12} 
	 {\mathcal D}_{13} } 
\label{6MI4} \\
& = & \left( \frac{\mu^{2}}{a} \right) ^{2 \epsilon} 
F^{(21)}_{0} + {\mathcal O} \left( 
\epsilon \right) , 
\eea
where:
\bea
a^2 F^{(21)}_{0} & = &  \frac{1}{x^2}   
            \Bigl[
            6 \zeta(2) H(1,1,x)
          + 6 H(1,r,r,0,x)
          - 2 H(1,0,0,-1,x) \nn\\
& & 
          + H(1,0,1,0,x)
          - 12 H(1,1,0,-1,x)
          + 4 H(1,1,0,0,x)
	   \Bigr] \, .
\eea
The masses of the vector bosons exchanged in the $t$ channel completely cut-off the
infrared singularities, so the above MI is IR (as well as UV) finite. As noted in our previous 
work \cite{UgoRo}, a non-zero mass on the outer boson line is already sufficient to completely
screen the IR singularities.

The above MI has been computed in \cite{fleischer} by fitting a small 
momentum expansion to an assumed form for the exact expression.
We have compared the first $15$ terms of the small momentum expansion 
in \cite{fleischer} with an analogous expansion of our expression, 
finding complete agreement. The first few terms of the small momentum 
expansion of $F^{(21)}$ are given in the next section. 
In \cite{smirnov} a leading-twist large-momentum expansion of the above MI has 
been presented, based on the separation of the loop space in leading IR
regions and a related approximation on the integrand. The above 
result is in agreement with a preliminary large momentum expansion of 
our expression \cite{inpreparation}.

\section{Reducible six-denominator amplitudes \label{red6den}}

\bea
\parbox{20mm}{\begin{fmfgraph*}(15,15)
\fmfleft{i1,i2}
\fmfright{o}
\fmfforce{0.2w,0.9h}{v2}
\fmfforce{0.2w,0.1h}{v1}
\fmfforce{0.2w,0.5h}{v3}
\fmfforce{0.8w,0.5h}{v5}
\fmf{photon}{i1,v1}
\fmf{photon}{i2,v2}
\fmf{plain}{v5,o}
\fmf{photon,tension=0}{v2,v5}
\fmf{photon,tension=0}{v3,v4}
\fmf{photon,tension=.4}{v1,v4}
\fmf{photon,tension=.4}{v4,v5}
\fmf{plain,tension=0}{v1,v3}
\fmf{plain,tension=0}{v2,v3}
\end{fmfgraph*}} & = & \mu^{2(4-D)} 
\int {\mathfrak D}^D k_1 {\mathfrak D}^D k_2
\frac{1}{{\mathcal D}_{4} 
         {\mathcal D}_{5} 
	 {\mathcal D}_{7} 
	 {\mathcal D}_{8} 
	 {\mathcal D}_{12} 
	 {\mathcal D}_{13} }  \\
& = & \left( \frac{\mu^{2}}{a} \right) ^{2 \epsilon} 
F^{(22)}_{0} + {\mathcal O} \left( 
\epsilon \right) , 
\eea
where:
\bea
a^2 F^{(22)}_{0} & = &  - \frac{1}{x}   
            \Bigl[
	    \zeta(2) H(0,1,x)
          + \zeta(2) H(1,1,x)
          - H(0,1,0,0,x) \nn\\
& & 
          + H(0,1,1,0,x)
          - H(1,1,0,0,x)
          + H(1,1,1,0,x)
	   \Bigr] \, .
\eea
In the above amplitude, the photon is emitted by the $W$ boson internally to the triangle.
It is therefore ``trapped'' and cannot propagate for large distances. 
The consequence is that the above amplitude
is IR finite, as expected on the basis of physical intuition.

\bea
\parbox{20mm}{\begin{fmfgraph*}(15,15)
\fmfleft{i1,i2}
\fmfright{o}
\fmfforce{0.2w,0.9h}{v2}
\fmfforce{0.2w,0.1h}{v1}
\fmfforce{0.2w,0.5h}{v3}
\fmfforce{0.8w,0.5h}{v5}
\fmf{photon}{i1,v1}
\fmf{photon}{i2,v2}
\fmf{plain}{v5,o}
\fmf{photon,tension=0}{v2,v5}
\fmf{plain,tension=0}{v3,v4}
\fmf{photon,tension=.4}{v1,v4}
\fmf{photon,tension=.4}{v4,v5}
\fmf{photon,tension=0}{v1,v3}
\fmf{plain,tension=0}{v2,v3}
\end{fmfgraph*}} & = & \mu^{2(4-D)} 
\int {\mathfrak D}^D k_1 {\mathfrak D}^D k_2
\frac{1}{{\mathcal D}_{4} 
         {\mathcal D}_{5} 
	 {\mathcal D}_{7} 
	 {\mathcal D}_{8} 
	 {\mathcal D}_{12} 
	 {\mathcal D}_{13} }  \\
& = & \left( \frac{\mu^{2}}{a} \right) ^{2 \epsilon} 
\sum_{i=-1}^{0} \epsilon^{i} F^{(23)}_{i} + {\mathcal O} \left( 
\epsilon \right) , 
\eea
where:
\bea
a^2 F^{(23)}_{-1} & = &  - \frac{1}{x}   
            \Bigl[
            2 H(0,0,-1,x)
          + H(0,1,0,x)
          + 2 H(1,0,-1,x) \nn\\
& & 
          + H(1,1,0,x)
	   \Bigr] \, , \\
a^2 F^{(23)}_{0} & = &  - \frac{1}{x}   
            \Bigl[
            4 \zeta(2) \bigl( H(0,1,x)
          + H(1,1,x) \bigr)
          - 8 H(0,0,-1,-1,x) \nn\\
& & 
          + H(0,0,1,0,x)
          - H(0,1,0,0,x)
          + 4 H(0,1,1,0,x) \nn\\
& & 
          - 8 H(1,0,-1,-1,x)
          + H(1,0,1,0,x)
          - H(1,1,0,0,x) \nn\\
& & 
          + 4 H(1,1,1,0,x)
	   \Bigr] \, .
\eea
The above amplitude corresponds to a photon emitted by a $W$ boson
outside the triangle. In this case, the photon {\it can} propagate to
large distances and a collinear singularity is generated, 
characterized by the presence of the simple $1/\epsilon$ pole.
As anticipated in the introduction, a qualitative analysis of the infrared 
singularities can be done by shrinking all the internal massive lines to a point:
in this limit the diagram factorizes into a massless bubble evaluated at 
the light-cone momentum $p_2$ times a massless bubble evaluated at the general
momentum $q$.
The collinear singularity originate from the former bubble,
representing the evolution of a jet formed by an initial particle.

\bea
\parbox{20mm}{\begin{fmfgraph*}(15,15)
\fmfleft{i1,i2}
\fmfright{o}
\fmfforce{0.2w,0.9h}{v2}
\fmfforce{0.2w,0.1h}{v1}
\fmfforce{0.2w,0.5h}{v3}
\fmfforce{0.8w,0.5h}{v5}
\fmf{photon}{i1,v1}
\fmf{photon}{i2,v2}
\fmf{plain}{v5,o}
\fmf{photon,tension=0}{v2,v5}
\fmf{plain,tension=0}{v3,v4}
\fmf{photon,tension=.4}{v1,v4}
\fmf{photon,tension=.4}{v4,v5}
\fmf{plain,tension=0}{v1,v3}
\fmf{photon,tension=0}{v2,v3}
\end{fmfgraph*}} & = & \mu^{2(4-D)} 
\int {\mathfrak D}^D k_1 {\mathfrak D}^D k_2
\frac{1}{{\mathcal D}_{4} 
         {\mathcal D}_{5} 
	 {\mathcal D}_{7} 
	 {\mathcal D}_{8} 
	 {\mathcal D}_{12} 
	 {\mathcal D}_{13} }  \\
& = & \left( \frac{\mu^{2}}{a} \right) ^{2 \epsilon} 
\sum_{i=-2}^{1} \epsilon^{i} F^{(24)}_{i} + {\mathcal O} \left( 
\epsilon \right) , 
\eea
where:
\bea
a^2 F^{(24)}_{-2} & = &  \frac{1}{x}  \, , \\
a^2 F^{(24)}_{-1} & = &  - \frac{1}{x}   
            \Bigl[
            1
          + H(0,x)
          - H(-1,x)
	   \Bigr]  + \frac{1}{x^2}   
            \Bigl[
            H(-1,x)
          + H(0,-1,x)
	   \Bigr]\, , \\
a^2 F^{(24)}_{0} & = &  - \frac{1}{x}   
            \Bigl[
            \zeta(2)
          - 2 H(0,x)
          + 2 H(-1,x)
          - H(0,0,x)
          + 4 H(-1,-1,x) \nn\\
& & 
          - 4 H(0,-1,x)
	   \Bigr] - \frac{1}{x^2}   
            \Bigl[
            2 H(-1,x)
          + 4 H(-1,-1,x)
          + 2 H(0,-1,x) \nn\\
& & 
          + 2 H(r,r,0,x)
          + 4 H(0,-1,-1,x)
          + 2 H(0,0,-1,x)
	   \Bigr]  \nn\\
& & 
	- \frac{4-x}{x \sqrt{x(4-x)}}   H(r,0,x) \, , \\
a^2 F^{(24)}_{1} & = &  \frac{1}{x}   
            \Bigl[
            4
          + 2 \zeta(2)
          - 2 \zeta(3)
          - 4 H(0,x)
          + \zeta(2) H(0,x)
          + 2 \zeta(2) H(1,x) \nn\\
& & 
          + 4 H(-1,x)
          + 2 \zeta(2) H(-1,x)
          - 2 H(0,0,x)
          - 8 H(0,-1,x) \nn\\
& & 
          + 8 H(-1,-1,x)
          - H(0,0,0,x)
          + 16 H(-1,-1,-1,x) \nn\\
& & 
          - 6 H(-1,0,-1,x)
          - 16 H(0,-1,-1,x)
          + 2 H(0,0,-1,x) \nn\\
& & 
          - 4 H(1,0,-1,x)
          + 3 H(r,r,0,x)
	   \Bigr]  
	   + \frac{1}{x^2}   
            \Bigl[
            4 H(-1,x)
          + 2 \zeta(2) H(-1,x) \nn\\
& & 
          - 2 \zeta(2) H(1,x)
          - 2 \zeta(2) H(0,1,x)
          + 4 H(0,-1,x)
          + 2 \zeta(2) H(0,-1,x) \nn\\
& & 
          - 2 \zeta(2) H(r,r,x)
          + 8 H(-1,-1,x)
          + 4 H(0,0,-1,x)
          + 4 H(1,0,-1,x) \nn\\
& & 
          + 16 H(-1,-1,-1,x)
          - 6 H(-1,0,-1,x)
          + 8 H(0,-1,-1,x) \nn\\
& & 
          + 2 H(r,r,0,0,x)
          + 2 H(r,0,r,0,x)
          - 4 H(r,4,r,0,x) \nn\\
& & 
          + 16 H(0,-1,-1,-1,x)
          - 6 H(0,-1,0,-1,x)
          + 8 H(0,0,-1,-1,x) \nn\\
& & 
          + 4 H(0,0,0,-1,x)
          + 4 H(0,1,0,-1,x)
	   \Bigr]
	- \frac{4-x}{x \sqrt{x(4-x)}}    
            \Bigl[
            \zeta(2) H(r,x) \nn\\
& & 
          - 2 H(r,0,x)
          - H(r,0,0,x)
          - H(0,r,0,x)
          + 2 H(4,r,0,x)
	   \Bigr]  
\, .
\eea
The above amplitude has a double pole in $\epsilon$ coming
from the following region: in the inner triangle a large momentum $k_2$
flows while in the external triangle a soft momentum $k_1$ flows, or
$k_1^2\ll k_2^2$.  
In this limit, the diagram factorizes into a massless triangle
with the well-known double IR pole (soft x collinear),
times a coefficient functions given by the inner triangle, 
which is effectively point-like.

\section{Small momentum expansion of six denominator amplitudes \label{PiccoliP} }

In this Section we present the small momentum expansions  $|s| \ll m^2$ of all the 
6-denomi\-na\-tor diagrams, i.e. expansions in powers of $x$ and 
$L=\log x$ up to first order in $x$ included. 
The expansion of the GHPLs for a small value of the argument $|x|\ll 1$ 
is obtained in the following way:
\begin{itemize}
\item
we explicitly write the GHPL as a repeated integration over the basis functions, as for
example:
\be
H(1,r,r,0;x)=\int_0^{x}\frac{dx_1}{1-x_1}\int_0^{x_1}\frac{dx_2}{\sqrt{x_2(4-x_2)}}
\int_0^{x_2}\frac{dx_3}{\sqrt{x_3(4-x_3)}} \log x_3;
\ee
\item
we expand the basis functions in powers of $x$ up to the required order 
in x\footnote{The factors $1/x$ and $1/\sqrt{x}$ are clearly not expanded.}, such as
for instance:
\be
\frac{1}{\sqrt{x(4-x)}} = \frac{1}{2\sqrt{x}} + \frac{\sqrt{x}}{16} + \cdots;
\ee
\item
we integrate term by term the expanded functions. This involves in general 
the integration of functions of the form $x^q\log x^k$, with $q$ integer or 
half integer and $k$ integer.
\end{itemize}
The expansions are given below.
\be
\parbox{20mm}{\begin{fmfgraph*}(15,15)
\fmfleft{i1,i2}
\fmfright{o}
\fmfforce{0.2w,0.9h}{v2}
\fmfforce{0.2w,0.1h}{v1}
\fmfforce{0.2w,0.5h}{v3}
\fmfforce{0.8w,0.5h}{v5}
\fmf{photon}{i1,v1}
\fmf{photon}{i2,v2}
\fmf{plain}{v5,o}
\fmf{photon,tension=0}{v2,v5}
\fmf{plain,tension=0}{v3,v4}
\fmf{photon,tension=.4}{v1,v4}
\fmf{photon,tension=.4}{v4,v5}
\fmf{plain,tension=0}{v1,v3}
\fmf{plain,tension=0}{v2,v3}
\end{fmfgraph*}} = \left( \frac{\mu^{2}}{a} \right) ^{2 \epsilon} 
\sum_{j=0}^{2} x^j
A^{0}_{j} + {\mathcal O} (x^{3}) , 
\ee
where:
\bea
a^2 A^{0}_{0} & = & 
            4
	   + \sqrt{3} H(r,0,1) 
          - L 
\, , \\
a^2 A^{0}_{1} & = & 
            \frac{19}{9}
          + \frac{1}{2} \zeta(2)
          + \frac{3 \sqrt{3}}{4} H(r,0,1)
          - \frac{13}{24} L 
\, , \\
a^2 A^{0}_{2} & = & 
            \frac{12509}{16200}
          + \frac{2}{3} \zeta(2)
          + \frac{\sqrt{3}}{2} H(r,0,1)
          - \frac{197}{540} L
\eea

\be
\parbox{20mm}{\begin{fmfgraph*}(15,15)
\fmfleft{i1,i2}
\fmfright{o}
\fmfforce{0.2w,0.9h}{v2}
\fmfforce{0.2w,0.1h}{v1}
\fmfforce{0.2w,0.5h}{v3}
\fmfforce{0.8w,0.5h}{v5}
\fmf{photon}{i1,v1}
\fmf{photon}{i2,v2}
\fmf{plain}{v5,o}
\fmf{photon,tension=0}{v2,v5}
\fmf{photon,tension=0}{v3,v4}
\fmf{plain,tension=.4}{v1,v4}
\fmf{photon,tension=.4}{v4,v5}
\fmf{photon,tension=0}{v1,v3}
\fmf{plain,tension=0}{v2,v3}
\end{fmfgraph*}} = \left( \frac{\mu^{2}}{a} \right) ^{2 \epsilon} 
\sum_{j=0}^{2} x^j B^{0}_{j} + {\mathcal O} (x^{3}) , 
\ee
where:
\bea
a^2 B^{0}_{0} & = & 3 
          - 2 L 
          + \frac{1}{2} L^2
\, , \\
a^2 B^{0}_{1} & = & 
           \frac{11}{16}
          - \frac{1}{2} \zeta(2)
          - \frac{1}{2} L
          + \frac{1}{8} L^2
\, , \\
a^2 B^{0}_{2} & = & 
            \frac{7}{8} 
          - \frac{1}{3} \zeta(2)
          - \frac{41}{108}  L
          + \frac{5}{36}  L^2
\, .
\eea

\be
\parbox{20mm}{\begin{fmfgraph*}(15,15)
\fmfleft{i1,i2}
\fmfright{o}
\fmf{photon}{i1,v1}
\fmf{photon}{i2,v2}
\fmf{plain}{v5,o}
\fmf{plain,tension=.3}{v2,v3}
\fmf{photon,tension=.3}{v3,v5}
\fmf{plain,tension=.3}{v1,v4}
\fmf{photon,tension=.3}{v4,v5}
\fmf{photon,tension=0}{v2,v1}
\fmf{photon,tension=0}{v4,v3}
\end{fmfgraph*}} = \left( \frac{\mu^{2}}{a} \right) ^{2 \epsilon} 
\sum_{j=0}^{2} x^{j}
C^{0}_{j} + {\mathcal O} (x^{\frac{9}{2}}) , 
\ee
where:
\bea
a^2 C^{0}_{0} & = & 
                    1
                  + 2 \zeta(2)
                  - L
                  + \frac{1}{2} L^2
\, , \\
a^2 C^{0}_{1} & = & 
          - \frac{17}{24}
          -  \zeta(2)
          + \frac{5}{12} L
          - \frac{1}{4} L^2
\, , \\
a^2 C^{0}_{2} & = & 
            \frac{827}{2160}
          + \frac{1}{2} \zeta(2)
          - \frac{61}{360} L
          + \frac{1}{8} L^2
\, .
\eea

\be
\parbox{20mm}{\begin{fmfgraph*}(15,15)
\fmfleft{i1,i2}
\fmfright{o}
\fmf{photon}{i1,v1}
\fmf{photon}{i2,v2}
\fmf{plain}{v5,o}
\fmf{photon,tension=.3}{v2,v3}
\fmf{photon,tension=.3}{v3,v5}
\fmf{photon,tension=.3}{v1,v4}
\fmf{photon,tension=.3}{v4,v5}
\fmf{plain,tension=0}{v2,v1}
\fmf{plain,tension=0}{v4,v3}
\end{fmfgraph*}} = \left( \frac{\mu^{2}}{a} \right) ^{2 \epsilon} 
\sum_{j=0}^{2} x^j E^{0}_{j} + {\mathcal O} (x^{3}) , 
\ee
where:
\bea
a^2 E^{0}_{0} & = & 
          - 4
          + 3 \zeta(2)
          - L
          + L^2
\, , \\
a^2 E^{0}_{1} & = & 
          - \frac{341}{72}
          + 3 \zeta(2)
          - \frac{5}{6} L
          + L^2
\, , \\
a^2 E^{0}_{2} & = & 
          - \frac{2617}{600}
          + \frac{11}{4} \zeta(2)
          - \frac{79}{120} L
          + \frac{11}{12} L^2
\, .
\eea

\be
\parbox{20mm}{\begin{fmfgraph*}(15,15)
\fmfleft{i1,i2}
\fmfright{o}
\fmfforce{0.2w,0.9h}{v2}
\fmfforce{0.2w,0.1h}{v1}
\fmfforce{0.2w,0.5h}{v3}
\fmfforce{0.8w,0.5h}{v5}
\fmf{photon}{i1,v1}
\fmf{photon}{i2,v2}
\fmf{plain}{v5,o}
\fmf{photon,tension=0}{v2,v5}
\fmf{photon,tension=0}{v3,v4}
\fmf{photon,tension=.4}{v1,v4}
\fmf{photon,tension=.4}{v4,v5}
\fmf{plain,tension=0}{v1,v3}
\fmf{plain,tension=0}{v2,v3}
\end{fmfgraph*}} = \left( \frac{\mu^{2}}{a} \right) ^{2 \epsilon} 
\sum_{j=0}^{2} x^j G^{0}_{j} + {\mathcal O} (x^{3}) , 
\ee
where:
\bea
a^2 G^{0}_{0} & = & 
            3
          - \zeta(2)
          - 2 L
          + \frac{1}{2} L^2
\, , \\
a^2 G^{0}_{1} & = & 
            \frac{25}{16}
          - \frac{3}{4} \zeta(2)
          - \frac{5}{4} L
          + \frac{3}{8} L^2
\, , \\
a^2 G^{0}_{2} & = & 
            \frac{251}{216}
          - \frac{11}{18} \zeta(2)
          - \frac{107}{108} L
          + \frac{11}{36} L^2
\, .
\eea

\be
\parbox{20mm}{\begin{fmfgraph*}(15,15)
\fmfleft{i1,i2}
\fmfright{o}
\fmfforce{0.2w,0.9h}{v2}
\fmfforce{0.2w,0.1h}{v1}
\fmfforce{0.2w,0.5h}{v3}
\fmfforce{0.8w,0.5h}{v5}
\fmf{photon}{i1,v1}
\fmf{photon}{i2,v2}
\fmf{plain}{v5,o}
\fmf{photon,tension=0}{v2,v5}
\fmf{plain,tension=0}{v3,v4}
\fmf{photon,tension=.4}{v1,v4}
\fmf{photon,tension=.4}{v4,v5}
\fmf{photon,tension=0}{v1,v3}
\fmf{plain,tension=0}{v2,v3}
\end{fmfgraph*}} = \left( \frac{\mu^{2}}{a} \right) ^{2 \epsilon} 
\sum_{j=0}^{2} x^j  \biggl[ 
\sum_{i=-1}^{0} \epsilon^{i} 
I^{i}_{j}  \biggr] + {\mathcal O} (x^{3}) , 
\ee
where:
\bea
a^2 I^{-1}_{0} & = & L
\, , \\
a^2 I^{0}_{0} & = & 
          - 6
          + 4 \zeta(2)
          + 3 L
          - \frac{1}{2} L^2
\, , \\
a^2 I^{-1}_{1} & = & 
          - \frac{1}{4}
          + \frac{3}{4} L
\, , \\
a^2 I^{0}_{1} & = & 
          - \frac{11}{2}
          + 3 \zeta(2)
          + \frac{21}{8} L
          - \frac{3}{8} L^2
\, , \\
a^2 I^{-1}_{2} & = & 
          - \frac{1}{12}
          + \frac{11}{18} L
\, , \\
a^2 I^{0}_{2} & = & 
          - \frac{859}{216}
          + \frac{22}{9} \zeta(2)
          + \frac{22}{9} L
          - \frac{11}{36} L^2
\, .
\eea

\be
\parbox{20mm}{\begin{fmfgraph*}(15,15)
\fmfleft{i1,i2}
\fmfright{o}
\fmfforce{0.2w,0.9h}{v2}
\fmfforce{0.2w,0.1h}{v1}
\fmfforce{0.2w,0.5h}{v3}
\fmfforce{0.8w,0.5h}{v5}
\fmf{photon}{i1,v1}
\fmf{photon}{i2,v2}
\fmf{plain}{v5,o}
\fmf{photon,tension=0}{v2,v5}
\fmf{plain,tension=0}{v3,v4}
\fmf{photon,tension=.4}{v1,v4}
\fmf{photon,tension=.4}{v4,v5}
\fmf{plain,tension=0}{v1,v3}
\fmf{photon,tension=0}{v2,v3}
\end{fmfgraph*}} = \left( \frac{\mu^{2}}{a} \right) ^{2 \epsilon} 
\sum_{j=-1}^{2} x^j  \biggl[ 
\sum_{i=-2}^{1} \epsilon^{i} 
J^{i}_{j}  \biggr] + {\mathcal O} (x^{3}) , 
\ee
where:
\bea
a^2 J^{-2}_{-1} & = & 
                 1
\, , \\
a^2 J^{-1}_{-1} & = & 
                  1 - L
\, , \\
a^2 J^{0}_{-1} & = & 
                  1 - \zeta(2) - L + \frac{1}{2} L^2
\, , \\
a^2 J^{1}_{-1} & = & 
            1
          - \zeta(2)
          - 2 \zeta(3)
          + \zeta(2) L
          - L
          + \frac{1}{2} L^2
          - \frac{1}{6} L^3
\, , \\
a^2 J^{-1}_{0} & = & 
            \frac{1}{4}
\, , \\
a^2 J^{0}_{0} & = & 
            \frac{35}{72}
          + \frac{1}{12} L
\, , \\
a^2 J^{1}_{0} & = & 
          - \frac{359}{432}
          + \frac{13}{12} \zeta(2)
          + \frac{7}{24} L
          - \frac{1}{24} L^2
\, , \\
a^2 J^{-1}_{1} & = & 
          - \frac{1}{18}
\, , \\
a^2 J^{0}_{1} & = & 
          - \frac{209}{675}
          + \frac{1}{180} L
\, , \\
a^2 J^{1}_{1} & = & 
          - \frac{141091}{162000}
          + \frac{1}{180} \zeta(2)
          + \frac{23}{1080} L
          - \frac{1}{360} L^2
\, , \\
a^2 J^{-1}_{2} & = & 
            \frac{1}{48}
\, , \\
a^2 J^{0}_{2} & = & 
            \frac{11063}{78400}
          + \frac{1}{1680} L
\, , \\
a^2 J^{1}_{2} & = & 
            \frac{105455939}{296352000}
          + \frac{47}{560} \zeta(2)
          + \frac{17}{6720} L
          - \frac{1}{3360} L^2
\, .
\eea

\section{Large momentum expansion of six denominator amplitudes \label{GrandiP}}

In this Section we give the asymptotic expansions for $|s| \gg m^2$ of
all the 6-denominator scalar integrals, i.e. the expansion in powers
of $1/x$ and $L=\log x$ up to the order $1/x^{4}$ included. 
These results are relevant for the study of the structure of the
infrared logarithms coming from multiple emission.
\be
\parbox{20mm}{\begin{fmfgraph*}(15,15)
\fmfleft{i1,i2}
\fmfright{o}
\fmfforce{0.2w,0.9h}{v2}
\fmfforce{0.2w,0.1h}{v1}
\fmfforce{0.2w,0.5h}{v3}
\fmfforce{0.8w,0.5h}{v5}
\fmf{photon}{i1,v1}
\fmf{photon}{i2,v2}
\fmf{plain}{v5,o}
\fmf{photon,tension=0}{v2,v5}
\fmf{photon,tension=0}{v3,v4}
\fmf{plain,tension=.4}{v1,v4}
\fmf{photon,tension=.4}{v4,v5}
\fmf{photon,tension=0}{v1,v3}
\fmf{plain,tension=0}{v2,v3}
\end{fmfgraph*}} = \left( \frac{\mu^{2}}{a} \right) ^{2 \epsilon} 
\sum_{j=1}^{4} \frac{1}{x^j} B^{0}_{-j} 
+ {\mathcal O} \left( \frac{1}{x^{5}} \right) , 
\ee
where:
\bea
a^2 B^{0}_{-1} & = &
          - 8 a_4
          + \frac{19}{4} \zeta^2(2)
          + 2 \zeta(2) \log^2{2}
          - \frac{1}{3} \log^4{2}
\, , \\
a^2 B^{0}_{-2} & = &
          - 3
          - 2 \zeta(2)
          + \zeta(3)
          - 2 \zeta(2) L
          - 3 L
          - \frac{3}{2} L^2
          - \frac{1}{2} L^3
\, , \\
a^2 B^{0}_{-3} & = &
          - \frac{21}{16}
          - \frac{1}{2} \zeta(2)
          + \frac{1}{2} \zeta(3)
          - L \zeta(2)
          + \frac{3}{8} L
          - \frac{1}{8} L^2
          - \frac{1}{4} L^3
\, , \\
a^2 B^{0}_{-4} & = & 
          - \frac{11}{24}
          + \frac{1}{9} \zeta(2)
          + \frac{1}{3} \zeta(3)
          + \frac{3}{4} L
          - \frac{2}{3} L \zeta(2)
          + \frac{1}{4} L^2
          - \frac{1}{6} L^3
\, ,
\eea
and where $a_4= {\rm Li}_4(1/2)$.

\be
\parbox{20mm}{\begin{fmfgraph*}(15,15)
\fmfleft{i1,i2}
\fmfright{o}
\fmfforce{0.2w,0.9h}{v2}
\fmfforce{0.2w,0.1h}{v1}
\fmfforce{0.2w,0.5h}{v3}
\fmfforce{0.8w,0.5h}{v5}
\fmf{photon}{i1,v1}
\fmf{photon}{i2,v2}
\fmf{plain}{v5,o}
\fmf{photon,tension=0}{v2,v5}
\fmf{photon,tension=0}{v3,v4}
\fmf{photon,tension=.4}{v1,v4}
\fmf{photon,tension=.4}{v4,v5}
\fmf{plain,tension=0}{v1,v3}
\fmf{plain,tension=0}{v2,v3}
\end{fmfgraph*}} = \left( \frac{\mu^{2}}{a} \right) ^{2 \epsilon} 
\sum_{j=1}^{4} \frac{1}{x^j} G^{0}_{-j} + {\mathcal O} 
\left( \frac{1}{x^{5}} \right) , 
\ee
where:
\bea
a^2 G^{0}_{-1} & = &
            \frac{27}{10} \zeta^2(2)
\, , \\
a^2 G^{0}_{-2} & = &
          - 2
          - \zeta(2)
          - \zeta(3)
          - L \zeta(2)
          - 2 L
          - L^2
          - \frac{1}{3} L^3
\, , \\
a^2 G^{0}_{-3} & = &
            \frac{3}{8}
          + \frac{1}{4} \zeta(2)
          - \frac{1}{2} \zeta(3)
          - \frac{1}{2} \zeta(2) L
          + \frac{3}{4} L
          + \frac{1}{4} L^2
          - \frac{1}{6} L^3
\, , \\
a^2 G^{0}_{-4} & = & 
          - \frac{13}{162}
          + \frac{7}{18} \zeta(2)
          - \frac{1}{3} \zeta(3)
          - \frac{1}{3} \zeta(2) L
          + \frac{55}{108} L
          + \frac{7}{18} L^2
          - \frac{1}{9} L^3
\, .
\eea

\be
\parbox{20mm}{\begin{fmfgraph*}(15,15)
\fmfleft{i1,i2}
\fmfright{o}
\fmfforce{0.2w,0.9h}{v2}
\fmfforce{0.2w,0.1h}{v1}
\fmfforce{0.2w,0.5h}{v3}
\fmfforce{0.8w,0.5h}{v5}
\fmf{photon}{i1,v1}
\fmf{photon}{i2,v2}
\fmf{plain}{v5,o}
\fmf{photon,tension=0}{v2,v5}
\fmf{plain,tension=0}{v3,v4}
\fmf{photon,tension=.4}{v1,v4}
\fmf{photon,tension=.4}{v4,v5}
\fmf{photon,tension=0}{v1,v3}
\fmf{plain,tension=0}{v2,v3}
\end{fmfgraph*}} = \left( \frac{\mu^{2}}{a} \right) ^{2 \epsilon} 
\sum_{j=2}^{4} \frac{1}{x^j}  \biggl[ 
\sum_{i=-2}^{2} \epsilon^{i} 
I^{0}_{-j} \biggr] + {\mathcal O} 
\left( \frac{1}{x^{5}} \right)  , 
\ee
where:
\bea
a^2 I^{-1}_{-1} & = & - 2 \zeta(3)
\, , \\
a^2 I^{0}_{-1} & = & \frac{1}{5} \zeta^2(2)
\, , \\
a^2 I^{-1}_{-2} & = &
            1
          + L
          + \frac{1}{2} L^2
\, , \\
a^2 I^{0}_{-2} & = &
          - 4
          + 2 \zeta(2)
          - 4 \zeta(3)
          + 2 \zeta(2) L
          - 4 L
          - 2 L^2
          - \frac{2}{3} L^3
\, , \\
a^2 I^{-1}_{-3} & = &
          - \frac{1}{8}
          + \frac{3}{4} L
          + \frac{1}{4} L^2
\, , \\
a^2 I^{0}_{-3} & = &
            \frac{41}{8}
          - \frac{3}{2} \zeta(2)
          - 2 \zeta(3)
          + \zeta(2) L
          + \frac{5}{4} L
          - \frac{7}{4} L^2
          - \frac{1}{3} L^3
\, , \\
a^2 I^{-1}_{-4} & = & 
            \frac{13}{108}
          + \frac{11}{18} L
          + \frac{1}{6} L^2
\, , \\
a^2 I^{0}_{-4} & = & 
            \frac{3019}{648}
          - \frac{16}{9} \zeta(2)
          - \frac{4}{3} \zeta(3)
          + \frac{2}{3} \zeta(2) L
          + \frac{23}{27} L
          - \frac{53}{36} L^2
          - \frac{2}{9} L^3
\, .
\eea

\section{Conclusions\label{concl}}

We have presented the exact analytic evaluation of the $25$ master integrals 
containing 2 and 3 massive propagators entering the planar amplitudes 
of the 2-loop electroweak form factor.
While the reduction to master integrals does not present any 
new element with respect to our
previous computation \cite{UgoRo} and is done with the same algorithm, 
the analytic evaluation of the master integrals requires a 
non-trivial extension of the harmonic polylogarithm theory. 
The presence of $2$ massive particles in the $s$ or in the $t$ channel
opens indeed thresholds and pseudothresholds in 
$s=\pm 4m^2$ respectively, in addition to the old ones in $s=0,\pm m^2$.

The generalization of the 1-dimensional harmonic polylogarithms 
has basically required:
\begin{itemize}
\item
the introduction of new basis functions, in addition to the
usual one, involving complex constants and radicals; 
\item
a set of recursion relations to take the integrals with 
semi-integer powers coming from the evaluation of the master 
integrals to a unique form fixed by the basis function choice.
\end{itemize}
The basic properties of the ordinary harmonic polylogarithms
are maintained by the generalization.

The small momentum expansion of all the 6-denominator 
amplitudes has been obtained by means of a series expansion 
of the basis functions.

We could also obtain the large momentum expansion of all the 
six-denominator amplitude involving only ordinary harmonic 
polylogarithms.

We compared our results with those present in the literature
usually in the form of resummed small momentum expansions
or truncated large momentum expansions, finding complete agreement.

In order to complete the evaluation of the master integrals, 
3 steps are still to be taken:
\begin{itemize}
\item
as explained in Section \ref{HPLs}, 
the transformation $x\rightarrow 1/x$ requires the knowledge of
all the $H(\vec{w};x)$'s in $x=1$. The $H(\vec{w};1)$'s have to 
be expressed in terms of a minimal set of transcendental 
constants \cite{inpreparation};
\item
the evaluation of the master integrals related to 
the crossed ladder topology. 
The reduction, using both the numerical-indices method
and the symbolic method, shows that this topology has 3 master 
integrals. The resulting system of 3 differential equations cannot 
be completely triangularized by means of the techniques discussed 
in this work, but can be split into a second-order
and a first-order differential equations \cite{inpreparation}.
\item
The numerical evaluation of the generalized harmonic polylogarithms,
which does not seem to have specific difficulties with respect to the
ordinary case.
\end{itemize}

\section{Acknowledgement}

We are grateful to J.~Vermaseren for his kind assistance in the use
of the algebra manipulating program {\tt FORM}~\cite{FORM}, by which
all our calculations were carried out.

We wish to thank E.~Remiddi for discussions and for use of the {\tt C} 
program {\tt SOLVE} \cite{SOLVE} to solve the linear systems generated 
by the ibp identities.

R.B. wishes to thank the Universit\`a di Roma  ``La Sapienza'' for hospitality 
during the final part of this work.

\appendix

\section{One-loop master integrals \label{app1}}

In this Appendix we present the results for the 
1-loop master integrals containing 2 massive propagators.
We have recomputed them with the method of the differential equations 
described in the main body of the paper in terms of GHPLs. In the case
of the bubble, we found that it was necessary to push the $\epsilon$ expansion
up to third order included.
In our previous work \cite{UgoRo} we gave the expressions of the 1-loop
master integrals containing at most 1 massive propagator.
The above amplitudes are necessary for the computation of the 
factorized 2-loop master integrals, which are presented in the next
section.

\subsection{Bubble}

\bea
\hspace{-5mm}
\parbox{15mm}{
\begin{fmfgraph*}(15,15)
\fmfleft{i}
\fmfright{o}
\fmf{plain}{i,v1}
\fmf{plain}{v2,o}
\fmf{plain,tension=.15,left}{v1,v2}
\fmf{plain,tension=.15,left}{v2,v1}
\end{fmfgraph*} } & = & 
\mu^{(4-D)} 
\int {\mathfrak D}^D k 
\frac{1}{(k^{2} +a) \, [(p-k)^{2}+a]} \nn\\
\hspace{-5mm}
& = & \left( \frac{\mu^{2}}{a} \right) ^{\epsilon} 
\sum_{i=-1}^{3} \epsilon^{i} B_{i}
+ {\mathcal O} \left( \epsilon^{4} \right) \, , 
\label{appb2}
\eea
where:
\bea
B_{-1} & = & 1 \, ,  \\
B_{0} & = &   2 - \frac{x+4}{\sqrt{x(x+4)}} H( - r,x) \, , \\
B_{1} & = &   4 
             - \frac{x+4}{\sqrt{x(x+4)}} \Bigl[
	         2 H( - r,x)
	       - H(-4, - r,x)
	        \Bigr]
\, , \\
B_{2} & = &  8 \! 
             - \frac{x+4}{\sqrt{x(x+4)}} \Bigl[
	         4 H( - r,x) \! 
	       -  \! 2 H(-4,  \! - r,x) \! 
	       +  \! H(-4,  \! -4,  \! - r,x) \! 
	        \Bigr] , \\
B_{3} & = &    16 
             - \frac{x+4}{\sqrt{x(x+4)}} \Bigl[
	         8 H( - r,x)
	       - 4 H(-4, - r,x)
	       + 2 H(-4, -4, - r,x) \nn\\
& & 
	       - H(-4, -4, -4, - r,x)
	        \Bigr]
\, .
\eea

\subsection{Vertex}

\bea
\hspace{-5mm} 
\parbox{15mm}{\begin{fmfgraph*}(15,15)
\fmfleft{i1,i2}
\fmfright{o}
\fmf{photon}{i1,v1}
\fmf{photon}{i2,v2}
\fmf{plain}{v3,o}
\fmf{plain,tension=.3}{v2,v3}
\fmf{plain,tension=.3}{v1,v3}
\fmf{photon,tension=0}{v2,v1}
\end{fmfgraph*} } & = & \mu^{(4-D)}
\int {\mathfrak D}^D k 
\frac{1}{k^{2} \, [(p_{1}-k)^{2}+a] \, [(p_{2}+k)^{2}+a]} \nn\\
\hspace{-5mm} 
& = & \left( \frac{\mu^{2}}{a} \right) ^{\epsilon} 
 \sum_{i=0}^{2} \epsilon^{i} V_{i} 
 + {\mathcal O} \left( \epsilon^{3} \right) ,
\label{appb6}
\eea
where:
\bea
a K_{0} & = &  \frac{2}{x} H( - r, - r,x)
\, , \\
a K_{1} & = &  - \frac{2}{x} \Bigl[
                 H( - r,-4, - r,x)
	       - H(0, - r, - r,x)
                 \Bigr]
\, , \\
a K_{2} & = &   \frac{2}{x} \Bigl[
                 H( - r,-4,-4, - r,x)
               - H(0, - r,-4, - r,x) \nn\\
& & 
	       + H(0,0, - r, - r,x)
                 \Bigr]
\, .
\eea

\section{Factorized  master integrals \label{app2}}

In this Appendix we give the expressions of the factorized
2-loop master integrals, i.e. of the MIs in which the 2 loops do not have
common propagators. One has only to multiply the 1-loop master
integrals representing the separated subdiagrams and convert products of
GHPLs $H(\vec{a};x)H(\vec{b};x)$ into linear combinations of $H$'s by
using the algebra-identity in \cite{Polylog}.

\subsection{ Topology $t=2$ }

\bea
\parbox{20mm}{\begin{fmfgraph*}(15,15)
\fmfleft{i}
\fmfright{o}
\fmf{phantom}{i,v1}
\fmf{phantom}{v1,o}
\fmf{plain,left}{v1,v1}
\fmf{plain,right}{v1,v1}
\end{fmfgraph*}} & = & \mu^{2(4-D)} 
\int {\mathfrak D}^D k_1 {\mathfrak D}^D k_2
\frac{1}{{\mathcal D}_{12} {\mathcal D}_{13} } \\
& = & \left( \frac{\mu^{2}}{a} \right) ^{2 \epsilon} 
\sum_{i=-2}^{2} \epsilon^{i} F^{(25)}_{i} + {\mathcal O} \left( 
\epsilon^{3} \right) , 
\eea
where:
\bea
\frac{F^{(25)}_{-2}}{a^2} & = & 1 \, , \\
\frac{F^{(25)}_{-1}}{a^2} & = & 2 \, , \\
\frac{F^{(25)}_{0}}{a^2} & = & 3 \, , \\
\frac{F^{(25)}_{1}}{a^2} & = & 4 \, , \\
\frac{F^{(25)}_{2}}{a^2} & = & 5 \, .
\eea

The above amplitude appears, in general, in the reduction of all
the amplitudes having at least 1 massive propagator
in anyone of the 2 loops.

\subsection{ Topology $t=4$  }

\bea
\parbox{20mm}{\begin{fmfgraph*}(15,15)
\fmfleft{i}
\fmfright{o}
\fmf{plain}{i,v1}
\fmf{plain}{v3,o}
\fmf{plain,tension=.2,left}{v1,v2}
\fmf{plain,tension=.2,right}{v1,v2}
\fmf{photon,tension=.2,left}{v2,v3}
\fmf{photon,tension=.2,right}{v2,v3}
\end{fmfgraph*}} & = & \mu^{2(4-D)} 
\int {\mathfrak D}^D k_1 {\mathfrak D}^D k_2
\frac{1}{{\mathcal D}_{2} 
         {\mathcal D}_{10} 
	 {\mathcal D}_{12} 
         {\mathcal D}_{20} } \\
& = & \left( \frac{\mu^{2}}{a} \right) ^{2 \epsilon} 
\sum_{i=-2}^{2} \epsilon^{i} F^{(26)}_{i} + {\mathcal O} \left( 
\epsilon^{3} \right) , 
\eea
where:
\bea
F^{(26)}_{-2} & = & 1 \, , \\
F^{(26)}_{-1} & = &  4 - H(0;x) - \frac{4+x}{\sqrt{x(4+x)}} H(-r;x)
 \, , \\
F^{(26)}_{0} & = & 12
          - \zeta(2) \! 
          - 4 H(0;x) \! 
          +  \! H(0,0;x) \! 
	  - \frac{(4+x)}{\sqrt{x(4+x)}} \Bigl[ 
          - 4 H( - r;x) \! 
          +  \! H( - r,0;x) \nn\\
& & 
          + H(-4, - r;x)
          + H(0, - r;x) \Bigr]
	  \, , \\
F^{(26)}_{1} & = & 32
          - 4 \zeta(2) \! 
          - 2 \zeta(3) \! 
          - 12 H(0;x) \! 
          + H(0;x) \zeta(2) \! 
          + \!  4 H(0,0;x) \! 
          - H(0,0,0;x) \nn\\
& & 
	  +  \! \frac{(4+x)}{\sqrt{x(4+x)}} \Bigl[ 
          - 12 H( - r;x) \! 
          +  \! \zeta(2) H( - r;x) \! 
          +  \! 4 H( - r,0;x) \! 
          - H( - r,0,0;x) \nn\\
& & 
          +  \! 4 H(-4, - r;x) \! 
          - H(-4, - r,0;x) \! 
          - H(-4,-4, - r;x) \! 
          - H(-4,0, - r;x) \nn\\
& & 
          + \!  4 H(0, - r;x) \! 
          - H(0, - r,0;x) \! 
          - H(0,-4, - r;x) \! 
          - H(0,0, - r;x) \Bigr]
	   \, , \\
F^{(26)}_{2} & = & 80
          -  \! 12 \zeta(2) \! 
          -  \! \frac{9}{10} \zeta^2(2) \! 
          - \!  8 \zeta(3) \! 
          -  \! 32 H(0;x) \! 
          +  \! 4 \zeta(2) H(0;x) \! 
          + \!  2 \zeta(3) H(0;x) \nn\\
& & 
          + 12 H(0,0;x)
          - \zeta(2) H(0,0;x)
          - 4 H(0,0,0;x)
          + H(0,0,0,0;x) \nn\\
& & 
	  + \frac{(4+x)}{\sqrt{x(4+x)}} \Bigl[ 
          -  \! 32 H( - r;x) \! 
          +  \! 2 \bigl( 2 \zeta(2)
          +  \! \zeta(3) \bigr) H( - r;x)
          +  12 H( - r,0;x) \nn\\
& & 
          - \zeta(2) H( - r,0;x)
          - 4 H( - r,0,0;x)
          + H( - r,0,0,0;x)
          + 12 H(-4, - r;x) \nn\\
& & 
          - \zeta(2) H(-4, - r;x)
          - 4 H(-4, - r,0;x)
          + H(-4, - r,0,0;x) \nn\\
& & 
          - 4 H(-4,-4, - r;x)
          + H(-4,-4, - r,0;x)
          + H(-4,-4,-4, - r;x) \nn\\
& & 
          + H(-4,-4,0, - r;x)
          - 4 H(-4,0, - r;x)
          + H(-4,0, - r,0;x) \nn\\
& & 
          + H(-4,0,-4, - r;x)
          + H(-4,0,0, - r;x)
          + 12 H(0, - r;x) \nn\\
& & 
          - \zeta(2) H(0, - r;x)
          - 4 H(0, - r,0;x)
          + H(0, - r,0,0;x)
          - 4 H(0,-4, - r;x) \nn\\
& & 
          + H(0,-4, - r,0;x)
          + H(0,-4,-4, - r;x)
          + H(0,-4,0, - r;x) \nn\\
& & 
          - 4 H(0,0, - r;x)
          + H(0,0, - r,0;x)
          + H(0,0,-4, - r;x) \nn\\
& & 
          + H(0,0,0, - r;x) \Bigr] \, .
\eea

\bea
\parbox{20mm}{\begin{fmfgraph*}(15,15)
\fmfleft{i1,i2}
\fmfright{o}
\fmf{photon}{i1,v1}
\fmf{photon}{i2,v2}
\fmf{plain}{v3,o}
\fmf{photon,tension=.3}{v2,v3}
\fmf{photon,tension=.3}{v1,v3}
\fmf{plain,tension=0}{v2,v1}
\fmf{plain,right=45}{v3,v3}
\end{fmfgraph*}} & = & \mu^{2(4-D)} 
\int {\mathfrak D}^D k_1 {\mathfrak D}^D k_2
\frac{1}{{\mathcal D}_{4} 
         {\mathcal D}_{5} 
	 {\mathcal D}_{12} 
	 {\mathcal D}_{13}} \\
& = & \left( \frac{\mu^{2}}{a} \right) ^{2 \epsilon} 
\sum_{i=-1}^{1} \epsilon^{i} F^{(27)}_{i} + {\mathcal O} \left( 
\epsilon^{3} \right) , 
\eea
where:
\bea
F^{(27)}_{-1} & = & \frac{1}{x} H(1,0,x) \, , \\
F^{(27)}_{0} & = & \frac{1}{x} \Bigl[
            \zeta(2) H(1,x)
          + H(1,0,x)
          + H(0,1,0,x)
          - H(1,0,0,x) \nn\\
& & 
          + H(1,1,0,x) \Bigr] 
\, , \\
F^{(27)}_{1} & = & \frac{1}{x} \Bigl[ 
            \Bigl( \zeta(2)
          + 2 \zeta(3) \Bigr) H(1,x)
          + \Bigl( 1
          - \zeta(2) \Bigr) H(1,0,x)
          + \zeta(2) H(1,1,x) \nn\\
& & 
          + \zeta(2) H(0,1,x)
          + H(1,1,0,x)
          + H(0,1,0,x)
          - H(1,0,0,x) \nn\\
& & 
          + H(0,0,1,0,x)
          - H(0,1,0,0,x)
          + H(0,1,1,0,x)
          + H(1,0,0,0,x) \nn\\
& & 
          + H(1,0,1,0,x)
          - H(1,1,0,0,x)
          + H(1,1,1,0,x) \Bigr]
\, .
\eea

\subsection{ Topology $t=5$ }

\bea
\parbox{20mm}{\begin{fmfgraph*}(15,15)
\fmfleft{i1,i2}
\fmfright{o}
\fmf{photon}{i1,v1}
\fmf{photon}{i2,v2}
\fmf{plain}{v4,o}
\fmf{plain,tension=.3}{v2,v3}
\fmf{plain,tension=.3}{v1,v3}
\fmf{photon,tension=0}{v2,v1}
\fmf{photon,tension=.2,left}{v3,v4}
\fmf{photon,tension=.2,right}{v3,v4}
\end{fmfgraph*}} & = & \mu^{2(4-D)} 
\int {\mathfrak D}^D k_1 {\mathfrak D}^D k_2
\frac{1}{{\mathcal D}_{1} 
         {\mathcal D}_{2} 
	 {\mathcal D}_{10} 
	 {\mathcal D}_{15} 
	 {\mathcal D}_{16} } \\
& = & \left( \frac{\mu^{2}}{a} \right) ^{2 \epsilon} 
\sum_{i=-1}^{1} \epsilon^{i} F^{(28)}_{i} + {\mathcal O} \left( 
\epsilon^{3} \right) , 
\eea
where:
\bea
aF^{(28)}_{-1} & = & \frac{2}{x} H( - r, - r,x)
\, , \\
aF^{(28)}_{0} & = & \frac{2}{x} \Bigl[ 
            2 H( - r, - r,x)
          - H( - r, - r,0,x)
          - H( - r,-4, - r,x) \nn\\
& & 
          - H( - r,0, - r,x)
		   \Bigr]
\, , \\
aF^{(28)}_{1} & = & \frac{2}{x} \Bigl[ 
             \bigl( 4 
          - \zeta(2) \bigr) H( - r, - r,x)
          - 2 H( - r, - r,0,x)
          - 2 H( - r,-4, - r,x) \nn\\
& & 
          - 2 H( - r,0, - r,x)
          + H( - r, - r,0,0,x)
          + H( - r,-4, - r,0,x) \nn\\
& & 
          + H( - r,-4,-4, - r,x)
          + H( - r,-4,0, - r,x)
          + H( - r,0, - r,0,x) \nn\\
& & 
          + H( - r,0,-4, - r,x)
          + H( - r,0,0, - r,x)
		   \Bigr]
\, .
\eea

\section{Reducible two-loop amplitudes \label{app3}}

In this Appendix we present the expressions of some 
interesting 2-loop amplitudes which can be reduced
to the MIs given in the present paper and in \cite{UgoRo}.

\subsection{Topology $t=3$}

This amplitude reduces to the product of tadpoles coming
from the contraction of its massless line.

\bea
\parbox{15mm}{\begin{fmfgraph*}(15,15)
\fmfleft{i}
\fmfright{o}
\fmf{photon}{i,v1}
\fmf{photon}{v2,o}
\fmf{plain,tension=.15,left}{v1,v2}
\fmf{plain,tension=.15}{v1,v2}
\fmf{photon,tension=.15,right}{v1,v2}
\end{fmfgraph*} }   & = & \mu^{2(4-D)} 
\int {\mathfrak D}^D k_1 {\mathfrak D}^D k_2
\frac{1}{{\mathcal D}_{8} 
         {\mathcal D}_{12} 
	 {\mathcal D}_{13} } \\
& = & \left( \frac{\mu^{2}}{a} \right) ^{2 \epsilon} 
\sum_{i=-2}^{2} \epsilon^{i} F^{(29)}_{i} + {\mathcal O} \left( 
\epsilon^{3} \right) , 
\eea
where:
\bea
\frac{F^{(29)}_{-2}}{a} & = & - 1 \, , \\
\frac{F^{(29)}_{-1}}{a} & = & - 3 \, , \\
\frac{F^{(29)}_{0}}{a} & = & - 7 \, , \\
\frac{F^{(29)}_{1}}{a} & = & - 15 \, , \\
\frac{F^{(29)}_{2}}{a} & = & - 31 \, .
\eea

\subsection{Topology $t=5$}

\bea
\parbox{15mm}{\begin{fmfgraph*}(15,15)
\fmfleft{i1,i2}
\fmfright{o}
\fmfforce{0.8w,0.5h}{v4}
\fmf{photon}{i1,v1}
\fmf{photon}{i2,v2}
\fmf{plain}{v4,o}
\fmf{photon,tension=.4}{v1,v3}
\fmf{photon,tension=.2}{v3,v4}
\fmf{photon,tension=.15}{v2,v4}
\fmf{plain,tension=0}{v2,v1}
\fmf{plain,tension=0,left=.5}{v3,v4}
\end{fmfgraph*}}   & = & \mu^{2(4-D)} 
\int {\mathfrak D}^D k_1 {\mathfrak D}^D k_2
\frac{1}{{\mathcal D}_{4} 
         {\mathcal D}_{5} 
	 {\mathcal D}_{8} 
         {\mathcal D}_{12} 
         {\mathcal D}_{13} } \\
& = & \left( \frac{\mu^{2}}{a} \right) ^{2 \epsilon} 
\sum_{i=-1}^{1} \epsilon^{i} F^{(30)}_{i} + {\mathcal O} \left( 
\epsilon^{2} \right) , 
\eea
where:
\bea
aF^{(30)}_{-1} & = &  - \frac{1}{x} H(1,0,x)
\, , \\
aF^{(30)}_{0} & = &  - \frac{1}{x}   \Bigl[
            2 \zeta(2) H(1,x)
          + H(0,-1,x)
          + H(0,1,0,x)
          + H(1,0,x) \nn\\
& & 
          - 2 H(1,0,-1,x)
          - H(1,0,0,x)
          + H(1,1,0,x)
          \Bigr] \nn\\
& & 
       + \frac{1}{(1-x)}   \Bigl[
            \zeta(2)
          - 2 H(0,-1,x)
          \Bigr] \, , \\
aF^{(30)}_{1} & = &  - \frac{1}{x} \Bigl[
            \zeta(2) H(1,x)
          + \zeta(3) H(1,x)
          + 3 H(0,-1,x)
          + H(1,0,x) \nn\\
& & 
          - \zeta(2) H(1,0,x)
          + \zeta(2) H(1,1,x)
          + 2 \zeta(2) H(0,1,x)
          + H(0,0,-1,x) \nn\\
& & 
          + H(0,1,0,x)
          - H(1,0,0,x)
          + H(1,1,0,x)
          - 4 H(0,-1,-1,x) \nn\\
& & 
          + H(0,0,1,0,x)
          - 2 H(0,1,0,-1,x)
          - H(0,1,0,0,x) \nn\\
& & 
          + H(0,1,1,0,x)
          + H(1,0,0,0,x)
          + H(1,0,1,0,x) \nn\\
& & 
          - H(1,1,0,0,x)
          + H(1,1,1,0,x)
          + 8 H(1,0,-1,-1,x)
          \Bigr] \nn\\
& & 
       + \frac{1}{(1-x)} \Bigl[
            3 \zeta(2)
          - \zeta(3)
          - 6 H(0,-1,x)
          + 8 H(0,-1,-1,x)
          \Bigr] \, .
\eea

\bea
\parbox{15mm}{\begin{fmfgraph*}(15,15)
\fmfleft{i1,i2}
\fmfright{o}
\fmfforce{0.2w,0.9h}{v2}
\fmfforce{0.2w,0.1h}{v1}
\fmfforce{0.2w,0.55h}{v3}
\fmfforce{0.2w,0.15h}{v5}
\fmfforce{0.8w,0.5h}{v4}
\fmf{photon}{i1,v1}
\fmf{photon}{i2,v2}
\fmf{plain}{v4,o}
\fmf{plain}{v2,v3}
\fmf{photon,left}{v3,v5}
\fmf{plain,right}{v3,v5}
\fmf{photon}{v1,v4}
\fmf{photon}{v2,v4}
\end{fmfgraph*}}   & = & \mu^{2(4-D)} 
\int {\mathfrak D}^D k_1 {\mathfrak D}^D k_2
\frac{1}{{\mathcal D}_{3} 
         {\mathcal D}_{4} 
         {\mathcal D}_{5} 
	 {\mathcal D}_{12} 
         {\mathcal D}_{13} } \\
& = & \left( \frac{\mu^{2}}{a} \right) ^{2 \epsilon} 
\sum_{i=-1}^{1} \epsilon^{i} F^{(31)}_{i} + {\mathcal O} \left( 
\epsilon^{2} \right) , 
\eea
where:
\bea
aF^{(31)}_{-1} & = &  - \frac{1}{x} H(1,0,x)
\, , \\
aF^{(31)}_{0} & = &  - \frac{1}{x}  \Bigl[
            \zeta(2) H(1,x)
          + 2 H(1,0,x)
          - H(1,0,0,x)
          + H(1,1,0,x)
          \Bigr]
\, , \\
aF^{(31)}_{1} & = &  - \frac{1}{x} \Bigl[
            2 \zeta(2) H(1,x)
          + 2 \zeta(3) H(1,x)
          + 4 H(1,0,x)
          + \zeta(2) H(1,1,x) \nn\\
& & 
          - 2 \zeta(2) H(0,1,x)
          - \zeta(2) H(1,0,x)
          + 2 H(1,1,0,x)
          - 2 H(1,0,0,x) \nn\\
& & 
          - 2 H(0,1,1,0,x)
          + H(1,0,0,0,x)
          + H(1,0,1,0,x) \nn\\
& & 
          - H(1,1,0,0,x)
          + H(1,1,1,0,x)
          \Bigr]
\, .
\eea
The above 2 amplitudes have a simple ultraviolet pole coming from the nested bubble, containing
1 massive denominator. The only difference is that the bubble is inserted in the $s$-channel
in the former diagram and in the $t$-channel in the latter. The coefficient of the simple pole
is indeed the same.

\bea
\parbox{15mm}{\begin{fmfgraph*}(15,15)
\fmfleft{i1,i2}
\fmfright{o}
\fmfforce{0.2w,0.9h}{v2}
\fmfforce{0.2w,0.1h}{v1}
\fmfforce{0.2w,0.55h}{v3}
\fmfforce{0.2w,0.15h}{v5}
\fmfforce{0.8w,0.5h}{v4}
\fmf{photon}{i1,v1}
\fmf{photon}{i2,v2}
\fmf{plain}{v4,o}
\fmf{photon}{v2,v3}
\fmf{plain,left}{v3,v5}
\fmf{plain,right}{v3,v5}
\fmf{photon}{v1,v4}
\fmf{photon}{v2,v4}
\end{fmfgraph*}}   & = & \mu^{2(4-D)} 
\int {\mathfrak D}^D k_1 {\mathfrak D}^D k_2
\frac{1}{{\mathcal D}_{1} 
         {\mathcal D}_{4} 
         {\mathcal D}_{5} 
	 {\mathcal D}_{13} 
         {\mathcal D}_{14} } \\
& = & \left( \frac{\mu^{2}}{a} \right) ^{2 \epsilon} 
\sum_{i=-3}^{1} \epsilon^{i} F^{(32)}_{i} + {\mathcal O} \left( 
\epsilon^{2} \right) , 
\eea
where:
\bea
aF^{(32)}_{-3} & = &  \frac{1}{x} 
\, , \\
aF^{(32)}_{-2} & = &  - \frac{1}{x} H(0,x)
\, , \\
aF^{(32)}_{-1} & = & - \frac{1}{x} \Bigl[
            \zeta(2)
          - H(0,0,x)
          \Bigr]
\, , \\
aF^{(32)}_{0} & = &   \frac{1}{x}   \Bigl[
            8 \! 
          - 2 \zeta(3) \! 
          +  \! H(r,r,0,x) \! 
          - 4 H(0,x) \! 
          +  \! \zeta(2) H(0,x) \! 
          - H(0,0,0,x) \! 
          \Bigr] \nn\\
& & 
       + \frac{2(4-x)}{x\sqrt{x(4-x)}} H(r,0,x)
\, , \\
aF^{(32)}_{1} & = &  - \frac{1}{x} \Biggl[
            16
          + 4 \zeta(2)
          + \frac{9}{10} \zeta^2(2)
          - 2 \zeta(3) H(0,x)
          - ( 4 - \zeta(2) ) H(0,0,x) \nn\\
& & 
          - \zeta(2) H(r,r,x)
          + 4 H(r,r,0,x)
          - H(0,0,0,0,x)
          + H(r,r,0,0,x) \nn\\
& & 
          + H(r,0,r,0,x)
          - 2 H(r,4,r,0,x)
          - H(0,r,r,0,x)
          \Biggr] \nn\\
& & 
       + \frac{2(4-x)}{x\sqrt{x(4-x)}} \Bigl[
            \zeta(2) H(r,x)
          - H(r,0,0,x)
          - H(0,r,0,x) \nn\\
& & 
          + 2 H(4,r,0,x)
          \Bigr]
\, .
\eea
In the IR limit, the above amplitude factorizes into the product of a massless
1-loop triangle times a vacuum bubble with 2 masses.
The triple pole is the product of a double IR pole coming from the triangle
and a simple UV pole coming from the bubble.

\bea
\parbox{15mm}{\begin{fmfgraph*}(15,15)
\fmfleft{i1,i2}
\fmfright{o}
\fmfforce{0.2w,0.9h}{v2}
\fmfforce{0.2w,0.1h}{v1}
\fmfforce{0.2w,0.5h}{v3}
\fmfforce{0.8w,0.5h}{v4}
\fmf{photon}{i1,v1}
\fmf{photon}{i2,v2}
\fmf{plain}{v4,o}
\fmf{photon,tension=0}{v1,v3}
\fmf{plain,tension=0}{v3,v4}
\fmf{photon,tension=0}{v2,v4}
\fmf{plain,tension=0}{v2,v3}
\fmf{photon,tension=0}{v1,v4}
\end{fmfgraph*}}   & = & \mu^{2(4-D)} 
\int {\mathfrak D}^D k_1 {\mathfrak D}^D k_2
\frac{1}{{\mathcal D}_{2} 
         {\mathcal D}_{4}  
         {\mathcal D}_{6} 
	 {\mathcal D}_{12} 
         {\mathcal D}_{14} } \\
& = & \left( \frac{\mu^{2}}{a} \right) ^{2 \epsilon} 
\sum_{i=-1}^{0} \epsilon^{i} F^{(33)}_{i} + {\mathcal O} \left( 
\epsilon \right) , 
\eea
where:
\bea
aF^{(33)}_{-1} & = & - \frac{1}{x} H(0,0,-1,x)
\, , \\
aF^{(33)}_{0} & = & - \frac{1}{x} \Bigl[
                 \zeta(2) H(0,1,x)
	      - H(0,0,0,-1,x)
	      - 4 H(0,0,-1,-1,x) \nn\\
& & 
	      - 2 H(0,1,0,-1,x)
                 \Bigr]
\, .
\eea
According to IR power counting, one can shrink all the massive lines in the amplitude.
This shows that the above amplitude has a simple collinear pole associated
to the evolution of the lower external line.

\section{Scalar diagrams of self-energy type \label{app4}}

In this Appendix we present the results for the scalar diagrams of self-energy insertion
type (see Fig.~\ref{fig1bis}). As explained in detail in Section \ref{compff}, 
these diagrams are effectively 5- and 4-denominator amplitudes and are all reducible to the MIs
by means of the ibps identities. 
Their analytic expressions are given for completeness.
\bea
\parbox{20mm}{\begin{fmfgraph*}(15,15)
\fmfleft{i1,i2}
\fmfright{o}
\fmfforce{0.2w,0.93h}{v2}
\fmfforce{0.2w,0.07h}{v1}
\fmfforce{0.8w,0.5h}{v5}
\fmfforce{0.2w,0.4h}{v9}
\fmfforce{0.5w,0.45h}{v10}
\fmfforce{0.2w,0.5h}{v11}
\fmf{photon}{i1,v1}
\fmf{photon}{i2,v2}
\fmf{plain}{v5,o}
\fmf{photon}{v2,v3}
\fmf{plain,tension=.25,right}{v3,v4}
\fmf{photon,tension=.25}{v3,v4}
\fmf{photon}{v4,v5}
\fmf{photon}{v1,v5}
\fmf{plain}{v2,v1}
\end{fmfgraph*}} & = & \mu^{2(4-D)} 
\int {\mathfrak D}^D k_1 {\mathfrak D}^D k_2
\frac{1}{{\mathcal D}_{4}^2 
         {\mathcal D}_{5} 
	 {\mathcal D}_{7} 
	 {\mathcal D}_{12} 
	 {\mathcal D}_{13} } 
\label{6red1} \\
& = & \left( \frac{\mu^{2}}{a} \right) ^{2 \epsilon} 
\sum_{i=-2}^{2} \epsilon^{i} F^{(34)}_{i} + {\mathcal O} \left( 
\epsilon^{3} \right) , 
\eea
where:
\bea
a^2 F^{(34)}_{-2} & = & - \frac{1}{x} \, , \\
a^2 F^{(34)}_{-1} & = & - \frac{1}{x} \Bigl[
            1
          - H(0,x)
          \Bigr]
       + \frac{1}{(1-x)} H(0,x)
\, , \\
a^2 F^{(34)}_{0} & = & \frac{1}{2x^2}  H(-1,x)
       - \frac{1}{x}  \Biggl[
            \frac{3}{2}
          - \zeta(2)
          - H(0,x)
          - H(-1,x)
          + H(0,0,x) \nn\\
& & 
          + \frac{1}{2} H(0,-1,x)
          - \frac{3}{2} H(1,0,x)
          \Biggr]
       + \frac{1}{(1-x)}  \Bigl[
            \zeta(2)
          + H(0,x)
          + H(-1,x) \nn\\
& & 
          - H(0,0,x)
          + H(1,0,x)
          \Bigr]
       + \frac{1}{(1-x)} \left[ 1- \frac{1}{(1-x)} \right]  \Biggl[
            \frac{1}{2} \zeta(2) \nn\\
& & 
          - H(0,-1,x)
          \Biggr] \, , \\
a^2 F^{(34)}_{1} & = & \frac{1}{x^2}  \Biggl[
            \frac{7}{4} H(-1,x)
          - 2 H(-1,-1,x)
          \Biggr]
       - \frac{1}{x}  \Biggl[
            \frac{11}{4}
          - \zeta(2)
          - 2 \zeta(3)
          - H(0,x)  \nn\\
& & 
          + \zeta(2) H(0,x)
          - \frac{7}{2} H(-1,x)
          - \frac{5}{2} \zeta(2) H(1,x)
          + H(0,0,x)
          - \frac{7}{4} H(1,0,x) \nn\\
& & 
          + 4 H(-1,-1,x)
          - \frac{5}{4} H(0,-1,x)
          - H(0,0,0,x)
          - 2 H(0,-1,-1,x) \nn\\
& & 
          + \frac{1}{2} H(0,0,-1,x)
          - \frac{3}{2} H(0,1,0,x)
          + 2 H(1,0,-1,x)
          + \frac{3}{2} H(1,0,0,x) \nn\\
& & 
          - \frac{3}{2} H(1,1,0,x)
          \Biggr]
       - \frac{1}{(1-x)}   \Bigl[
            \zeta(2)
          - 2 \zeta(3)
          - (1 - \zeta(2) )H(0,x) \nn\\
& & 
          - \zeta(2) H(1,x)
          - \frac{7}{2} H(-1,x)
          + H(0,0,x)
          - H(1,0,x)
          + 4 H(-1,-1,x) \nn\\
& & 
          - 4 H(0,-1,x)
          - H(0,0,0,x)
          - H(0,1,0,x)
          + H(1,0,0,x) \nn\\
& & 
          - H(1,1,0,x)
          \Bigr]
       + \frac{1}{(1-x)} \left[ 1- \frac{1}{(1-x)} \right]   \Biggl[
            \frac{7}{4} \zeta(2)
          - \frac{1}{2} \zeta(3) \nn\\
& & 
          - \frac{7}{2} H(0,-1,x)
          + 4 H(0,-1,-1,x)
          \Biggr] \, , \\
a^2 F^{(34)}_{2} & = &  \frac{1}{x^2}  \Biggl[
            \frac{35}{8} H(-1,x) \! 
          +  \! \zeta(2) H(-1,x) \! 
          -  \! 7 H(-1, \! -1,x) \! 
          +  \! 8 H(-1, \! -1, \! -1,x) \nn\\
& & 
          - 3 H(-1,0,-1,x)
          \Biggr]
       + \frac{1}{x} \Biggl[
          - \frac{43}{8}
          + \frac{9}{10} \zeta(2)^2
          + 2 \zeta(3)
          + (1 - \zeta(2)  \nn\\
& & 
          - 2 \zeta(3) ) H(0,x)
          + \frac{35}{4} H(-1,x)
          + 2 \zeta(2) H(-1,x)
          + \frac{9}{4} \zeta(2) H(1,x) \nn\\
& & 
          + 2 \zeta(3) H(1,x)
          - (1 - \zeta(2) ) H(0,0,x)
          - 14 H(-1,-1,x) \nn\\
& & 
          + \frac{49}{8} H(0,-1,x)
          - \zeta(2) H(0,-1,x)
          + \frac{5}{2} \zeta(2) H(0,1,x)
          + \frac{15}{8} H(1,0,x) \nn\\
& & 
          - \frac{3}{2} \zeta(2) H(1,0,x)
          + \frac{3}{2} \zeta(2) H(1,1,x)
          + H(0,0,0,x)
          - 5 H(0,-1,-1,x) \nn\\
& & 
          + \frac{5}{4} H(0,0,-1,x)
          + \frac{7}{4} H(0,1,0,x)
          - H(1,0,-1,x)
          - \frac{7}{4} H(1,0,0,x) \nn\\
& & 
          + 16 H(-1,-1,-1,x)
          - 6 H(-1,0,-1,x)
          + \frac{7}{4} H(1,1,0,x) \nn\\
& & 
          - H(0,0,0,0,x)
          + 2 H(0,0,-1,-1,x)
          - 8 H(0,-1,-1,-1,x) \nn\\
& & 
          + 3 H(0,-1,0,-1,x)
          - \frac{1}{2} H(0,0,0,-1,x)
          + \frac{3}{2} H(0,0,1,0,x) \nn\\
& & 
          - 2 H(0,1,0,-1,x)
          - \frac{3}{2} H(0,1,0,0,x)
          + \frac{3}{2} H(0,1,1,0,x) \nn\\
& & 
          + 8 H(1,0,-1,-1,x)
          + \frac{3}{2} H(1,0,0,0,x)
          + \frac{3}{2} H(1,0,1,0,x) \nn\\
& & 
          - \frac{3}{2} H(1,1,0,0,x)
          + \frac{3}{2} H(1,1,1,0,x)
          \Biggr]
       - \frac{1}{(1-x)}  \Biggl[
            6 \zeta(2)
          - \frac{9}{10} \zeta^2(2) \nn\\
& & 
          - 4 \zeta(3)
          - H(0,x)
          + \zeta(2) H(0,x)
          + 2 \zeta(3) H(0,x)
          - 2 \zeta(3) H(1,x) \nn\\
& & 
          - \frac{35}{4} H(-1,x)
          - 2 \zeta(2) H(-1,x)
          + H(0,0,x)
          - \zeta(2) H(0,0,x) \nn\\
& & 
          - H(1,0,x)
          + \zeta(2) H(1,0,x)
          - \zeta(2) H(0,1,x)
          - \zeta(2) H(1,1,x) \nn\\
& & 
          + 14 H(-1,-1,x)
          - 14 H(0,-1,x)
          - H(0,0,0,x)
          - H(0,1,0,x) \nn\\
& & 
          - 2 H(1,0,-1,x)
          + H(1,0,0,x)
          - H(1,1,0,x)
          + 16 H(0,-1,-1,x) \nn\\
& & 
          - 2 H(0,0,-1,x)
          - 16 H(-1,-1,-1,x)
          + 6 H(-1,0,-1,x) \nn\\
& & 
          + H(0,0,0,0,x)
          - H(0,0,1,0,x)
          + H(0,1,0,0,x) \nn\\
& & 
          - H(0,1,1,0,x)
          - H(1,0,0,0,x)
          - H(1,0,1,0,x) \nn\\
& & 
          + H(1,1,0,0,x)
          - H(1,1,1,0,x)
          \Biggr] \nn\\
& & 
       + \frac{1}{(1-x)} \left[ 1- \frac{1}{(1-x)} \right]
         \Biggl\{
            \frac{35}{8} \zeta(2) \! 
          + \!  \frac{9}{10} \zeta^2(2) \! 
          - \frac{7}{4} \zeta(3) \! 
          +  \! \zeta(2) H(0,1,x) \nn\\
& & 
          - \frac{35}{4} H(0,-1,x)
          - 2 \zeta(2) H(0,-1,x)
          + 14 H(0,-1,-1,x) \nn\\
& & 
          - 16 H(0,-1,-1,-1,x)
          + 6 H(0,-1,0,-1,x)
          - 2 H(0,0,0,-1,x) \nn\\
& & 
          - 2 H(0,1,0,-1,x)
          \Biggr\}  \, .
\eea

\bea
\parbox{20mm}{\begin{fmfgraph*}(15,15)
\fmfleft{i1,i2}
\fmfright{o}
\fmfforce{0.2w,0.93h}{v2}
\fmfforce{0.2w,0.07h}{v1}
\fmfforce{0.2w,0.3h}{v3}
\fmfforce{0.2w,0.7h}{v4}
\fmfforce{0.8w,0.5h}{v5}
\fmf{photon}{i1,v1}
\fmf{photon}{i2,v2}
\fmf{plain}{v5,o}
\fmf{photon}{v2,v5}
\fmf{photon}{v1,v3}
\fmf{photon}{v2,v4}
\fmf{photon}{v1,v5}
\fmf{plain,right}{v4,v3}
\fmf{plain,right}{v3,v4}
\end{fmfgraph*}} & = & \mu^{2(4-D)} 
\int {\mathfrak D}^D k_1 {\mathfrak D}^D k_2
\frac{1}{{\mathcal D}_{1}^2 
         {\mathcal D}_{4} 
	 {\mathcal D}_{5} 
	 {\mathcal D}_{13} 
	 {\mathcal D}_{14} } 
\label{6red2} \\
& = & \left( \frac{\mu^{2}}{a} \right) ^{2 \epsilon} 
\sum_{i=-2}^{2} \epsilon^{i} F^{(35)}_{i} + {\mathcal O} \left( 
\epsilon^{3} \right) , 
\eea
where:
\bea
a^2 F^{(35)}_{-2} & = &  - \frac{2}{x^2} 
       - \frac{1}{6x} 
\, , \\
a^2 F^{(35)}_{-1} & = & \frac{2}{x^2} \Bigl[
            1
          + H(0,x)
          \Bigr]
       + \frac{1}{6x} H(0,x)
\, , \\
a^2 F^{(35)}_{0} & = & - \frac{1}{x^2}   \Biggl[
            \frac{14}{3}
          - 2 \zeta(2)
          + \frac{2}{3} H(0,x)
          + 2 H(0,0,x)
          \Biggr]
       + \frac{1}{x}   \Biggl[
            \frac{26}{27}
          + \frac{1}{6} \zeta(2) \nn\\
& & 
          - \frac{4}{9} H(0,x)
          - \frac{1}{6} H(0,0,x)
          \Biggr]
       - \frac{1}{3 \sqrt{x(4-x)}}   \Biggl[
         \frac{8}{x^2} 
       - \frac{4}{x} 
       + \frac{1}{2} \Biggr] H(r,0,x)
\, , \\
a^2 F^{(35)}_{1} & = & - \frac{1}{x^2} \Biggl[
            \frac{14}{9}
          + \frac{2}{3} \zeta(2)
          - 4 \zeta(3)
          - \frac{58}{9} H(0,x)
          + 2 \zeta(2) H(0,x)
          - \frac{2}{3} H(0,0,x) \nn\\
& & 
          - 2 H(0,0,0,x)
           \Biggr]
       - \frac{1}{x}   \Biggl[
            \frac{320}{81}
          + \frac{4}{9} \zeta(2)
          - \frac{1}{3} \zeta(3)
          - \Biggl( \frac{26}{27}
          - \frac{1}{6} \zeta(2) \Biggr) H(0,x) \nn\\
& & 
          - \frac{4}{9} H(0,0,x)
          - \frac{1}{6} H(0,0,0,x)
          + \frac{1}{2} H(r,r,0,x)
          \Biggr]
- \frac{1}{3 \sqrt{x(4-x)}}   \Biggl\{ \Biggl[
         \frac{8}{x^2}   \nn\\
& & 
       - \frac{4}{x}
       + \frac{1}{2} \Biggr] \Biggl[
            \zeta(2) H(r,x)
          -  H(r,0,0,x)
          -  H(0,r,0,x)
          + 2 H(4,r,0,x) \Biggr] \nn\\
& & 
	  + \frac{4}{3} \Biggl[
         \frac{20}{x^2} 
       - \frac{1}{x} 
       - 1 \Biggr] H(r,0,x) \Biggr\}
\, , \\
a^2 F^{(35)}_{2} & = &  \frac{1}{x^2}  \Biggl[
            \frac{10}{27}
          + \frac{58}{9} \zeta(2)
          + \frac{9}{5} \zeta(2)^2
          - \frac{4}{3} \zeta(3)
          + \Biggl( \frac{10}{27} 
          + \frac{2}{3} \zeta(2) 
          - 4 \zeta(3) \Biggr) H(0,x) \nn\\
& & 
          - \Biggl( \frac{58}{9}
          - 2 \zeta(2) \Biggr) H(0,0,x)
          - \frac{2}{3} H(0,0,0,x)
          + 4 H(r,r,0,x) \nn\\
& & 
          - 2 H(0,0,0,0,x)
          \Biggr]
       + \frac{1}{x}    \Biggl[
            \frac{968}{81}
          + \frac{26}{27} \zeta(2)
          + \frac{3}{20} \zeta(2)^2
          - \frac{8}{9} \zeta(3) \nn\\
& & 
          - \Biggl( \frac{160}{81}
          - \frac{4}{9} \zeta(2)
          + \frac{1}{3} \zeta(3) \Biggr) H(0,x)
          - \Biggl( \frac{26}{27}
          - \frac{1}{6} \zeta(2) \Biggr) H(0,0,x) \nn\\
& & 
          - \frac{1}{2} \zeta(2) H(r,r,x)
          - \frac{4}{9} H(0,0,0,x)
          + \frac{4}{3} H(r,r,0,x)
          - \frac{1}{6} H(0,0,0,0,x) \nn\\
& & 
          + \frac{1}{2} H(r,r,0,0,x)
          + \frac{1}{2} H(r,0,r,0,x)
          - \frac{1}{2} H(0,r,r,0,x) \nn\\
& & 
          - H(r,4,r,0,x)
          \Biggr]
       - \frac{1}{x^2} \frac{1}{\sqrt{x(4-x)}} \Biggl[
            \frac{80}{9} \zeta(2) H(r,x)
          + \frac{16}{3} \zeta(3) H(r,x) \nn\\
& & 
          + \frac{128}{27} H(r,0,x)
          - \frac{8}{3} \zeta(2) H(r,0,x)
          - \frac{8}{3} \zeta(2) H(0,r,x)
          + \frac{16}{3} \zeta(2) H(4,r,x) \nn\\
& & 
          - \frac{80}{9} H(r,0,0,x)
          - \frac{80}{9} H(0,r,0,x)
          + \frac{160}{9} H(4,r,0,x)
          + 8 H(r,r,r,0,x) \nn\\
& & 
          + \frac{8}{3} H(r,0,0,0,x)
          + \frac{8}{3} H(0,r,0,0,x)
          + \frac{8}{3} H(0,0,r,0,x) \nn\\
& & 
          - \frac{16}{3} H(0,4,r,0,x)
          - \frac{16}{3} H(4,r,0,0,x)
          - \frac{16}{3} H(4,0,r,0,x) \nn\\
& & 
          + \frac{32}{3} H(4,4,r,0,x)
          \Biggr]
       + \frac{1}{x} \frac{1}{\sqrt{x(4-x)}}  \Biggl[
            \frac{4}{9} \zeta(2) H(r,x)
          + \frac{8}{3} \zeta(3) H(r,x) \nn\\
& & 
          + \frac{136}{27} H(r,0,x)
          - \frac{4}{3} \zeta(2) H(r,0,x)
          - \frac{4}{3} \zeta(2) H(0,r,x)
          + \frac{8}{3} \zeta(2) H(4,r,x) \nn\\
& & 
          - \frac{4}{9} H(r,0,0,x)
          - \frac{4}{9} H(0,r,0,x)
          + \frac{8}{9} H(4,r,0,x)
          + 4 H(r,r,r,0,x) \nn\\
& & 
          + \frac{4}{3} H(r,0,0,0,x)
          + \frac{4}{3} H(0,r,0,0,x)
          + \frac{4}{3} H(0,0,r,0,x) \nn\\
& & 
          - \frac{8}{3} H(0,4,r,0,x)
          - \frac{8}{3} H(4,r,0,0,x)
          - \frac{8}{3} H(4,0,r,0,x) \nn\\
& & 
          + \frac{16}{3} H(4,4,r,0,x)
          \Biggr]
       + \frac{1}{\sqrt{x(4-x)}}  \Biggl[
            \frac{4}{9} \zeta(2) H(r,x)
          - \frac{1}{3} \zeta(3) H(r,x) \nn\\
& & 
          - \frac{26}{27} H(r,0,x)
          + \frac{1}{6} \zeta(2) H(r,0,x)
          + \frac{1}{6} \zeta(2) H(0,r,x)
          - \frac{1}{3} \zeta(2) H(4,r,x) \nn\\
& & 
          - \frac{4}{9} H(r,0,0,x)
          - \frac{4}{9} H(0,r,0,x)
          + \frac{8}{9} H(4,r,0,x)
          - \frac{1}{2} H(r,r,r,0,x) \nn\\
& & 
          - \frac{1}{6} H(r,0,0,0,x)
          - \frac{1}{6} H(0,r,0,0,x)
          - \frac{1}{6} H(0,0,r,0,x) \nn\\
& & 
          + \frac{1}{3} H(0,4,r,0,x)
          + \frac{1}{3} H(4,r,0,0,x)
          + \frac{1}{3} H(4,0,r,0,x) \nn\\
& & 
          - \frac{2}{3} H(4,4,r,0,x)
          \Biggr] \, .
\eea

\bea
\parbox{20mm}{\begin{fmfgraph*}(15,15)
\fmfleft{i1,i2}
\fmfright{o}
\fmfforce{0.2w,0.93h}{v2}
\fmfforce{0.2w,0.07h}{v1}
\fmfforce{0.2w,0.3h}{v3}
\fmfforce{0.2w,0.7h}{v4}
\fmfforce{0.8w,0.5h}{v5}
\fmf{photon}{i1,v1}
\fmf{photon}{i2,v2}
\fmf{plain}{v5,o}
\fmf{photon}{v2,v5}
\fmf{plain}{v3,v1}
\fmf{plain}{v2,v4}
\fmf{photon}{v1,v5}
\fmf{photon,right}{v4,v3}
\fmf{photon,right}{v3,v4}
\end{fmfgraph*}} & = & \mu^{2(4-D)} 
\int {\mathfrak D}^D k_1 {\mathfrak D}^D k_2
\frac{1}{{\mathcal D}_{2} 
         {\mathcal D}_{3} 
	 {\mathcal D}_{4} 
	 {\mathcal D}_{5} 
	 {\mathcal D}_{12}^2 } 
\label{6red3} \\
& = & \left( \frac{\mu^{2}}{a} \right) ^{2 \epsilon} 
\sum_{i=-1}^{2} \epsilon^{i} F^{(36)}_{i} + {\mathcal O} \left( 
\epsilon^{3} \right) , 
\eea
where:
\bea
a^2 F^{(36)}_{-1} & = & - \frac{1}{(1-x)} H(0,x)
\, , \\
a^2 F^{(36)}_{0} & = & - \frac{3}{x}  H(1,0,x)
       - \frac{1}{(1-x)}  \Bigl[
            \zeta(2)
          + 2 H(0,x)
          - 2 H(0,0,x) \nn\\
& & 
          + H(1,0,x)
          \Bigr]
\, , \\
a^2 F^{(36)}_{1} & = & - \frac{1}{x} \Bigl[
            3 \zeta(2) H(1,x)
          + 6 H(1,0,x)
          + 3 H(0,1,0,x)
          - 6 H(1,0,0,x) \nn\\
& & 
          + 3 H(1,1,0,x)
          \Bigr]
       - \frac{1}{(1-x)} \Bigl[
            2 \zeta(2)
          + 3 \zeta(3)
          + 4 H(0,x)
          + \zeta(2) H(1,x) \nn\\
& & 
          - 4 H(0,0,x)
          + 2 H(1,0,x)
          + 4 H(0,0,0,x)
          + H(0,1,0,x) \nn\\
& & 
          - 2 H(1,0,0,x)
          + H(1,1,0,x)
          \Bigr]
\, , \\
a^2 F^{(36)}_{2} & = &  - \frac{1}{x} \Bigl[
            6 \zeta(2) H(1,x)
          + 9 \zeta(3) H(1,x)
          + 12 H(1,0,x)
          + 3 \zeta(2) H(0,1,x) \nn\\
& & 
          + 3 \zeta(2) H(1,1,x)
          + 6 H(0,1,0,x)
          + 6 H(1,1,0,x)
          - 12 H(1,0,0,x) \nn\\
& & 
          + 3 H(0,0,1,0,x)
          - 6 H(0,1,0,0,x)
          + 3 H(0,1,1,0,x) \nn\\
& & 
          + 12 H(1,0,0,0,x)
          + 3 H(1,0,1,0,x)
          - 6 H(1,1,0,0,x) \nn\\
& & 
          + 3 H(1,1,1,0,x)
          \Bigr]
       - \frac{1}{(1-x)} \Biggl[
            4 \zeta(2)
          + \frac{21}{5} \zeta^2(2)
          + 6 \zeta(3)
          + 8 H(0,x) \nn\\
& & 
          - 8 \zeta(3) H(0,x)
          + 2 \zeta(2) H(1,x)
          + 3 \zeta(3) H(1,x)
          - 8 H(0,0,x) \nn\\
& & 
          + 4 H(1,0,x)
          + \zeta(2) H(1,1,x)
          + \zeta(2) H(0,1,x)
          + 8 H(0,0,0,x) \nn\\
& & 
          + 2 H(0,1,0,x)
          - 4 H(1,0,0,x)
          + 2 H(1,1,0,x)
          - 8 H(0,0,0,0,x) \nn\\
& & 
          + H(0,0,1,0,x)
          - 2 H(0,1,0,0,x)
          + H(0,1,1,0,x) \nn\\
& & 
          + 4 H(1,0,0,0,x)
          + H(1,0,1,0,x)
          - 2 H(1,1,0,0,x) \nn\\
& & 
          + H(1,1,1,0,x)
          \Biggr]
\, .
\eea

The following diagram is reducible to a combination of two different 
5-denominator diagrams, as explained in Eqs. (5,6) of \cite{UgoRo}, by
simple partial fractioning.

\bea
\parbox{20mm}{\begin{fmfgraph*}(15,15)
\fmfleft{i1,i2}
\fmfright{o}
\fmfforce{0.2w,0.93h}{v2}
\fmfforce{0.2w,0.07h}{v1}
\fmfforce{0.2w,0.3h}{v3}
\fmfforce{0.2w,0.7h}{v4}
\fmfforce{0.8w,0.5h}{v5}
\fmf{photon}{i1,v1}
\fmf{photon}{i2,v2}
\fmf{plain}{v5,o}
\fmf{photon}{v2,v5}
\fmf{plain}{v3,v1}
\fmf{photon}{v2,v4}
\fmf{photon}{v1,v5}
\fmf{plain,right}{v4,v3}
\fmf{plain,right}{v3,v4}
\end{fmfgraph*}} & = & \mu^{2(4-D)} 
\int {\mathfrak D}^D k_1 {\mathfrak D}^D k_2
\frac{1}{{\mathcal D}_{1} 
         {\mathcal D}_{4} 
	 {\mathcal D}_{5} 
	 {\mathcal D}_{12} 
	 {\mathcal D}_{13} 
	 {\mathcal D}_{14} } 
\label{6red4} \\
& = & \left( \frac{\mu^{2}}{a} \right) ^{2 \epsilon} 
\sum_{i=-3}^{1} \epsilon^{i} F^{(37)}_{i} + {\mathcal O} \left( 
\epsilon^{2} \right) , 
\eea
where:
\bea
a^2 F^{(37)}_{-3} & = &  \frac{1}{x}
\, , \\
a^2 F^{(37)}_{-2} & = &  - \frac{1}{x} H(0,x)
\, , \\
a^2 F^{(37)}_{-1} & = & - \frac{1}{x} \Bigl[
            \zeta(2)
          - H(0,0,x)
          - H(1,0,x)
          \Bigr]
\, , \\
a^2 F^{(37)}_{0} & = & \frac{1}{x}  \Bigl[
            8
          - 2 \zeta(3)
          - \Bigl( 4
          - \zeta(2) \Bigr) H(0,x)
          + \Bigl( \zeta(2)
	  + \sqrt{3} H(r,0;1) \Bigr) H(1,x) \nn\\
& & 
          + 2 H(1,0,x)
          - H(0,0,0,x)
          + H(0,1,0,x)
          - H(1,0,0,x) \nn\\
& & 
          + H(1,1,0,x)
          - 3 H( 1+r,r,0,x)
          \Bigr]
       + \frac{2(4-x)}{x\sqrt{x(4-x)}} H(r,0,x)
\, , \\
a^2 F^{(37)}_{1} & = & - \frac{1}{x} \Biggl\{
            16
          + 4 \zeta(2)
          + \frac{9}{10} \zeta^2(2)
          - 2 \zeta(3) H(0,x)
          - 2 \Bigl( \zeta(2)
          + \zeta(3) \Bigr) H(1,x) \nn\\
& & 
          - \Bigl( 4 \! 
          -  \! \zeta(2) \Bigr) 
	      \Bigl( H(0,0,x)  \! 
	           +  \! H(1,0,x) \Bigr) \! 
          -  \! \zeta(2) H(0,1,x) \! 
          -  \! \zeta(2) H(1,1,x) \nn\\
& & 
          + 3 \zeta(2) H( 1+r,r,x)
          - 2 H(0,1,0,x)
          + 2 H(1,0,0,x)
          - 2 H(1,1,0,x) \nn\\
& & 
          + 6 H( 1+r,r,0,x)
          + 6 H(r,r,0,x)
          - 3 H( 1+r,r,0,0,x) \nn\\
& & 
          - 3 H( 1+r,0,r,0,x)
          + 6 H( 1+r,4,r,0,x)
          + 3 H(0, 1+r,r,0,x) \nn\\
& & 
          - H(0,0,0,0,x)
          - H(0,0,1,0,x)
          + H(0,1,0,0,x)
          - H(0,1,1,0,x) \nn\\
& & 
          + 3 H(1, 1 \! + \! r,r,0,x) \! 
          -  \! H(1,0,0,0,x) \! 
          -  \! H(1,0,1,0,x) \! 
          +  \! H(1,1,0,0,x) \nn\\
& & 
          - H(1,1,1,0,x)
       - \sqrt{3} \Bigl[
            \Bigl( 
            H(r,0,0;1)
          + 2 H(r,0;1)
          + H(4,r,0;1) \Bigr) H(1,x) \nn\\
& & 
          + H(r,0;1) H(0,1,x)
          + H(r,0;1) H(1,1,x)
          \Bigr] \Biggr\} \nn\\
& & 
       + \frac{2(4-x)}{x \sqrt{x(4-x)}} \Bigl[
            \zeta(2) H(r,x)
          - H(r,0,0,x)
          - H(0,r,0,x) \nn\\
& & 
          + 2 H(4,r,0,x)
          \Bigr]
\, .
\eea

\bea
\parbox{20mm}{\begin{fmfgraph*}(15,15)
\fmfleft{i1,i2}
\fmfright{o}
\fmfforce{0.2w,0.93h}{v2}
\fmfforce{0.2w,0.07h}{v1}
\fmfforce{0.2w,0.3h}{v3}
\fmfforce{0.2w,0.7h}{v4}
\fmfforce{0.8w,0.5h}{v5}
\fmf{photon}{i1,v1}
\fmf{photon}{i2,v2}
\fmf{plain}{v5,o}
\fmf{photon}{v2,v5}
\fmf{plain}{v3,v1}
\fmf{plain}{v2,v4}
\fmf{photon}{v1,v5}
\fmf{plain,right}{v4,v3}
\fmf{photon,right}{v3,v4}
\end{fmfgraph*}} & = & \mu^{2(4-D)} 
\int {\mathfrak D}^D k_1 {\mathfrak D}^D k_2
\frac{1}{{\mathcal D}_{3} 
         {\mathcal D}_{4} 
	 {\mathcal D}_{5} 
	 {\mathcal D}_{12}^2 
	 {\mathcal D}_{13} } 
\label{6red5} \\
& = & \left( \frac{\mu^{2}}{a} \right) ^{2 \epsilon} 
\sum_{i=-1}^{1} \epsilon^{i} F^{(38)}_{i} + {\mathcal O} \left( 
\epsilon^{2} \right) , 
\eea
where:
\bea
a^2 F^{(38)}_{-1} & = & - \frac{1}{(1-x)} H(0,x)
\, , \\
a^2 F^{(38)}_{0} & = & \frac{1}{x}  \Bigl[
            \zeta(2) H(1,x)
          + H(0,1,0,x)
          - 2 H(1,0,x)
          + H(1,1,0,x)
          \Bigr] \nn\\
& & 
       - \frac{1}{(1-x)}  \Bigl[
            \zeta(2)
          + 2 H(0,x)
          - H(0,0,x)
          + H(1,0,x)
          \Bigr]
\, , \\
a^2 F^{(38)}_{1} & = & - \frac{1}{x} \Bigl[
            \zeta(3) H(1,x)
          + 4 H(1,0,x)
          - 4 \zeta(2) H(0,1,x)
          - 4 \zeta(2) H(1,1,x) \nn\\
& & 
          - 2 H(1,0,0,x)
          - H(0,0,1,0,x)
          + H(0,1,0,0,x)
          - 4 H(0,1,1,0,x) \nn\\
& & 
          - H(1,0,1,0,x) \! 
          +  \! H(1,1,0,0,x)
          - 4 H(1,1,1,0,x)
          \Bigr]
       - \frac{1}{(1-x)} \Bigl[
            2 \zeta(2) \nn\\
& & 
          + 2 \zeta(3) \! 
          +  \! \Bigl( 4
          - \zeta(2) \Bigr) H(0,x) \! 
          +  \! \zeta(2) H(1,x)
          - 2 H(0,0,x) \! 
          +  \! 2 H(1,0,x) \nn\\
& & 
          + H(0,0,0,x)
          + H(0,1,0,x)
          - H(1,0,0,x)
          + H(1,1,0,x)
          \Bigr]
\, .
\eea

\bea
\parbox{20mm}{\begin{fmfgraph*}(15,15)
\fmfleft{i1,i2}
\fmfright{o}
\fmfforce{0.2w,0.93h}{v2}
\fmfforce{0.2w,0.07h}{v1}
\fmfforce{0.2w,0.3h}{v3}
\fmfforce{0.2w,0.7h}{v4}
\fmfforce{0.8w,0.5h}{v5}
\fmf{photon}{i1,v1}
\fmf{photon}{i2,v2}
\fmf{plain}{v5,o}
\fmf{photon}{v2,v5}
\fmf{plain}{v3,v1}
\fmf{plain}{v2,v4}
\fmf{photon}{v1,v5}
\fmf{plain,right}{v4,v3}
\fmf{plain,right}{v3,v4}
\end{fmfgraph*}} & = & \mu^{2(4-D)} 
\int {\mathfrak D}^D k_1 {\mathfrak D}^D k_2
\frac{1}{{\mathcal D}_{4} 
         {\mathcal D}_{5} 
	 {\mathcal D}_{12}^2 
	 {\mathcal D}_{13} 
	 {\mathcal D}_{14} } 
\label{6red6} \\
& = & \left( \frac{\mu^{2}}{a} \right) ^{2 \epsilon} 
\sum_{i=-1}^{1} \epsilon^{i} F^{(39)}_{i} + {\mathcal O} \left( 
\epsilon^{2} \right) , 
\eea
where:
\bea
a^2 F^{(39)}_{-1} & = & - \frac{1}{(1-x)} H(0,x)
\, , \\
a^2 F^{(39)}_{0} & = & - \frac{1}{x}  \Biggl[
            \frac{2 \sqrt{3}}{3} H(r,0;1) H(1,x)
          + 3 H(1,0,x)
          - 2 H( 1+r,r,0,x)
          \Biggr] \nn\\
& & 
       - \frac{1}{(1-x)}  \Bigl[
            \zeta(2)
	  + \sqrt{3} H(r,0;1)
          + 2 H(0,x)
          - H(0,0,x)
          + H(1,0,x)
          \Bigr] \nn\\
& & 
       + \frac{1}{\sqrt{x(4-x)}}  \Biggl[ 1 + \frac{3}{(1-x)} \Biggr]
          H(r,0,x)
\, , \\
a^2 F^{(39)}_{1} & = & - \frac{1}{x} \Biggl\{
            3 \zeta(2) H(1,x)
          + 6 H(1,0,x)
          - 3 H(r,r,0,x)
          - 2 \zeta(2) H( 1+r,r,x) \nn\\
& & 
          + 3 H(0,1,0,x)
          - 3 H(1,0,0,x)
          + 3 H(1,1,0,x)
          - 9 H( 1+r,r,0,x) \nn\\
& & 
          + 2 H( 1+r,r,0,0,x)
          + 2 H( 1+r,0,r,0,x)
          - 4 H( 1+r,4,r,0,x) \nn\\
& & 
          - 2 H(0, 1+r,r,0,x)
          - 2 H(1, 1+r,r,0,x)
       + \sqrt{3} \Biggl[
            \frac{2}{3} H(r,0;1) H(0,1,x) \nn\\
& & 
          + \frac{2}{3} H(r,0,0;1) H(1,x)
          + 3 H(r,0;1) H(1,x)
          + \frac{2}{3} H(4,r,0;1) H(1,x) \nn\\
& & 
          + \frac{2}{3} H(r,0;1) H(1,1,x)
          \Biggr] \Biggr\}
       - \frac{1}{(1-x)} \Biggl\{ 
            2 \zeta(2)
          + 2 \zeta(3) \nn\\
& & 
          + 4 H(0,x)
          - H(0,x) \zeta(2)
          + H(1,x) \zeta(2)
          - 2 H(0,0,x) \nn\\
& & 
          + 2 H(1,0,x)
          + H(0,0,0,x)
          + H(0,1,0,x)
          - H(1,0,0,x) \nn\\
& & 
          + H(1,1,0,x)
          - 3 H( 1+r,r,0,x)
       + \sqrt{3} \Bigl[
            H(r,0,0;1)
          + 2 H(r,0;1) \nn\\
& & 
          + H(r,0;1) H(1,x)
          + H(4,r,0;1)
          \Bigr] \Biggr\} \nn\\
& & 
       + \frac{1}{\sqrt{x(4-x)}} 
          \Biggl[ 1 + \frac{3}{(1-x)} \Biggr] 
           \Bigl[
            \zeta(2) H(r,x)
          + 2 H(r,0,x)
          - H(r,0,0,x) \nn\\
& & 
          - H(0,r,0,x)
          + 2 H(4,r,0,x)
           \Bigr]
\, . 
\eea

The following two diagrams are reducible to a combination of two 
different 4-denominator diagrams, as explained in Eqs. (5,6) of 
\cite{UgoRo}, by simple partial fractioning.

\bea
\parbox{20mm}{\begin{fmfgraph*}(15,15)
\fmfleft{i1,i2}
\fmfright{o}
\fmfforce{0.2w,0.9h}{v2}
\fmfforce{0.2w,0.1h}{v1}
\fmfforce{0.2w,0.5h}{v3}
\fmfforce{0.8w,0.5h}{v5}
\fmf{photon}{i1,v1}
\fmf{photon}{i2,v2}
\fmf{plain}{v5,o}
\fmf{photon,tension=0}{v2,v5}
\fmf{plain,left=90}{v3,v3}
\fmf{photon,tension=0}{v1,v5}
\fmf{plain,tension=0}{v3,v1}
\fmf{photon,tension=0}{v2,v3}
\end{fmfgraph*}} & = & \mu^{2(4-D)} 
\int {\mathfrak D}^D k_1 {\mathfrak D}^D k_2
\frac{1}{{\mathcal D}_{1} 
         {\mathcal D}_{4} 
	 {\mathcal D}_{5} 
	 {\mathcal D}_{12} 
	 {\mathcal D}_{13} } 
\label{6red7} \\
& = & \left( \frac{\mu^{2}}{a} \right) ^{2 \epsilon} 
\sum_{i=-3}^{1} \epsilon^{i} F^{(40)}_{i} + {\mathcal O} \left( 
\epsilon^{2} \right) , 
\eea
where:
\bea
a F^{(40)}_{-3} & = & - \frac{1}{x} 
\, , \\
a F^{(40)}_{-2} & = & - \frac{1}{x} \Bigl[
            1
          - H(0,x)
                     \Bigr]
\, , \\
a F^{(40)}_{-1} & = & - \frac{1}{x} \Bigl[
            1
          - \zeta(2)
          - H(0,x)
          + H(0,0,x)
          + H(1,0,x)
                     \Bigr]
\, , \\
a F^{(40)}_{0} & = & - \frac{1}{x} \Bigl[
            1
          - \zeta(2)
          - 2 \zeta(3)
          - \Bigl( 1 
          - \zeta(2) \Bigr) H(0,x)
          + \zeta(2) H(1,x)
          + H(0,0,x) \nn\\
& & 
          - H(0,0,0,x)
          + H(0,1,0,x)
          + H(1,0,x)
          - H(1,0,0,x) \nn\\
& & 
          + H(1,1,0,x)
                     \Bigr]
\, , \\
a F^{(40)}_{1} & = & - \frac{1}{x} \Biggl[
            1
          - \zeta(2)
          - \frac{9}{10} \zeta^2(2)
          - 2 \zeta(3)
          - H(0,x)
          + \Bigl( \zeta(2) 
	  + 2 \zeta(3) \Bigr) H(0,x) \nn\\
& & 
          + \Bigl( \zeta(2) 
          + 2 \zeta(3) \Bigr) H(1,x)
          + \Bigl( 1
          - \zeta(2) \Bigr) \Bigl( 
	    H(0,0,x) 
	  + H(1,0,x) \Bigr) \nn\\
& & 
          + H(0,1,x) \zeta(2)
          + \zeta(2) H(1,1,x)
          - H(0,0,0,x)
          + H(0,1,0,x) \nn\\
& & 
          - H(1,0,0,x)
          + H(1,1,0,x)
          + H(0,0,0,0,x)
          + H(0,0,1,0,x) \nn\\
& & 
          - H(0,1,0,0,x)
          + H(0,1,1,0,x)
          + H(1,0,0,0,x) \nn\\
& & 
          + H(1,0,1,0,x)
          - H(1,1,0,0,x)
          + H(1,1,1,0,x)
                     \Biggr]
\, .
\eea

\bea
\parbox{20mm}{\begin{fmfgraph*}(15,15)
\fmfleft{i1,i2}
\fmfright{o}
\fmfforce{0.2w,0.9h}{v2}
\fmfforce{0.2w,0.1h}{v1}
\fmfforce{0.2w,0.5h}{v3}
\fmfforce{0.8w,0.5h}{v5}
\fmf{photon}{i1,v1}
\fmf{photon}{i2,v2}
\fmf{plain}{v5,o}
\fmf{photon,tension=0}{v2,v5}
\fmf{plain,left=90}{v3,v3}
\fmf{photon,tension=0}{v1,v5}
\fmf{plain,tension=0}{v3,v1}
\fmf{plain,tension=0}{v2,v3}
\end{fmfgraph*}} & = & \mu^{2(4-D)} 
\int {\mathfrak D}^D k_1 {\mathfrak D}^D k_2
\frac{1}{{\mathcal D}_{4} 
         {\mathcal D}_{5} 
	 {\mathcal D}_{12}^2 
	 {\mathcal D}_{13} } 
\label{6red8} \\
& = & \left( \frac{\mu^{2}}{a} \right) ^{2 \epsilon} 
\sum_{i=-1}^{2} \epsilon^{i} F^{(41)}_{i} + {\mathcal O} \left( 
\epsilon^{3} \right) , 
\eea
where:
\bea
a F^{(41)}_{-1} & = & \frac{1}{(1-x)} H(0,x)
\, , \\
a F^{(41)}_{0} & = & \frac{2}{x} H(1,0,x)
       + \frac{1}{(1-x)}  \Bigl[
            \zeta(2)
          + H(0,x)
          - H(0,0,x)
          + H(1,0,x)
          \Bigr]
\, , \\
a F^{(41)}_{1} & = & \frac{1}{x} \Bigl[
            2 \zeta(2) H(1,x)
          + 2 H(1,0,x)
          + 2 H(0,1,0,x)
          - 2 H(1,0,0,x) \nn\\
& & 
          + 2 H(1,1,0,x)
          \Bigr]
       + \frac{1}{(1-x)} \Bigl[
            \zeta(2)
          + 2 \zeta(3)
          + H(0,x)
          - \zeta(2) H(0,x) \nn\\
& & 
          + \zeta(2) H(1,x)
          - H(0,0,x)
          + H(1,0,x)
          + H(0,0,0,x)
          + H(0,1,0,x) \nn\\
& & 
          - H(1,0,0,x)
          + H(1,1,0,x)
          \Bigr]
\, , \\
a F^{(41)}_{2} & = &  \frac{2}{x} \Bigl[
            \zeta(2) H(1,x)
          + 2 \zeta(3) H(1,x)
          + H(1,0,x)
          - \zeta(2) H(1,0,x) \nn\\
& & 
          + \zeta(2) H(1,1,x)
          + \zeta(2) H(0,1,x)
          + H(0,1,0,x)
          - H(1,0,0,x) \nn\\
& & 
          + H(1,1,0,x)
          + H(0,0,1,0,x)
          - H(0,1,0,0,x)
          + H(0,1,1,0,x) \nn\\
& & 
          + H(1,0,0,0,x)
          + H(1,0,1,0,x)
          - H(1,1,0,0,x)
          + H(1,1,1,0,x)
          \Bigr] \nn\\
& & 
       + \frac{1}{(1-x)} \Biggl[
            \zeta(2)
          + \frac{9}{10} \zeta^2(2)
          + 2 \zeta(3)
          + \Bigl( 1 
          - \zeta(2) 
          - 2 \zeta(3) \Bigr) H(0,x) \nn\\
& & 
          + \Bigl( \zeta(2)
          + 2 \zeta(3) \Bigr) H(1,x)
          - \Bigl( 1 
          - \zeta(2) \Bigr) H(0,0,x)
          + \zeta(2) H(0,1,x) \nn\\
& & 
          + H(1,0,x)
          - \zeta(2) H(1,0,x)
          + \zeta(2) H(1,1,x)
          + H(0,0,0,x) \nn\\
& & 
          + H(0,1,0,x)
          - H(1,0,0,x)
          + H(1,1,0,x)
          - H(0,0,0,0,x) \nn\\
& & 
          + H(0,0,1,0,x)
          - H(0,1,0,0,x)
          + H(0,1,1,0,x)
          + H(1,0,0,0,x) \nn\\
& & 
          + H(1,0,1,0,x)
          - H(1,1,0,0,x)
          + H(1,1,1,0,x)
          \Biggr]
\, .
\eea

%
%
%
%

\end{fmffile}

\end{document}